\pdfoutput=1
\documentclass[12pt]{iopart}
\usepackage{iopams}
\usepackage{graphicx}
\usepackage{amsfonts}
\usepackage{amssymb}
\usepackage[breaklinks=true,colorlinks=false,urlbordercolor={1 1 1}]{hyperref}
\usepackage{subfigure}
\usepackage{braket}
\DeclareGraphicsExtensions{.pdf} 

\usepackage[square,numbers,sort&compress]{natbib}

\def\bs#1{\boldsymbol{#1}}
\def\ba#1{\left(\begin{array}{#1}}
	\def\ea{\end{array}\right)}
\def\bsm{\left(\begin{smallmatrix}}
	\def\esm{\end{smallmatrix}\right)}
\def\unit#1{\, \mathrm{#1}}
\def\da{^{\dagger}}
\def\datwo{^{\dagger 2}}
\def\dathree{^{\dagger 3}}
\def\daN{^{\dagger N}}
\begin{document}

\title[Exploring the unification of QT and GR with a Bose-Einstein condensate]{Exploring the unification of quantum theory and general relativity with a Bose-Einstein condensate}

\author{Richard Howl,$^1$ Roger Penrose,$^2$ and
	Ivette Fuentes$^1$}
\address{$^1$ School of Mathematical Sciences,  University of Nottingham, University Park, Nottingham, NG7~2RD, UK}
\address{$^2$ Mathematical Institute, Andrew Wiles Building, University of Oxford, Radcliffe Observatory Quarter, Woodstock Road, 
	Oxford, OX2~6GG, UK}
\ead{richard.howl1@nottingham.ac.uk}

\begin{abstract}
	Despite almost a century's worth of study, it is still unclear how general relativity (GR) and quantum theory (QT) should be unified into a consistent theory.  The conventional approach is to retain the foundational principles of QT, such as the superposition principle, and modify GR.  This is referred to as `quantizing gravity', resulting in a theory of `quantum gravity'. The opposite approach is `gravitizing QT' where we attempt to keep the principles of GR, such as the equivalence principle, and consider how this leads to modifications of QT. What we are most lacking in understanding which route to take, if either, is experimental guidance. Here we consider using a Bose-Einstein condensate (BEC) to search for clues. In particular, we study how a single BEC in a superposition of two locations could test a gravitizing QT proposal where wavefunction collapse emerges from a unified theory as an objective process, resolving the measurement problem of QT.    Such a modification to QT due to general relativistic principles is testable near the Planck mass scale, which is much closer to  experiments than the Planck length scale where quantum,  general relativistic effects are traditionally anticipated in quantum gravity theories.    Furthermore, experimental tests of this proposal should be simpler to perform than recently suggested experiments  that would   test  the quantizing gravity approach in the Newtonian gravity limit by searching for   entanglement between two  massive systems that are both in a superposition of two locations.
\end{abstract}

\maketitle

\section{Introduction}
\subsection{Motivation and background}
At the turn of the previous century, Newtonian mechanics
was advanced by two  revolutionary theories, quantum
theory (QT) and general relativity (GR). Both theories have
transformed our view of physical phenomena, with QT accurately predicting the results of low-mass experiments, and GR correctly describing observations for large masses. However, it remains unclear how QT and GR should be unified into a consistent theory. The  conventional approach, where we `quantize gravity', is to try to treat gravity like any other force as much as possible, and  formulate a `quantum gravity' theory, such as string theory.   The opposite approach, however, is to `gravitize QT' \cite{Penrose1965,Penrose1993,Penrose1996,RoadToReality,Penrose2009,Penrose2014}. The idea  here is that GR not only provides a unique and universal role for gravity among physical processes but also, given that it cannot straightforwardly be quantized as with other physical processes, requires the current framework of QT to be modified.

An additional  motivation behind  this  alternative approach is that it can resolve the measurement problem of QT and, therefore, arguably make the theory consistent and provide a well-defined classical limit, which is not possible for a conventional quantum gravity theory.\footnote{Of course proponents of certain interpretations of QT, such as many worlds \cite{ManyWorlds} and relational quantum mechanics \cite{RQM}, would argue that there is no problem.}  Since it is around the boundary of QT and GR (i.e.\ at macroscopic mass scales) that we have not observed quantum superposition, it is possible to modify QT such that quantum state reduction (QSR) is a (non-unitary) process that objectively occurs in nature due to gravitational influences,   without impacting on the accuracy of QT in its tested domain \cite{karolyhazy1966gravitation,karolyhazy1974,karolyhazy1986,Kibble1980,DIOSI1984199,DIOSI1987377,DisoiLukas87,GRWGravity,Diosi1989,Diosi1990Relativistic,Percival503,Penrose1996}. 

This predicted modification to QT also allows for tests of a unified theory of GR and QT that are far more achievable than probing the Planck length scale where quantum, general relativistic effects are predicted to occur in conventional quantum gravity theories. This may seem, at first, unimaginable since it is often stated that the gravitational force is absolutely insignificant when compared with the electromagnetic force that dominates the normal structural and dynamical behaviour of material bodies. Thus, the influence that GR has on the quantum behaviour of physical systems must be of a different character from the mere incorporation of gravitational forces. Indeed, it is argued that there is a certain profound tension between the foundational principles of QT and GR such that we must demand a time limitation on  unitary evolution, and that this is reciprocally related to the gravitational self-energy of the difference between mass distributions of pairs of states in quantum superposition \cite{Penrose1965,Penrose1993,Penrose1996,Penrose2009,RoadToReality,Penrose2014} (compare also \cite{DIOSI1984199,DIOSI1987377}). Quantum superposition is then an approximation to a more general process of a unified theory of GR and QT, and this approximation is very good for the low-mass systems that we study in quantum experiments, but poor for the large-mass systems that we observe in our macroscopic world.
\newpage

For example, taking a sphere of mass $M$ and radius $R$ in a superposition of two locations of separation $b$, the average lifetime of the superposition state is estimated to be $\tau = 5 \hbar R / (6 G M^2)$ when $b \gg R$ and a free parameter $\gamma$ in the theory is set to $1 / (8 \pi)$ \cite{Penrose2014} (see also Section \ref{sec:UniformSpheres}).  Quantum, general relativistic effects are often considered to  occur near the Planck length scale, which is proportional to $\sqrt{\hbar G}$ and far out of reach of current particle accelerators. However, here we have the ratio of two small quantities, $\hbar / G$, coming from the square of the Planck mass, which brings the effects of a unified theory of GR and QT much closer to current experiments. 

This ratio is also found in lab-based proposals for testing whether the gravitational field obeys quantum superposition. Such  tests were first suggested by Feynman who proposed using a Stern-Gerlach experiment to place a macroscopic ball in a quantum superposition, which, in principle, could place its gravitational field in a quantum superposition, and then use a second ball and, possibly, an inverse  Stern-Gerlach to determine whether the field is in a superposition or not \cite{FeynmanQG}. This has inspired many theoretical and experimental studies  (for a review see e.g.\ \cite{DanCarneyReview}) and would test an important prediction of the quantizing gravity approach in the Newtonian  gravity limit (the testable prediction can be derived when just considering applying QT to gravity in its non-relativistic limit, where the theories would be expected to be compatible  in the conventional approach).\footnote{Although the testable prediction can be derived using gravity in its non-relativistic limit, gravity is, as far as we know, best described using GR and so it can be enlightening to consider the experiments from a  GR-like point-of-view \cite{RovelliQGExp}.}

Most recently, modern versions of Feynman's experiment have been proposed   where measuring entanglement generated between two massive spheres, both in a superposition of two locations,  would prove that the field is also in a quantum superposition \cite{BoseQGExp,VedralQGExp}.    Assuming the conventional quantizing gravity approach, the state of the two-body system would be non-separable due to the relative phases $\phi_1 = G M^2 t b / [\hbar d (d-b)]$ and $\phi_2 = -G M^2 t b/ [\hbar d (d+b)]$, where $d$ is the separation of the two systems, and it is assumed that $b \gg R$ and $d - b \gg R$ \cite{BoseQGExp}.\footnote{It is assumed that  $d -b \gg R$ so that the Casimir force can be neglected for realistic masses \cite{BoseQGExp,CPInteraction}:
	\begin{equation}
	d -b \geq \Big[\frac{23 \hbar c}{0.1 \times 4 \pi G M^2} \Big(\frac{\epsilon_r-1}{\epsilon_r+2} \Big)^2 \Big]^{1/6} R 
	\end{equation} 
	where $\epsilon_r$ is the  relative permittivity of the material.} For the proposed experimental parameters  $M \sim 10^{-14} \unit{kg}$, $d \sim 200 \unit{\mu m}$, $b \sim 250\unit{\mu m}$, $R \sim 1 \unit{\mu m}$ and an interaction time of $t \approx 2.5\unit{s}$, the sum of the phases is $\mathcal{O}(1)$ and the entanglement is considered measurable \cite{BoseQGExp}. This test of the quantum superposition of gravity appears far more achievable than those based on how the position of one test mass is affected by the other due to quantum, gravitational interactions \cite{BoseQGExp}. However, for the above experimental parameters, gravitationally-induced quantum state reduction (GQSR) is predicted to occur, on average, around $0.01\unit{s}$ in this experiment, and so no entanglement would be observed if GQSR takes place. 
\newpage

This does not necessarily mean that entanglement cannot be generated in this two-body system with the GQSR proposal considered here, but it would be very challenging to observe: either competing effects must be reduced so that shorter times than $0.01\unit{s}$ can be probed, or the mass of each system must be increased by over an order of magnitude.\footnote{Note that $0.01\unit{s}$ is the average time that it will take for either of the massive superposition states to decay. Therefore, since this is an average time, there is still a probability that  entanglement could be measured  after $2.5\unit{s}$. In Section \ref{sec:PenroseCollapse}, we consider that GQSR is a Poisson process, in which case there would be an absolutely imperceptible chance of observing entanglement here.}  In the GQSR proposal, there is nothing necessarily preventing a gravitational field from being in a quantum superposition, only that there must, at least, be a time limitation for this that is dependent on the mass distribution of the system. This is in contrast to other proposed theories, such as a fundamental semi-classical gravity theory, where gravity is necessarily a classical effect, and  no entanglement can ever be generated \cite{VlatkoClassical,DanCarneyReview}.

The fact that GQSR occurs, on average, at a much earlier time scale than that required to see entanglement in the Feynman-inspired experiments, illustrates that GQSR could be observed with much lighter systems. Indeed, the mass could be reduced by an order of magnitude in these experiments. Furthermore, experiments of GQSR would only require \emph{one} massive system to be in a superposition of two locations rather than the \emph{two} systems for the above experiments. Effects such as the Casimir force between two systems clearly no longer have to be considered, drastically improving the experimental feasibility. Additionally, the distance between the superposition states can also be shorter in tests of GQSR since the average superposition lifetime  has a non-trivial dependence on $b$ and $R$ \cite{Penrose2014} (see \eref{eq:EGUniformSphere}) such that, for example, it does not change significantly from $b = 2R$ to $b \gg R$, in contrast to the gravitational potential that changes as the reciprocal of  the distance  between  two spherical systems. 

Evidence of GQSR would rule in the gravitizing QT approach and thus rule out the conventional quantizing gravity approach (since QT must be modified). In contrast, if entanglement is observed in the Feynman-inspired experiments then, although this would be a remarkable and significant result, this does not rule out the gravitizing QT approach since QT could still be modified, for example via a GQSR at some other scale such as the Planck length scale.  This is because  the  tested effect derives in the non-relativistic limit of quantum gravity, and so arguably the experiments cannot provide the specifics of how GR should be modified in order to be consistent with QT in the conventional quantizing gravity approach (see  \cite{RovelliDiscreteTime} for a possibility of extending the experiments with much heavier masses to achieve this). The GQSR process considered here, however, has been primarily motivated from conflicts between GR and QT \cite{DIOSI1987377,Penrose1996}. 

If, on the other hand, entanglement were not observed in the Feynman-inspired experiments then this would suggest that we should adopt the gravitizing QT approach. However, as illustrated above, much simpler tests, such as those of the GQSR proposal considered here, would already be able to provide evidence of this approach. Therefore, tests of GQSR based on the quantum superposition of a single massive system could be performed first and, if no deviations from QT are found, we could then consider predictions of the  conventional quantizing gravity approach, such as searching for entanglement between two massive quantum systems.

\subsection{Experimental approaches}
In general, GQSR  could be experimentally demonstrated   by preparing a superposition state of a single system that is massive enough to produce a non-negligible gravitational field while being sufficiently small enough for control in the quantum regime. For example, an optomechanical system could be used where a tiny mirror consisting of $10^{14}$ atoms is placed in a spatial superposition due to interactions with photons that are travelling through an interferometer \cite{Penrose2000}. If the mirror stays coherent  then there is quantum interference at the output, whereas, if the mirror state reduces then  so does the photon's, and there is a classical output. This type of experiment has been constructed using a Michelson interferometer with optical cavities \cite{MirrorSuperposition,TrampolineResonators,SideBandCooling}. However, the separation of the mirror superposition  can reach, at most, about one picometre, which may not be  enough to observe GQSR \cite{Adler2007}.

Another possibility is to send the massive system itself through a (matter-wave) interferometer. Typically these experiments use nano or micrometre sized spheres, rods or discs, which we will generally refer to as  nano/micro-objects, that are synthesized from metals or conducting materials and are cooled  such that their centre-of-mass motion approaches its quantum ground state. For example, in \cite{SuperconductingMicrosphere}  it is argued that a superconducting micro-sphere of mass $10^{13}\unit{a.m.u.} $ could be prepared in a spatial superposition of the order of half-a-micrometer in the near future.

Such matter-wave experiments could also be performed, in principle, using  ultracold atoms, and recently it has been suggested that Bose-Einstein condensates (BECs) confined to a double-well potential  would be effective systems for studying GQSR \cite{FuentesPenrose2018}.\footnote{Ultracold atoms have also  been considered recently for distinguishing between the conventional quantizing gravity approach and theories of classical gravity \cite{Haine}.}  BECs  are the coldest objects in the Universe that we know of, and experiments offer high-levels of control, such as the ability to tune the effective interaction strength between the atoms.  To date, coherent superposition states of a BEC consisting of around $10^5$ atoms over a distance of $54\,\mathrm{cm}$ and with decoherence times of around $1\,\mathrm{s}$ have been achieved \cite{KasevichHalfMetre}. It has been argued that BECs are less promising systems than nano/micro-object experiments for testing objective QSR since they will only demonstrate single-particle effects. However,  BECs often have non-negligible, effective interactions strengths and thus display effects that cannot be characterized with  a single-particle wavefunction. For example, when constrained to a box trap, BECs can have an effectively constant density \cite{UniformBECTrap} and, to generate macroscopic superposition states, the interactions generally play a significant role \cite{CatStatesBECs1998,CatsAttractiveBECs,CatStatesBECs,NOONRepulsive}. 

In all these experiments, the average lifetime of GQSR needs to be short enough to be seen above environmental decoherence. The most mathematically straightforward approach to decreasing the lifetime is to increase the mass of the system, which is a significant experimental challenge. However, different shapes of objects will also change the gravitational self-energy, suggesting an alternative approach to decreasing the lifetime that could be simpler to implement in the laboratory. As far as we are aware, only the quantum superposition lifetime of a uniform sphere has  been considered, with the exception of a uniform cube when the displacement is only very small \cite{Adler2007}.\footnote{See also \cite{EGOtherShapes} for an attempt of large separations of uniform cylinders and plates, but which were implicitly assumed to be of infinite extent.} In Section \ref{sec:UniformSpheroid} we generalize the spherical case to the quantum superposition of uniform spheroids, which can be generated in nano/micro-object experiments and approximates rods and discs at high values of ellipticity; finding that the associated time-scale of GQSR can be shorter for certain spheroidal configurations. Furthermore, we predict how this time-scale changes with the ellipticity and size of the superposition, potentially allowing for distinguishably from other models of objective QSR.   

In contrast to typical nano/micro-object experiments, BECs generally have non-uniform mass distributions, which are set by the trapping potential that constrains the BEC system, together with the atom-atom interactions. Often a Gaussian or quadratic density profile is assumed, which may also be applicable to other, non-BEC systems. An harmonic trap, which is the most common type of trap, can  generate spherical and spheroidal BECs, and prolate spheroidal (cigar-shaped) BECs are often used. We calculate the rate of GQSR for spherical and spheroidal BECs (with Gaussian and quadratic density profiles) and, conjecturing that GQSR follows a Poisson process, we also consider what experimental parameters  are required to observe GQSR over  prominent channels of environmental decoherence in BEC experiments (extending the analysis of \cite{FuentesPenrose2018}).

\subsection{Outline}
The rest of this paper is organized as follows: in Section \ref{sec:PenroseCollapse}, we provide a  derivation of  GQSR by considering a certain conflict between the superposition principle of QT and the equivalence principle of GR. We also review the GQSR process for   displaced, uniform spherical mass distributions (Section \ref{sec:UniformSpheres}) and generalize this to displaced, uniform spheroidal mass distributions (Section \ref{sec:UniformSpheroid}), which  can be generated in nano/micro-object experiments. In Section \ref{sec:BECTests}, we consider testing GQSR using a BEC, calculating the rate of GQSR for displaced, non-uniform BEC spheres and spheroids, and comparing this to prominent channels of environmental decoherence. Finally, in Section \ref{sec:Conclusions}, we summarize our findings and consider future prospects.

\newpage

\setcounter{footnote}{0}

\section{Gravitationally-induced state reduction from conflicts between general relativity and quantum theory} \label{sec:PenroseCollapse}

Here we consider how GQSR can arise due to a  conflict between the superposition principle of QT and  equivalence principle of GR.\footnote{Alternative approaches for identifying conflicts between QT and GR include, for example, how principles of GR might affect the uncertainty relation of QT (see e.g.\ \cite{PhysRev.135.B849,GUP}) and how measuring a classical gravitational field using an apparatus obeying QT could lead to a universal bound on the optimal precision  of the measurement \cite{DIOSI1987377}.} More detail can be found in \cite{RoadToReality,Penrose2014,FuentesPenrose2018}. Also, see \cite{Penrose1996,FuentesPenrose2018} for how the same proposed state reduction can be derived using the principle of covariance rather than the  principle of equivalence. 

\begin{figure*}[t!]
	\begin{center}
		\includegraphics[width=0.6\textwidth]{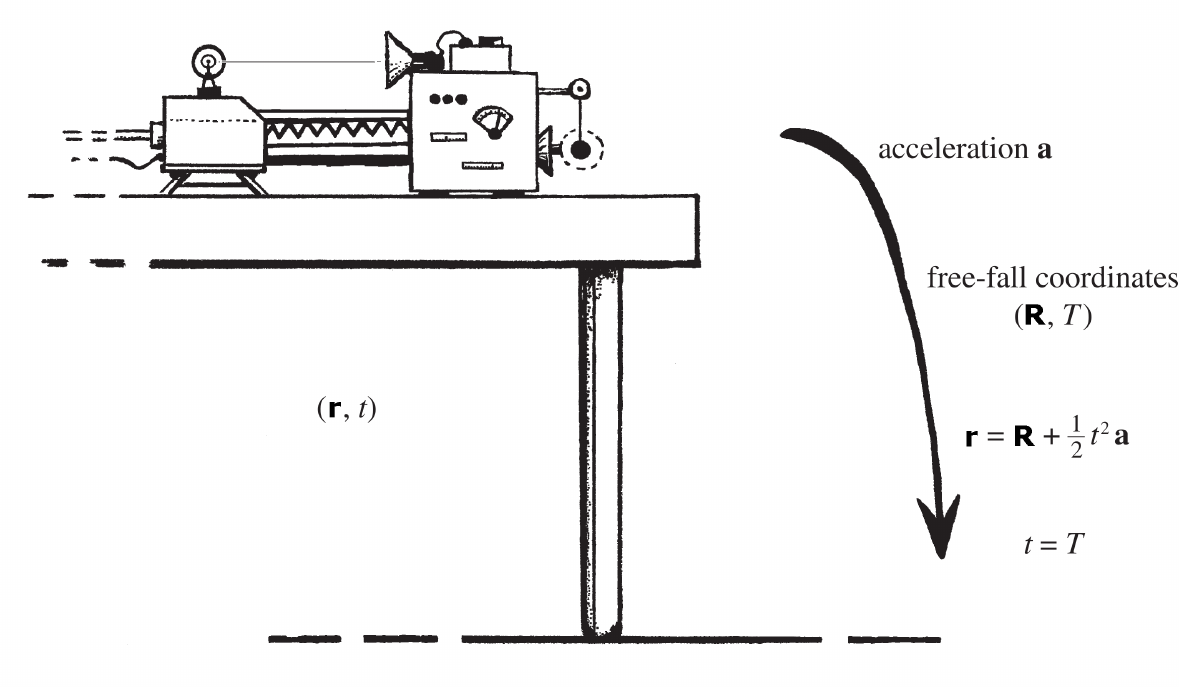}
	\end{center}
	\vspace{-0.5cm}
	\caption{An imagined quantum experiment for which the Earth's gravitational field is to be taken into consideration. The Newtonian perspective uses the laboratory coordinates $(\bs{r},t)$, while the Einsteinian perspective uses the free-fall coordinates $(\bs{R},T)$.} \label{fig:QMExp}
\end{figure*}

Let us first consider a simple situation of a tabletop quantum experiment where it is required that the Earth's gravitational field is to be taken into consideration (see Figure \ref{fig:QMExp}). There are basically two different ways to incorporate the Earth's field in this experiment (which is to be considered as constant, both spatially and temporally, and to be treated non-relativistically). The first, the \emph{Newtonian perspective}, would simply be to incorporate a term in the Hamiltonian representing the Newtonian potential (this being the normal prescription that most physicists would adopt), and use standard Cartesian coordinates $(x,y,z,t)$, or rather $(\bs{r},t)$ in 3-vector form. The second, the \emph{Einsteinian perspective}, would be to adopt a freely falling reference system $(\bs{R},T)$, in accordance with which the Earth's gravitational field \emph{vanishes}. The relation between the two is:

\begin{equation}
\bs{R} = \bs{r} - \frac{1}{2} t^2 \bs{a},\hspace{1cm} T = t,
\end{equation}

where the constant 3-vector $\bs{a}$ is the acceleration due to the Earth's gravity. We denote the wavefunction in the $(\bs{r},t)$ system, using the Newtonian perspective, by $\psi$, whereas for the $(\bs{R},T)$ system, using the Einsteinian perspective, we use $\Psi$. For a free particle of mass $m$, we have, according to the Newtonian perspective, the Schr\"{o}dinger equation

\begin{equation}
i \hbar \frac{\partial \psi}{\partial t} = - \frac{\hbar^2}{2m} \nabla^2 \psi - m \bs{r}.\bs{a} \psi,
\end{equation}

whereas, according to the Einsteinian perspective

\begin{equation}
i \hbar \frac{\partial \Psi}{\partial t} = - \frac{\hbar^2}{2m} \nabla^2 \Psi,
\end{equation}

the operator  $\nabla^2$ being the same in both coordinate systems. To get consistency between the two perspectives, we need to relate $\psi$ to $\Psi$ by a phase factor \cite{CoherenceNeutron,NonRelCOW,Rosu1999,RoadToReality,Penrose2009,Penrose2014}

\begin{equation}
\Psi=e^{i \frac{m}{\hbar} (\frac{1}{6} t^3 a^2 - t \bs{r}.\bs{a})} \psi.
\end{equation}

For a quantum experiment involving many particles of total mass $\bar{m}$ and centre of mass $\bar{\bs{r}}$ (or $\bar{\bs{R}}$ in the Einstein system), this generalizes to 

\begin{equation}\label{key}
\Psi=e^{i \frac{\bar{m}}{\hbar} (\frac{1}{6} t^3 a^2 - t \bar{\bs{r}}.\bs{a})} \psi.
\end{equation}

Since the difference between the Newtonian and Einsteinian perspectives is merely a phase factor, one might form the opinion that it makes no difference which perspective is adopted. Indeed, the famous experiment by Colella, Overhauser and Werner \cite{COW} (see also \cite{AfterCOW,COWPhase,NeutronInterferometryBook}) performed originally in 1975, did provide some impressive confirmation of the consistency (up to a certain point) of QT with Einstein's principle of equivalence.

However, it is important to note that the phase factor that is encountered here is not at all harmless, as it contains the time-dependence involved in the term

\begin{equation}\label{key}
\frac{1}{6} t^3 a^2,
\end{equation}

in the exponent, which affects the splitting of field amplitudes into positive and negative frequencies. In other words, the Einsteinian and Newtonian wavefunctions belong to different Hilbert spaces, corresponding to different quantum field theoretic vacua. In fact, this situation is basically just a limiting case of the familiar relativistic (Fulling-Davies-)Unruh effect \cite{UnruhEffectFulling,UnruhEffectDavies,UnruhEffect,CoherenceNeutron,PenroseCollapseClock,NonRelCOW}, where in a uniformly accelerating (Rindler) reference frame, we get a non-trivial thermal vacuum of temperature

\begin{equation}\label{key}
\frac{\hbar a}{2 \pi k c},
\end{equation}

where $a$ is the magnitude of acceleration, $k$ being Boltzmann's constant and $c$, the speed of light. In the current situation, we are considering the Newtonian approximation $c\rightarrow \infty$, so the temperature goes to zero. Nevertheless, as a direct calculation shows, the Unruh vacuum actually goes over to the Einsteinian one in the limit $c\rightarrow \infty$, in agreement with what has been shown above, and is thus still different from the Newtonian one even though the temperature difference goes to zero in this  limit.

At this stage, we could still argue that it makes no difference whether the Newtonian or Einsteinian perspective is adopted, so long as one sticks consistently to one perspective or the other overall (since the formalism is maintained within a single Hilbert space). However, the situation becomes \emph{radically different} when one considers the gravitating body, in this example the Earth, to be in a \emph{quantum superposition} between pairs of states in which the gravitational fields differ. If we were to adopt the Newtonian perspective for our quantum experiment then we would encounter no problem with the formalism of QT, the standard linear framework of unitary evolution applying as well to the Newtonian gravitational field as it does to electromagnetism, or to any other standard field theory of physics. But it is another matter altogether if we insist on adopting the Einsteinian perspective. Our standpoint here is that, owing to the enormous support that observations have now provided for GR in macroscopic physics, one must try to respect the Einsteinian perspective as far as possible, in quantum systems, especially in view of the foundational role that the principle of equivalence has for GR (see \cite{RoadToReality,Penrose2009,Penrose2014}).

Let us now replace the Earth with a small rock and try to imagine the quantum description of the physics taking place in some small region in the neighbourhood of the rock, where we consider that the rock can persist for some while in a superposition of two different locations, and we label the respective states as $\ket{L}$ and $\ket{R}$. We are not now trying to compare the Einsteinian perspective with a Newtonian one, since our point of view will be that the latter is not relevant to our current purposes, as we regard the Einsteinian perspective to be closer to nature's ways. Instead, we attempt to adopt an Einsteinian perspective for a quantum experiment in the neighbourhood of the rock that is in a quantum superposition of two locations, $\alpha \ket{L} + \beta \ket{R}$.  What we now have to contend with is a superposition of two different Einsteinian wavefunctions for the quantum experiment, each inhabiting a Hilbert space that will turn out to be incompatible with the other.

However, the preceding discussion does not hold exactly, because for each of the two components of the superposition of rock locations $\ket{L}$ and $\ket{R}$, the gravitational field of the rock is not completely uniform. Nevertheless, we shall consider, first, that we are examining the nature of the wavefunction in some spatial region that is small by comparison with the rock itself, so that we can assume that the gravitational field of each component of the superposition can be taken to be spatially uniform to a good approximation. Adopting the Einsteinian perspective, what we are now confronted with is the fact that the gravitational acceleration fields for the two rock locations will be different from each other, so that the difference between these local acceleration fields $\bs{a}$ and $\bs{a}'$ will lead to a difference between the Einsteinian vacuum for each rock location. In the neighbourhood of each spatial point, there will be a phase difference between the two states of our quantum experiment that are in superposition:

\begin{equation}
e^{i \frac{\bar{m}}{\hbar} (\frac{1}{6} t^3 (\bs{a}-\bs{a}')^2 - t \bar{\bs{r}}.(\bs{a}-\bs{a}'))}.
\end{equation}

Although the presence of the  $\frac{1}{6} t^3 (\bs{a}-\bs{a}')^2$ term tells us, strictly speaking, that when $\bs{a} \neq \bs{a}'$, the superposition is illegal (the states belonging to different Hilbert spaces), we adopt the view that this incompatibility takes some time to cause trouble (as would eventually become manifest in divergent scalar products, etc.). The idea is that in order to resolve this incompatibility of Hilbert spaces, the superposed state must eventually reduce to one alternative or the other, this incompatibility taking a while to build up. We compare the troublesome term $\frac{1}{6} t^3 (\bs{a}-\bs{a}')^2$ with the harmless one $ t \bar{\bs{r}}.(\bs{a}-\bs{a}')$, the latter ($\times \bar{m}/\hbar$) being linear in $t$ and therefore not altering the vacuum but, in effect, just corresponds to incorporating the Newtonian gravitational potential term into the Hamiltonian. We take the view that so long as $t$ is small enough, the trouble arising from $t^3$ remains insignificantly small, where the measure of this smallness comes from comparing  $\frac{1}{6} t^3 (\bs{a}-\bs{a}')^2$ with the harmless $ t \bar{\bs{r}}.(\bs{a}-\bs{a}')$. Thus, we take the coefficient  $\frac{1}{6} t^3 (\bs{a}-\bs{a}')^2$ as some kind of measure of the propensity for the state to reduce, as a contribution to the overall reduction process. To get our measure of total \textit{error}, or ``uncertainty'' $\Delta$, we integrate this expression over the whole of (coordinate) 3-space:
\begin{eqnarray}\label{eq:aa'}
\Delta &:= \gamma \int (\bs{a} - \bs{a}')^2 d^3 \bs{r},\\
&=\gamma  \int (\bs{\nabla} \phi - \bs{\nabla} \phi')^2 d^3 \bs{r},\\
&=\gamma   \int [\bs{\nabla} (\phi - \phi')]. [\bs{\nabla} \phi - \bs{\nabla} \phi']d^3 \bs{r},\\
&= - \gamma  \int (\phi - \phi') (\nabla^2 \phi - \nabla^2 \phi') d^3 \bs{r},
\end{eqnarray}

(assuming appropriate falloff at spatial infinity), where $\gamma$ is some positive constant, and   $\phi$ and $ \phi'$  are the respective gravitational potentials for the  states of the rock,  where we are  adopting a Newtonian approximation for estimating the required error ($\bs{a}=-\bs{\nabla} \phi$ and $\bs{a}'= -\bs{\nabla} \phi'$). By Poisson's formula ($G$ being Newton's gravitational constant)

\begin{equation}
\nabla^2 \phi = 4 \pi G \rho,
\end{equation}

we get

\begin{equation} \label{eq:Delta}
\Delta = - 4 \pi G \gamma \int (\phi - \phi') (\rho - \rho') d^3 \bs{r},
\end{equation}

where $\rho$ and $\rho'$ are the respective mass densities of the two states, and we shall take these mass densities in the sense of expectation values for the respective quantum states. Using the formula

\begin{equation} \label{eq:phi}
\phi(\bs{r}) \equiv - G \int \frac{\rho(\bs{r'})}{|\bs{r} - \bs{r'}|} d^3 \bs{r'},
\end{equation}

we obtain \cite{Penrose1996}:

\begin{equation} \label{eq:Deltarho}
\Delta = 4 \pi G^2 \gamma \int \int \frac{[\rho(\bs{r}) - \rho'(\bs{r})] [\rho(\bs{r'}) - \rho'(\bs{r'})]}{|\bs{r}-\bs{r'}|} d^3 \bs{r} d^3 \bs{r'} .
\end{equation}

Defining $E_G := \Delta / G$, we have a quantity that is proportional to (depending on the value of $\gamma$) the gravitational self-energy of the difference between the mass distributions of each of the two states
\begin{eqnarray} \label{eq:EG1}
E_G &= 4 \pi \gamma \int (\phi - \phi') (\rho' - \rho) d^3 \bs{r}\\  \label{eq:EG2}
&= 4 \pi G \gamma \int \int \frac{[\rho(\bs{r}) - \rho'(\bs{r})] [\rho(\bs{r'}) - \rho'(\bs{r'})]}{|\bs{r}-\bs{r'}|} d^3 \bs{r} d^3 \bs{r'}.
\end{eqnarray}

The quantity  $\Delta$ can be considered as a measure of a limitation to regarding the quantum superposition of the rock $\alpha \ket{L} + \beta \ket{R}$ as being a stationary state, in accordance with principles of GR. Thus, we may take it to be a reasonable inference from general-relativistic principles to regard $\Delta^{-1}$ as providing some kind of measure of a limit to the length of time that the superposition might persist, the shorter that this time-scale should presumably be, the larger the value $\Delta$ is found to have. This conclusion comes  from considerations of  GR, as applied simply to the notion of a quantum superposition of states, no consideration of quantum dynamics being involved except for the quantum notion of stationarity. Moreover, no actual measure of a time-scale for a ``lifetime'' of the superposition has yet been provided by these considerations.

However, a significant clue is provided by Heisenberg's time-energy uncertainty principle, where we note that the quantity $E_G = \Delta /G $ is an energy. In QT, the lifetime of an unstable particle is reciprocally related to an energy uncertainty, which can be regarded as a manifestation of Heisenberg's time-energy uncertainty principle. In a similar way, we propose that $E_G$ should be treated as a \emph{fundamental}  uncertainty in the energy of the superposition $\alpha \ket{L} + \beta \ket{R}$. We then take the view that the  ``energy uncertainty'' $E_G$ is reciprocally related to a lifetime of this superposition between the states $\ket{L}$ and $\ket{R}$, and we can, therefore, regard the macroscopic superposition  as having an average lifetime $\tau$ that is roughly given by
\begin{equation} \label{eq:ETuncertainty}
\tau \sim \frac{\hbar}{E_G},
\end{equation}
upon which time (on average) the state  $\alpha \ket{L} + \beta \ket{R}$ spontaneously ``decays'' into one or the other of $\ket{L}$ or $\ket{R}$. This decay process cannot be derived from considerations of QT alone, and instead we are assuming the invocation of a higher theory from which QT and GR are limiting cases. The energy uncertainty in  \eref{eq:ETuncertainty} arises due to a conflict between the general-relativistic and quantum principles that are being appealed to in relation to the description of stationary gravitational fields in quantum linear superposition, and there would be no need for such an  energy uncertainty if we had just assumed a Newtonian description of gravity without the philosophy of GR.  Similarly, if we had considered a contribution from the electromagnetic interaction of a (say charged) rock in addition to its gravitational field, then there would be no conflict with QT from electromagnetic effects (there being no equivalence principle for electromagnetism) and we would not be led to consider any energy uncertainty from electromagnetic effects contributing to the decay of the state to either  $\ket{L}$ or $\ket{R}$.

Taking the analogy with particle decay further, we could assume that the probability of, a presumed spontaneous, state reduction is an exponential function of time $t$:
\begin{eqnarray} \label{eq:Ps}
P_s(t) = e^{- t / \tau} = e^{- E_G t / \hbar},\\ \label{eq:Pd}
P_d(t) = 1- e^{- t / \tau} = 1- e^{- E_G t / \hbar},
\end{eqnarray}

where $P_s(t)$ and $P_d(t)$ are, respectively, the probability of survival and decay of the superposition state.  Here we are assuming, as with particle decay, that the  decay is memoryless,  which would seem the simplest assumption for describing the decay process given that there is not, at present, a full theory. Equation \eref{eq:Pd}  illustrates that it should not be necessary to wait for a time $\tau = E_G / \hbar$ in order to observe collapse of the wavefunction, and we can estimate how often collapse will occur at a given time $t$ without  having to appeal to a full dynamical model.

A few clarifying remarks should be made on our above derivation of $E_G$. We have considered  a rock to be in a superposition of two locations (similar to Schr\"{o}dinger's cat being in two locations). However, rather than a rock (which we have assumed to be a continuous mass distribution), we could  have  considered a point-like object in the superposition of two locations, which,  naively, would be a superposition of delta-function wavefunctions in position space. This would lead to an infinite value for $E_G$. The problem is that a delta function is not a stationary solution to the Schr\"{o}dinger equation since the position wavefunction would instantly spread out (there is infinite uncertainty in momentum).  However, the stationary solution of the Schr\"{o}dinger equation would be that where the state is spread out over the Universe (there is infinite uncertainty in position), which is clearly not satisfactory either. One might imagine that, in a full theory of GQSR, a spreading state keeps reducing by GQSR. For now, a systematic procedure would be to modify the Schr\"{o}dinger equation and use the Schr\"{o}dinger-Newton equation \cite{penrose1998quantum} to obtain stationary states \cite{Penrose2014,Moroz_1998}.  For a point-like object, the stationary solution is then a `smeared-out' delta function, and the position is no longer defined at a point.  To calculate $E_G$ for this state, we could think of each `smeared-out' delta function as representing point-like objects in a superposition of continuous positions with differing weights, and follow the procedure outlined with equations 
\eref{eq:aa'}-\eref{eq:EG2}.  In the continuous limit, this results in the same expression for $E_G$ as before, \eref{eq:EG2}, with $\rho(\bs{r})$ and $\rho'(\bs{r})$ representing the average mass density of the defined stationary states, which would be   $m |\psi(\bs{r})|^2$  and  $m |\psi'(\bs{r})|^2$  in our case, where $\psi$ and $\psi'$ are the normalised wavefunctions of the stationary states, and $m$ is the mass of the object.  The superposition state should then decay into one of these stationary states as outlined in \eref{eq:ETuncertainty}-\eref{eq:Pd}. For our rock,  we have assumed that the stationary states $\ket{L}$ and $\ket{R}$ should be close to the `classical' rock states since we do not see rocks spreading out across the Universe.\footnote{Rather than using the Schr\"{o}dinger-Newton equation to obtain stationary states, we could adopt the procedure of factoring out, and ignoring, the centre of mass and only considering relative distances \cite{Penrose2014}.}  The mass profile $\rho(\bs{r})$ of the stationary state of the rock should then be close to its classical mass distribution, which we have approximated as a continuous mass distribution.

The above calculation of $E_G$, \eref{eq:aa'}-\eref{eq:EG2}, has been carried out entirely within the framework of Newtonian mechanics since  we are considering the masses involved as being rather small  and moving  slowly, so that general-relativistic corrections can be ignored to a very good approximation. We can, therefore, also just use Schr\"{o}dinger's equation rather than, for example, the full framework of quantum field theory. However, the principles of GR still apply to gravity, such as the equivalence principle, and the fact that $E_G$ is to be regarded as an energy uncertainty is coming from considerations of general-relativistic principles and QT. The use of Newtonian mechanics for calculating an expression for $E_G$, while nevertheless retaining much of the basic philosophy of Einstein's theory, is perhaps most clearly expressed with the Newton-Cartan theory of gravity \cite{RoadToReality}.

\subsection{Gravitationally-induced state reduction for uniform spherical mass distributions} \label{sec:UniformSpheres}

\begin{figure*}[t!]
	\begin{center}
		\subfigure{%
			\put(120,140){(a)}
			\label{fig:EGSphere}
			\includegraphics[width=0.465\textwidth]{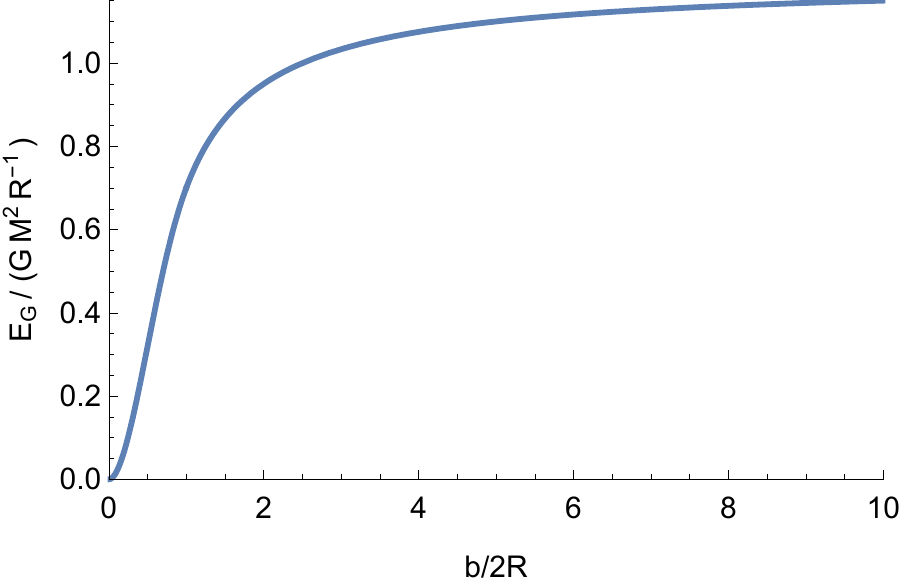}            
		}%
		\hspace{0.3cm}
		\subfigure{%
			\put(120,140){(b)}
			\label{fig:EGRateOfChangeSphere}
			\includegraphics[width=0.465\textwidth]{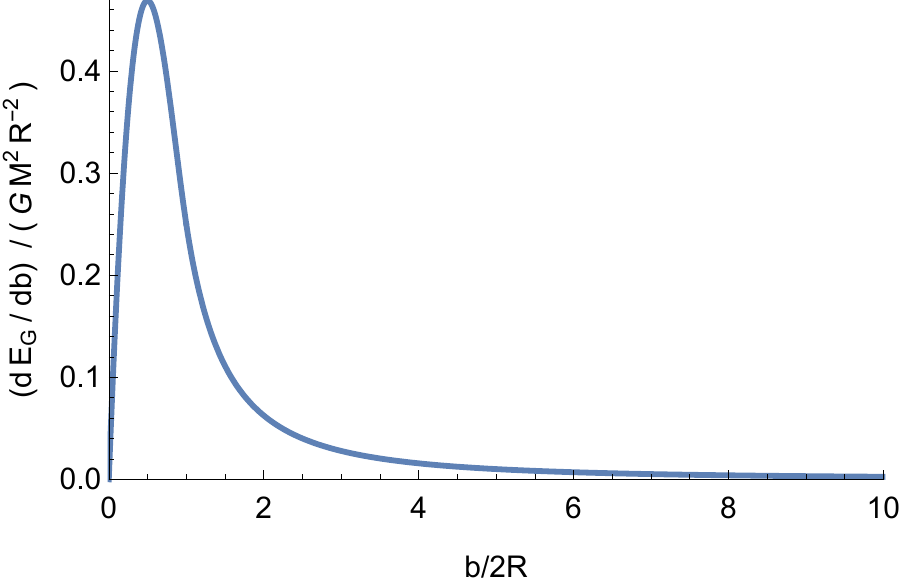}
		}\\ 
	\end{center}
	\vspace{-0.5cm}
	\caption{On the left is the gravitational self-energy of the difference $E_G$ between  displaced uniform spherical mass distributions (divided by $ G M^2 R^{-1}$) against $b/(2R)$, where $R$ is the radius of the sphere, $M$ is the mass, $G$ is the gravitational constant, and $b$ is the distance between the centres of the sphere states. On the right is $d E_G / d b$ (divided by $ G M^2 R^{-2}$) against $b/(2R)$  for the same uniform sphere.}
	\label{fig:Sphere}
\end{figure*}  

In order to get an impression of the role of $E_G$, we can first think of the case of a solid spherical ball of radius $R$, made from some uniform massive material of total mass $M$, where the ball is in a superposition of two locations, differing by a distance $b$. The quantity $E_G$ in this case is (see, for example, \cite{Penrose2014} and \ref{app:UniformSphere}):

\begin{equation} \label{eq:EGUniformSphere}
E_G = \cases{ \frac{6G M^2}{5R} \Big(  \frac{5}{3} \lambda^2 - \frac{5}{4} \lambda^3 + \frac{1}{6} \lambda^5\Big)  &if  $0 \leq \lambda \leq 1$,  \\
	\frac{6G M^2}{5R} \Big( 1 -  \frac{5}{12 \lambda}\Big) &if  $\lambda \geq 1$,}
\end{equation}
where  $\lambda := b / (2R)$, and we have taken $\gamma = 1 / (8 \pi)$ in \eref{eq:EG1}.  See Figure \ref{fig:EGSphere} for an illustration of $E_G$, and Figure \ref{fig:EGRateOfChangeSphere} for the rate of change of $E_G$, with separation $b$.  The only point of particular relevance is the fact that, for a displacement such that the two spheres are touching,  the value of $E_G$ is already nearly $\frac{2}{3}$ of the value it would reach for a displacement all the way out to infinity. Thus, for a uniformly solid body like this, we do not gain much by considering displacements in which the two instances of the body are moved apart by a distance of more than roughly its own diameter.

\newpage

\subsection{Gravitationally-induced state reduction for uniform spheroidal mass distributions} \label{sec:UniformSpheroid}

The above case of two uniform spherical mass distributions is that which is generically considered in the literature, apart from a study of two uniform cubes at only very small displacement $b$ \cite{Adler2007}.\footnote{See also \cite{EGOtherShapes} for an attempt of large separations of uniform cylinders and plates, but which were implicitly assumed to be of infinite extent.} In this section,  we generalize  to uniform spheroidal mass distributions and consider whether this can lead to an increase in $E_G$, and thus a faster rate of state reduction. 

Now that we no longer have spherical symmetry, there are various configurations for the displacement of the spheroids. Here we consider  four possible configurations: a) an oblate spheroid  displaced along its symmetry axis, b) a prolate spheroid also displaced along its symmetry axis, c) an oblate spheroid displaced along an equatorial (semi-major) axis, and d) a  prolate spheroid also displaced along an equatorial (now semi-minor) axis.  See Figure \ref{fig:Configurations} for a visual illustration of all these configurations. Although analytical solutions can be obtained for a general expression of $E_G$ for these cases (i.e.\ $E_G$ for a general equatorial or polar displacement $b$ between the spheroid states), the results are rather cumbersome and here we instead provide the results for the two cases a) and b) in the limit of high ellipticity $e$ (see  \ref{app:UniformSpheroid} for more detail). Defining  $\epsilon$ as $\epsilon := \sqrt{1-e^2}$, such that:
\begin{equation}  \label{eq:epsilon}
\epsilon := \cases{
	c/a &for $an~ oblate~ spheroid~(a>c)$, \\
	a/c &for $a~ prolate~ spheroid~(c>a)$,}
\end{equation}
where $a$ and $c$  are the equatorial and polar radii respectively\footnote{Note that $a$ and $c$ are respectively the semi-major and semi-minor axes for an oblate spheroid but semi-minor and semi-major axes for the prolate spheroid.}, then, when $e \approx 1$ we have $\epsilon \ll 1$. For the extreme ($\epsilon \ll 1$) prolate (\emph{spindle}-like) spheroid in configuration  b), we find  that, to first order in $\epsilon$:

\begin{figure*}[t!]
	\begin{center}
		\includegraphics[width=0.75\textwidth]{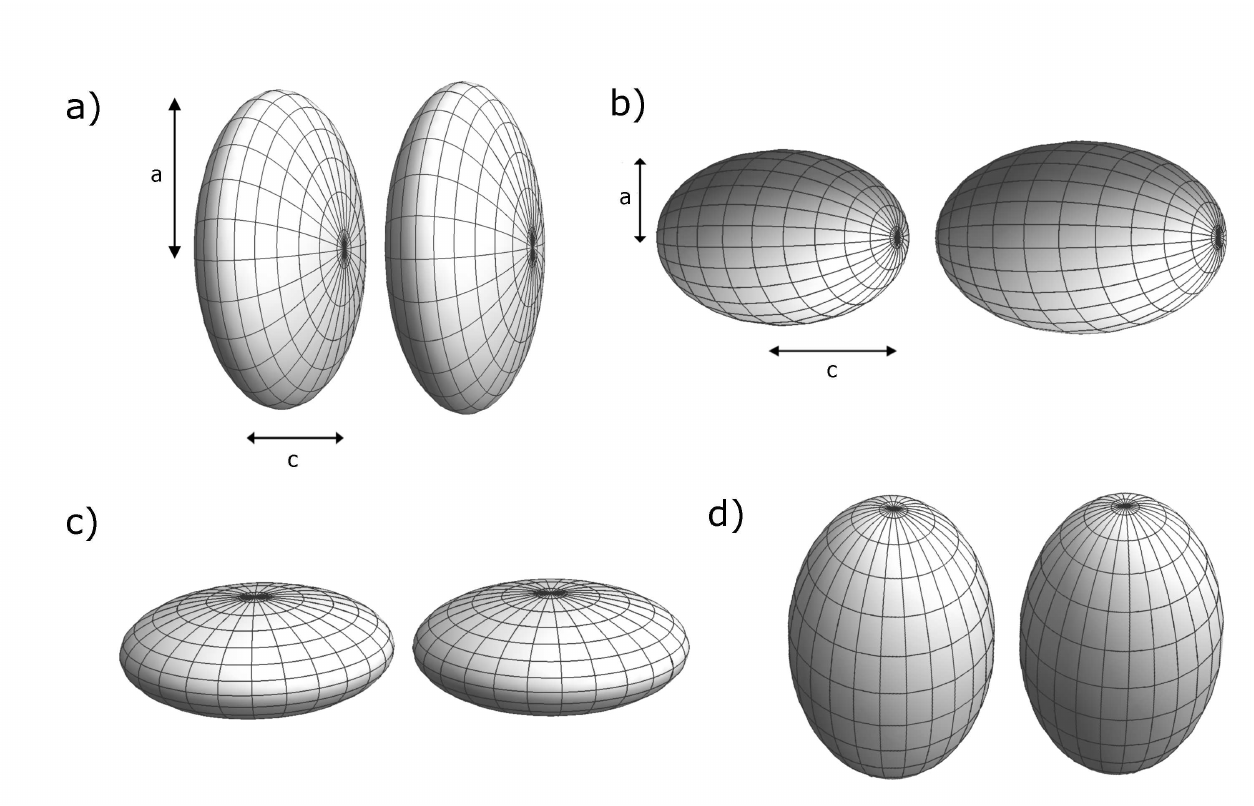}
	\end{center}
	\vspace{-0.5cm}
	\caption{The different spheroidal superposition configurations considered in Section \ref{sec:UniformSpheroid} and \ref{sec:BECSpheroids}: a) oblate and b) prolate spheroids displaced along their symmetry axis; and c) oblate and d) prolate spheroids displaced along an equatorial axis.} \label{fig:Configurations}
\end{figure*}

\begin{equation} \label{eq:EGUniformProlate}
E_G = \cases{
	\frac{6 G M^2}{5 c} \Big( A/4 - B \ln  \epsilon\Big)   &if $0 \leq b \leq 2c$, \\
	\frac{6 G M^2}{5 c} \Big(C/4 - \ln  \epsilon\Big) &if $ b \geq 2c$,}
\end{equation}
where
\begin{eqnarray}\nonumber
A &:= 2\lambda [1 + \ln (1 + \lambda)] + \lambda^2 (20 - 21 \ln 2) - \lambda^3\Big[21 - 10 \ln \frac{(1 + \lambda)^2}{4 \lambda}\Big] \\&~~~~~~~~~~+ 77\lambda^4 - 2 \lambda^5 \Big[35 + \ln \frac{1 + \lambda}{4 \lambda}\Big] +  20 \lambda^6, \\
B&:=  5 \lambda^2 - 5 \lambda^3 + \lambda^4,\\
C&:=4 \ln 2 -11 \lambda - 4(1 - 5\lambda^2) \coth^{-1} \lambda   + 2 \lambda^3 \Big[(\lambda^2-5) \ln \frac{\lambda^2}{1 - \lambda^2}-1\Big]
\end{eqnarray}
with $\lambda$ now defined as  $\lambda := b / (2 c)$.

\begin{figure*}[t!]
	\begin{center}
		\hspace{-0.65cm}
		\subfigure{%
			\put(120,140){(a)}
			\label{fig:EGSphereAllSpheroidse05}
			\includegraphics[trim={1cm 0 0 0},clip,width=0.55\textwidth]{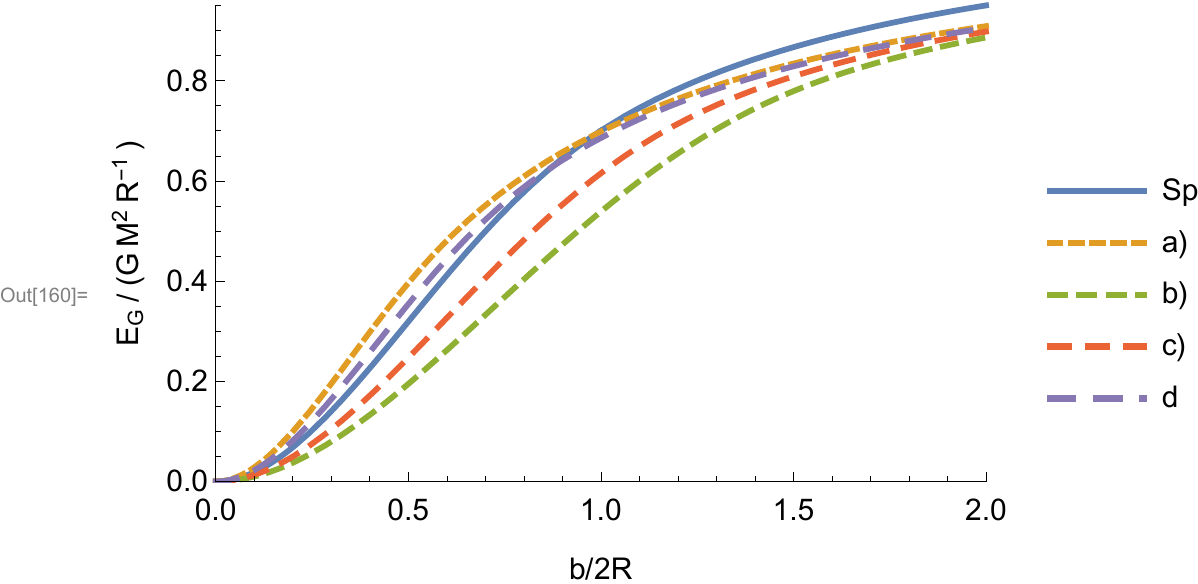}            
		}%
		\subfigure{%
			\put(120,140){(b)}
			\label{fig:EGSphereAllSpheroidse001}
			\includegraphics[width=0.45\textwidth]{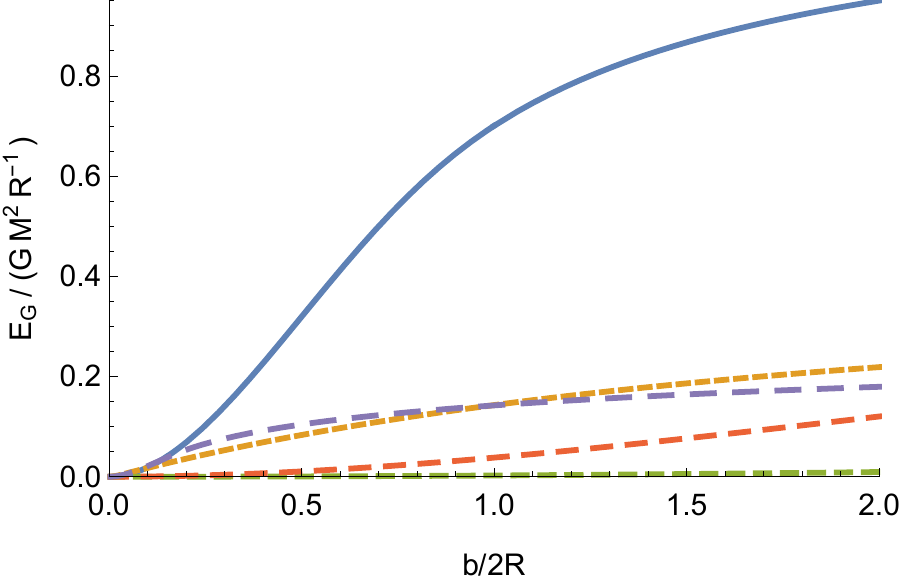}
		}\\ 
	\end{center}
	\vspace{-0.5cm}
	\caption{Both plots are of the gravitational self-energy of the difference between  displaced uniform spherical and spheroidal mass distributions, $E_G$, against $b/(2R)$, where $R$ is the radius of the sphere and $b$ is the distance between the centres of the states. All mass distributions have the same total mass and volume. The solid line is for the spherical case, and the various dashed and dotted lines are for the a), b), c) and d) spheroidal configurations illustrated in Figure \ref{fig:Configurations}. The left plot is for $\epsilon = 0.5$ (ellipticity $e=0.87$), and the right plot is for  $\epsilon = 0.01$ (ellipticity $e=0.99995$).}
	\label{fig:EGSphereAllSpheroids}
\end{figure*}

On the other hand,  for the extreme ($\epsilon \ll 1$) oblate (\emph{pancake}-like) spheroid in configuration a), we find that $E_G$ can be approximated by:

\begin{equation} \label{eq:EGUniformOblate}
E_G = \cases{
	\frac{6 G M^2}{5 a} \mathcal{A}  &if $0 \leq b \leq 2c$, \\
	\frac{6 G M^2}{5 a} \Big(\mathcal{C}/4 - \epsilon \Big) &if $ b \geq 2c$,}
\end{equation}
where $A$ and $C$ are defined as:
\begin{eqnarray}\label{key}
\mathcal{A} &:= 
5 \beta^2 \Big(\frac{1}{\epsilon} + \frac{\pi}{2}\Big) - \frac{5 \beta^3}{ \epsilon} \Big( \frac{\pi}{32}-\frac{1}{2 \epsilon}  \Big)+ \frac{\beta^5}{4 \epsilon^3} \Big( \frac{9 \pi}{8} + \frac{1}{\epsilon}\Big),\\
\mathcal{C} &:=  2 \pi + 11 \beta - 2 \beta^3 - 4(1 + 5 \beta^2) \cot^{-1} \beta + 2 \beta^3 (5 + \beta^2) \ln \frac{1 + \beta^2}{\beta^2},
\end{eqnarray}

with $\beta := b / (2a)$. In Figures \ref{fig:EGSphereAllSpheroidse05} and \ref{fig:EGSphereAllSpheroidse001}, we provide $E_G$ for the sphere and  the above four spheroidal cases a), b), c) and d) when $\epsilon = 0.5$ ($e = 0.87$) and $\epsilon =0.01$ ($e=0.99995$).\footnote{Formulas \eref{eq:EGUniformOblate}-\eref{eq:EGUniformProlate} are less reliable in the former case for configurations a) and b), and so we use general expressions for the plots.} 
In all cases we take the volume and mass (and so density) of the objects to be the same. These figures illustrate  that, for configurations a) and d), $E_G$ can be greater than that of the sphere (with the same volume and density) at certain displacements, although, at infinite displacement the sphere has the greatest $E_G$. Indeed, at $b=\infty$, we find
\begin{eqnarray}\label{eq:EGSphereInfty}
E^{sphere}_G &= \frac{6 G M^2}{5R},\\ \label{eq:EGProlateInfty}
E^{prolate}_G &= \frac{6 G M^2}{5l} \tanh^{-1} e\equiv \frac{6 G M^2}{5l} \cosh^{-1} \epsilon, \\\label{eq:EGOblateInfty}
E^{oblate}_G &= \frac{6 G M^2}{5l} \sin^{-1} e \equiv \frac{6 G M^2}{5l} \sec^{-1} \epsilon ,
\end{eqnarray}

for the three cases of sphere, prolate and oblate, irrespective of how they are displaced with respect to each other, and $l$ is the focal distance of the spheroids, which is $\sqrt{a^2-c^2}$ for the oblate, and $\sqrt{c^2-a^2}$ for the prolate spheroid. Equations \eref{eq:EGSphereInfty}-\eref{eq:EGOblateInfty} are valid for any value of $e$ and $\epsilon$ between $0$ and $1$, and no constraints are placed on the size of volume and density.  However,  taking all the objects to have the same volume and mass, and assuming low ellipticity $e \ll 1$,  we  find
\begin{eqnarray}\label{key}
E^{prolate}_G   &\approx \frac{6 G M^2}{5R} \Big(1 - \frac{1}{45 } e^4 - \frac{64}{2835} e^6 -\cdots\Big),\\
E^{oblate}_G &\approx \frac{6 G M^2}{5R} \Big(1 - \frac{1}{45 } e^4 - \frac{62}{2835 } e^6    -\cdots\Big ),
\end{eqnarray}
such that $E_G$ of the  the prolate and oblate is always less than that of the sphere at infinite separation of the two objects. In the same way, it is possible to also show that, for cases b) and c), $E_G$  is less than that of the sphere for any value of $b$.  

\begin{figure*}[t!]
	\begin{center}
		\subfigure{%
			\put(67,140){(a)}
			\label{fig:EGOblateVsSphere}
			\includegraphics[width=0.325\textwidth]{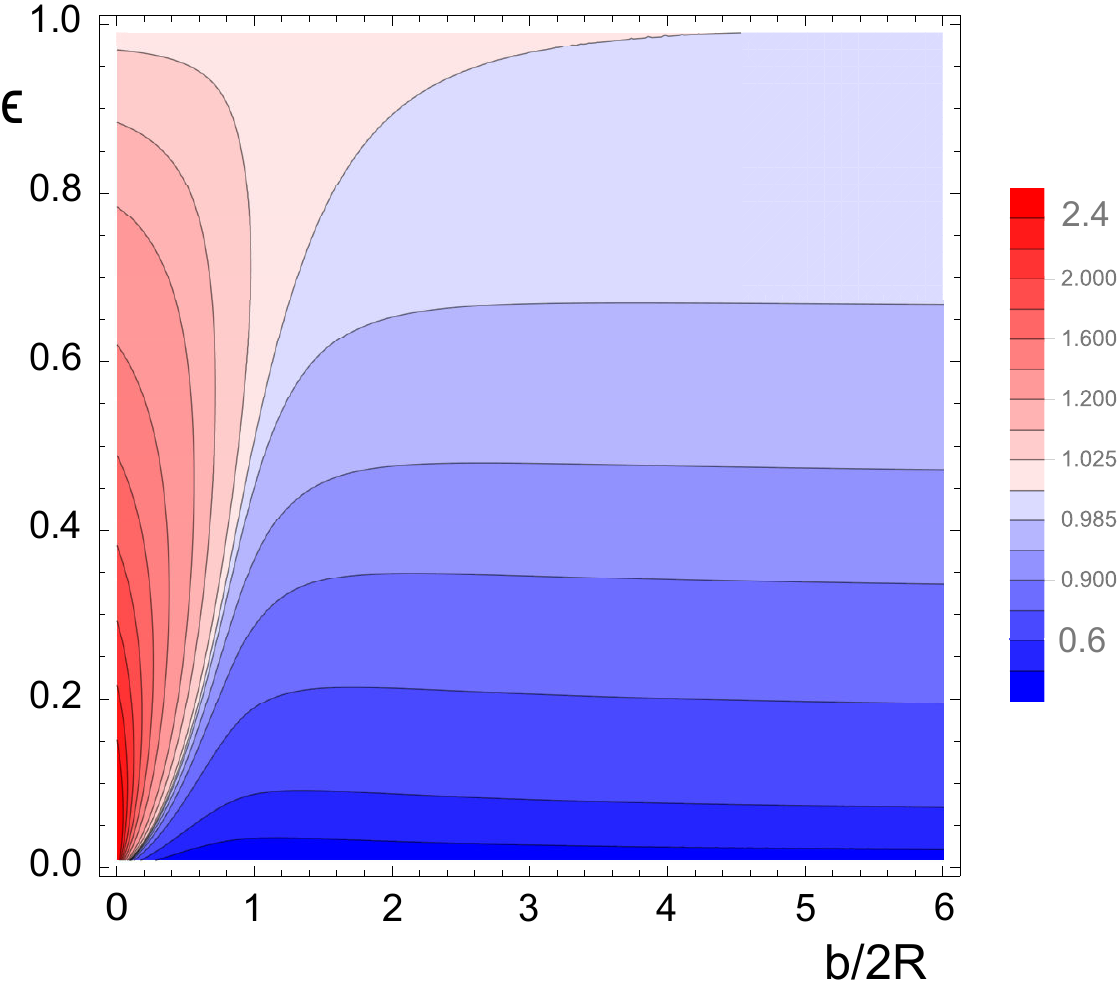}            
		}%
		\subfigure{%
			\put(67,140){(b)}
			\label{fig:EGParProlateVsSphere}
			\includegraphics[width=0.325\textwidth]{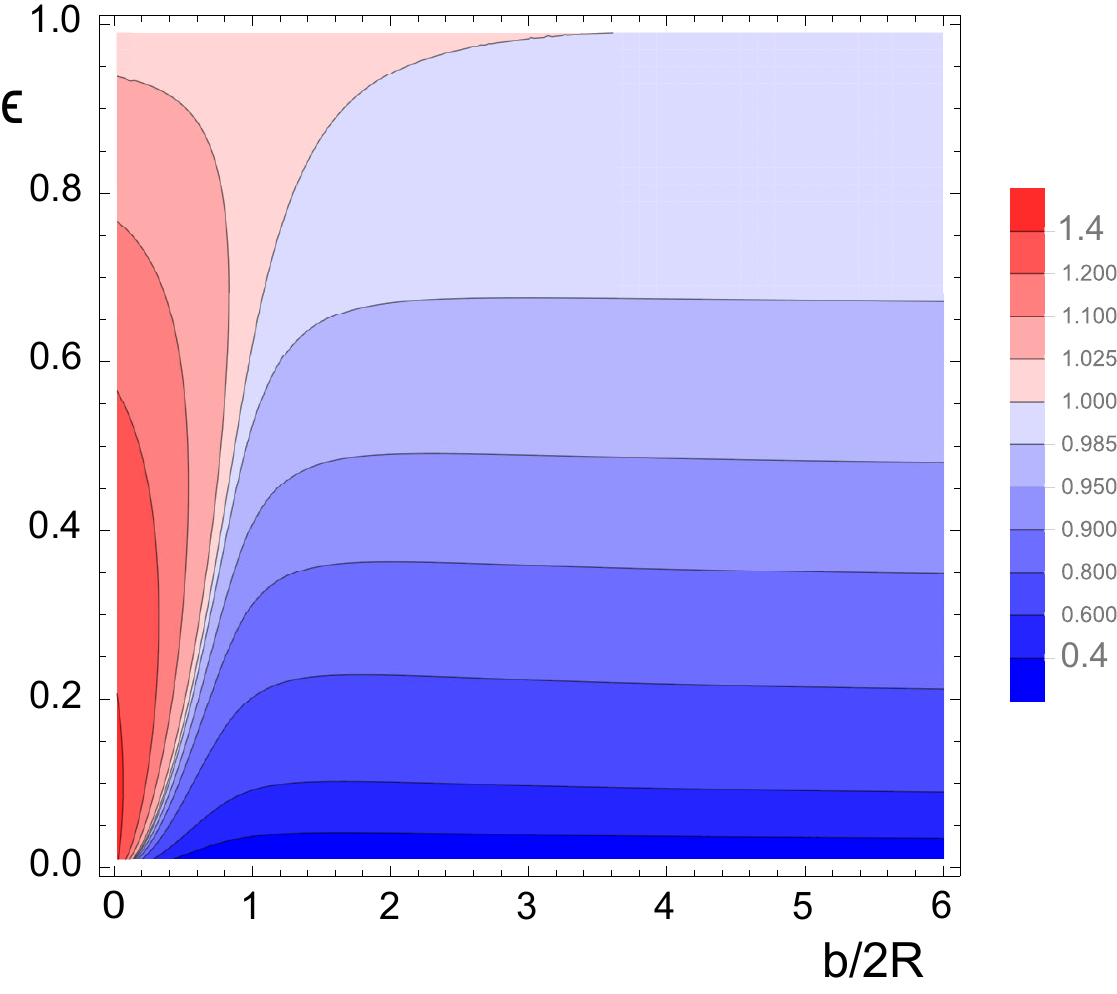}
		}%
		\subfigure{%
			\put(67,140){(c)}
			\label{fig:EGOblateVsParProlate}
			\includegraphics[width=0.325\textwidth]{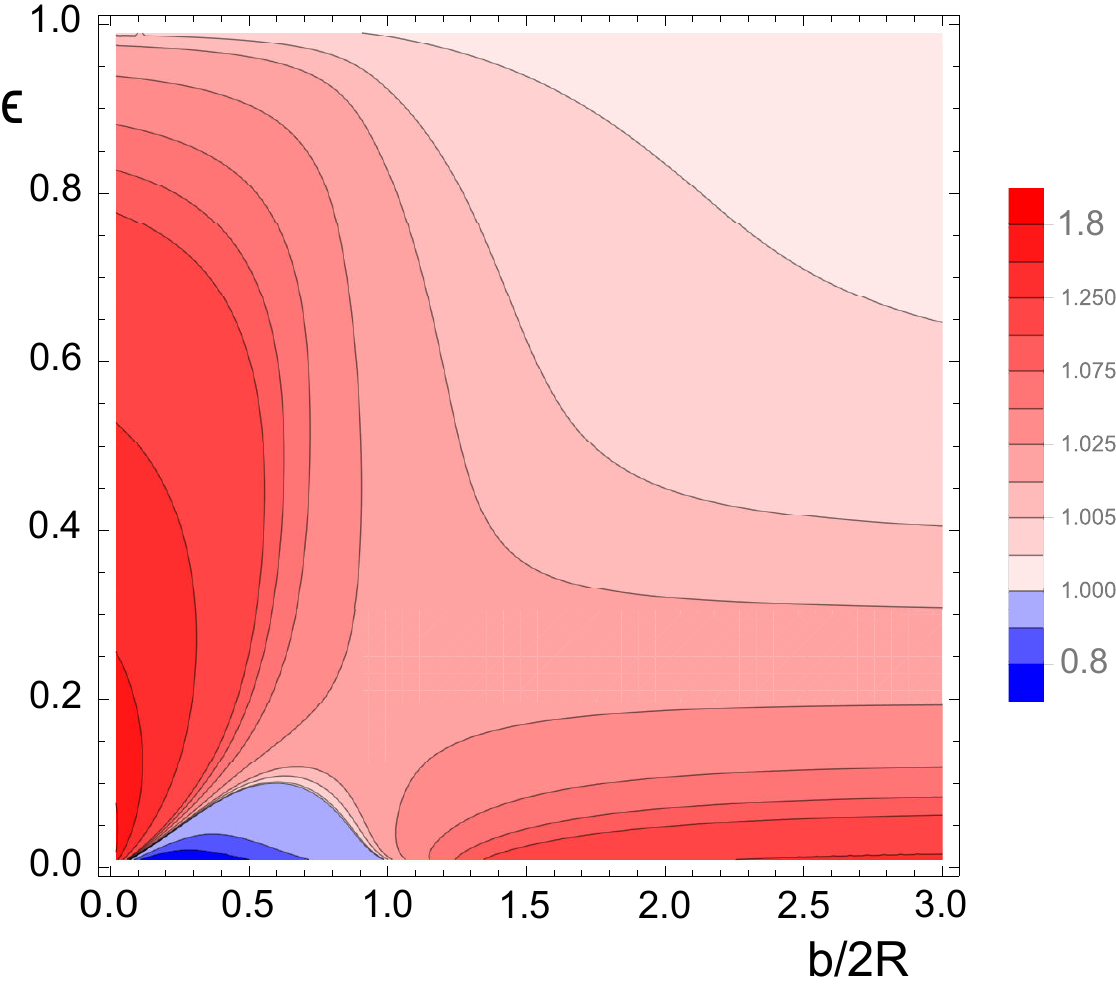}
		}\\ 
	\end{center}
	\vspace{-0.5cm}
	\caption{On the left is a contour plot of the gravitational self-energy of the difference between  displaced, uniform oblate spheroidal  mass distributions (displaced along their symmetry axes) over  that of   displaced, uniform spherical mass distributions i.e.\ $E^{a)}_G/E^{sphere}_G$. The x-axis is the distance $b$ between the centres of the states divided by twice the radius of the sphere, and the y-axis is the value of $\epsilon$, the ratio of the semi-major to semi-minor axes, for the spheroids. The middle plot is as the left but using the  equatorial-displaced prolate  spheroidal  mass distributions rather than oblate ones i.e.\ $E^{d)}_G/E^{sphere}_G$, and the right plot is of $E^{a)}_G/E^{d)}_G$.}
	\label{fig:EGOblateVsSpheroid}
\end{figure*}

However, as stated above, this is not true for the other cases - it is possible for the value of $E_G$ for the spheroidal configurations a) and d) to be greater than that of the sphere. This is further illustrated in Figures \ref{fig:EGOblateVsSphere} and \ref{fig:EGParProlateVsSphere}, which are contour plots of  $E^{a)}_G/E^{sphere}_G$ and  $E^{b)}_G/E^{sphere}_G$ for values of $\epsilon$ ranging from $0$ to $1$ (i.e.\ any ellipticity) and for the displacement $b$ ranging from $0$ to $12R$.   In particular, for small displacements it is possible for the  spheroidal $E_G$  to be a factor  greater than the spherical case. Taking the oblate spheroid and sphere to have the same volume and mass, then in the limit of $\epsilon \ll 1$ and $b \ll R$, we find, using \eref{eq:EGUniformSphere} and \eref{eq:EGUniformOblate}, that $E^{a)}_G = 3 \times E^{sphere}_G$.\footnote{The reason that the spheroidal configurations a) and d) can have a value of  $E_G$ that is greater than the spherical case is because these objects are displaced along a semi-minor axis, which will be shorter than the radius of the corresponding sphere.} Such a factor  would already be approximately satisfied when  $\epsilon \approx 0.01$ and $d \approx  0.01 R$, which could be possible in near-future experiments. For example,  $b \ll R$ is satisfied in the proposed nano-sphere experiment \cite{BoseMatterWave}, such that, taking  an oblate spheroid with $\epsilon \approx 0.01$ rather than a sphere would increase $E_G$ accordingly.

The findings here suggest that, for tests of the GQSR process, it may be preferable to use spheroidal rather than spherical mass distributions in nano/micro-object experiments. Figure \ref{fig:EGOblateVsParProlate} also provides a comparison of $E^{a)}_G$ against $E^{d)}_G$, illustrating that, although $E^{a)}_G$ is generically larger than  $E^{d)}_G$, there is a certain region of parameter space where the opposite is the case. Since a prolate and oblate spheroid can be used to  approximate, respectively, a rod or disc for high ellipticity, the self-energy of the difference of these objects could be used to approximate that which could be observed in nano/micro-rod and -disc experiments.  Note also that, if an experiment were able to observe state reduction in disagreement with standard QT, then comparing the results for different spheroidal geometries could be used to distinguish the GQSR considered here from other collapse models  since we 
have a direct prediction for how $E_G$ changes with just the ellipticity of the object ($e=0$ for a sphere).

\setcounter{footnote}{0}

\section{Testing with a Bose-Einstein condensate} \label{sec:BECTests}

In addition to nano/micro-object experiments, it may also be possible to test the GQSR process considered here using BECs.  Advantages of these systems  include the fact that they are highly controllable systems and  have large coherence times due to their extremely low temperatures and high isolation from their environments.  Certain superposition states have also already been observed for these experiments, such as a coherent state separated by over $0.5\,\mathrm{m}$, and there are several suggested techniques for creating macroscopic superposition states (see Section \ref{sec:DoubleWellBEC}). 

In Section \ref{sec:BECSpheroids}, we calculate the  self-energy of the difference between  spherical and spheroidal BEC mass distributions, which are created using harmonic trapping potentials. We then compare, in Section \ref{sec:EnvDecoherence}, the corresponding  rate of state reduction to the decoherence rate of prominent channels of environmental decoherence in BEC experiments, providing estimates for the values of experimental parameters, such as temperature and scattering length, that would be required to test the GQSR process.

\subsection{Gravitational self-energy of the difference between BEC mass distributions} \label{sec:BECSpheroids}

In Sections \ref{sec:UniformSpheres} and \ref{sec:UniformSpheroid}, we calculated $E_G$ for uniform spherical and spheroidal mass distributions.  Although such distributions can be created in nano/micro-object experiments, spherical and spheroidal  BEC distributions are generically non-uniform. This non-uniformity is due to the trapping potential that constrains the BEC. For example, to create a spherical BEC, the potential is $V(r) = \frac{1}{2} m \omega_0^2 r^2$ where $m$ is the atomic mass, $\omega_0$ is the trapping frequency, and $r = \sqrt{x^2+y^2+z^2}$ is the radial distance from the centre of the trap. 

Taking the BEC to obey the time-independent Gross-Pitaevskii equation \cite{Pitaevskii1961,Gross61}:

\begin{equation}\label{eq:GP}
\Big[ - \frac{\hbar^2}{2m } \nabla^2 + V(\bs{r}) + g n(\bs{r}) \Big] \psi_0(\bs{r}) = \mu \psi_0(\bs{r}) , 
\end{equation}

where $\psi_0(\bs{r})$ is the BEC wavefunction, $\mu$ is the chemical potential of the condensate, $g = 4 \pi \hbar^2 a_s /m$ is the $s$-wave interaction coupling constant with $a_s$ the $s$-wave scattering length,  and $n(\bs{r}) = |\psi_0(\bs{r})|^2$ is the condensate number density; we can solve for the density of the BEC at zero temperature.  Here we consider two analytical limiting cases: 1) the Gaussian approximation where we assume that the wavefunction $\psi_0$ is Gaussian, which is exact for an ideal Bose gas when we neglect the interaction term, and can also be used in describing repulsive BECs with low effective interaction strength, as well as attractive ($a_s < 0$) BECs \cite{GaussianAnsatzAttractive,GaussianAnsatzAttractiveShi,GaussianAnsatzAttractiveStoof,GaussianAnsatzParola,GaussianAnsatzFetter,BECReview}; and 2) the Thomas-Fermi (TF) approximation \cite{TFThomas,TFFermi}, which is most appropriate for repulsive BECs ($a_s > 0$) with large numbers of atoms,  where we neglect the kinetic term of \eref{eq:GP} in comparison to the interaction term.

\subsubsection{\small{Density in the Gaussian approximation}} \label{sec:GaussianDensity} \hfill \break

In this section, we consider the Gaussian approximation for BECs, which is useful for characterizing BECs with very low interaction strengths, as well as attractive interactions, as described above. When the interaction term is entirely neglected in the Gross-Pitaevskii equation, we have an ideal Bose gas, and the solution for a general harmonic trapping potential $V(\bs{r}) = \frac{1}{2} m (\omega^2_x x^2 + \omega^2_y y^2 + \omega^2_z z^2)$ is: 
\begin{equation}\label{eq:GuassianPsi}
\psi_0(\bs{r}) = \sqrt{N}  \Big(\frac{m \omega_0}{\hbar \pi}\Big)^{3/4} e^{-\frac{1}{2} m (\omega_x x^2 + \omega_y y^2 + \omega_z z^2)/\hbar},
\end{equation} 
where $N$ is the number of condensate atoms, $\omega_0 := (\omega_x \omega_y \omega_z)^{1/3}$ is the geometric average of the trapping frequencies, and the chemical potential is $\mu = \frac{1}{2} \hbar (\omega_x+\omega_y+\omega_z)$. Taking a spherical trap ($\omega_0 := \omega_x = \omega_y = \omega_z$), the mass density $\rho_0(\bs{r}) := m n (\bs{r})$ of the condensate is then
\begin{equation}\label{eq:GuassianRho}
\rho^{sphere}_0(\bs{r}) = \frac{4}{3 \sqrt{\pi}} \rho^{sphere}_0 e^{-r^2/R^2_0},
\end{equation}
where $\rho^{sphere}_0 := M/((4/3) \pi R^3_0)$, with $M = m N$ the total mass,  and  
\begin{equation}\label{eq:R0}
R_0 := \sqrt{\hbar / (m \omega_0)}
\end{equation}
is the width of the Gaussian wavefunction \eref{eq:GuassianPsi}.  To describe a BEC with attractive forces we can use a variational approach with the ansatz that the ground state is  of Gaussian form but  we now replace $R_0$ in \eref{eq:GuassianPsi} with
\begin{equation}\label{eq:Rprimed}
R'_0 := \alpha_R R_0,
\end{equation}
where $\alpha_R$ is a dimensionless variational parameter which fixes the width of the condensate for a given interaction strength \cite{GaussianAnsatzAttractive,GaussianAnsatzAttractiveShi,GaussianAnsatzAttractiveStoof,GaussianAnsatzParola,GaussianAnsatzFetter,BECReview}. The density for such a spherical BEC will then still be approximated by \eref{eq:GuassianRho} but with $R_0$ replaced with $R'_0$, where $R'_0 < R_0$ for a BEC with attractive forces.

To generate  a spheroidal BEC, the harmonic trapping potential must be of the form

\begin{equation} \label{eq:SpheroidalTrap}
V(\bs{r}) = \frac{1}{2} m [\omega_r^2 r^2_{\rho} + \omega_z^2 z^2]
\equiv  \frac{1}{2} m \omega_r^2 [r^2_{\rho} + \lambda_{\omega}^2 z^2],
\end{equation}

where  $r_{\rho} = \sqrt{x^2+y^2}$ is the radial cylindrical coordinate,  $\omega_r$ is the radial trapping frequency, $\omega_z$ is the axial trapping frequency, and $\lambda_{\omega} := \omega_z/\omega_r$ (which is sometimes referred to as the `asymmetry parameter'). For a prolate spheroid (often called a `cigar' BEC), $\lambda_{\omega} <0$, whereas, for an oblate spheroid (often called a `pancake' BEC),  $\lambda_{\omega} >0$. In the Gaussian approximation, the density of the BEC is given by

\begin{equation} \label{eq:OblateDensityGauss}
\rho^{spheroid}_0(\bs{r})= \frac{4}{3 \sqrt{\pi}} \rho^{spheroid}_{0} e^{-r^2_{\rho} /a^{'2}_0 - z^2/c^{'2}_0},
\end{equation}

where $\rho^{spheroid}_{0} := M / ((4/3) \pi a^{'2}_0 c'_0)$,  $c'_0 := \alpha_c c_0$ and $a'_0 = \alpha_a a_0$, with $c_0:= \sqrt{\hbar / (m \omega_z)}$ and $a_0 := \sqrt{\hbar / (m \omega_r)}$. Similar to the spherical case, the factors $\alpha_a$ and $\alpha_c$ control the size of the condensate for the given interaction strength.

For a BEC with attractive interactions, the condensate becomes unstable if the number of atoms exceeds a critical value. For a harmonic trap at zero temperature, this critical value can be estimated as  \cite{AttractiveSphericalCollapse}

\begin{equation}\label{eq:CritNum}
N_c \approx k_c \frac{s_0}{|a_s|}, 
\end{equation}
where $s_0$ is the width of the ground state Gaussian wavefunction of an atom in a parabolic potential well $\sqrt{\hbar / (m \omega_0)}$, with $\omega_0 := (\omega_x \omega_y \omega_z)^{1/3}$, and $k_c$ is a constant, which is estimated as $\approx 0.6$ for a single-well spherical trap. See \cite{CriticNumberDoubleWell} for values of $k_c$ for a double-well trap and, for example, \cite{CritNumberAnalytic} for an analytical expression for $k_c$.  Note that \eref{eq:CritNum}   is not applicable with the TF approximation since the kinetic term is  required to stabilize the system here.

\subsubsection{\small{Density in the Thomas-Fermi approximation}} \hfill \break

\vspace{-4mm}

In this section, we consider the TF approximation for BECs, which is useful for characterizing BECs with strong repulsive interactions, as described in Section \ref{sec:BECSpheroids} above. Assuming a spherical trap and that the kinetic term can be neglected in comparison to the interaction term, the solution $\psi_0$ of the Gross-Pitaevskii equation can be used to find the mass density of the BEC in this TF approximation:

\begin{equation} \label{eq:SphereDensity}
\rho^{sphere}_0(\bs{r})= \frac{5}{2} \rho^{sphere}_0 (1-r^2 /R^2),
\end{equation} 

where $R$ the radius of the spherical BEC (we assume $\rho(\bs{r})$ is vanishing here), $\rho^{sphere}_0 := M/((4/3) \pi R^3)$, and $M$  the total mass of the condensate:

\begin{equation}
M := m N=\int_0^{2 \pi} \int_{0}^{\pi} \int_0^R  \rho(\bs{r}) r^2 \sin \theta dr d\theta d \phi.
\end{equation}

\begin{figure*}[t!]
	\begin{center}
		\subfigure{%
			\put(42,100){(a)}
			\label{fig:TFSphereDensity}
			\includegraphics[width=0.2\textwidth]{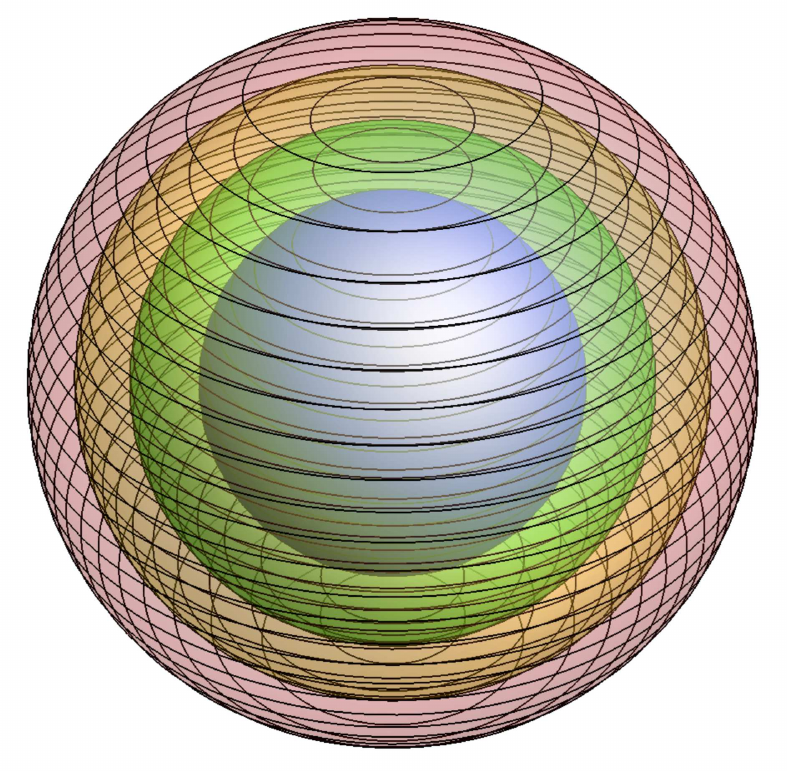}            
		}%
		\hspace{0.3cm}
		\subfigure{%
			\put(30,100){(b)}
			\label{fig:TFProlateDensity}
			\includegraphics[width=0.15\textwidth]{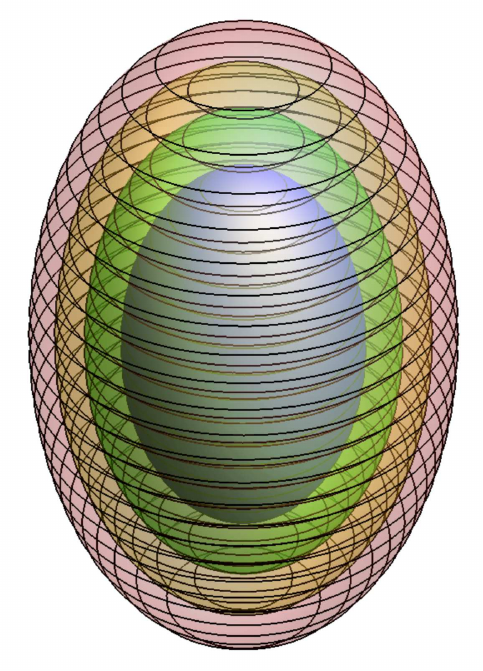}
		} 
		\hspace{0.3cm}
		\subfigure{%
			\put(48,100){(c)}
			\label{fig:TFOblateDensity}
			\includegraphics[width=0.225\textwidth]{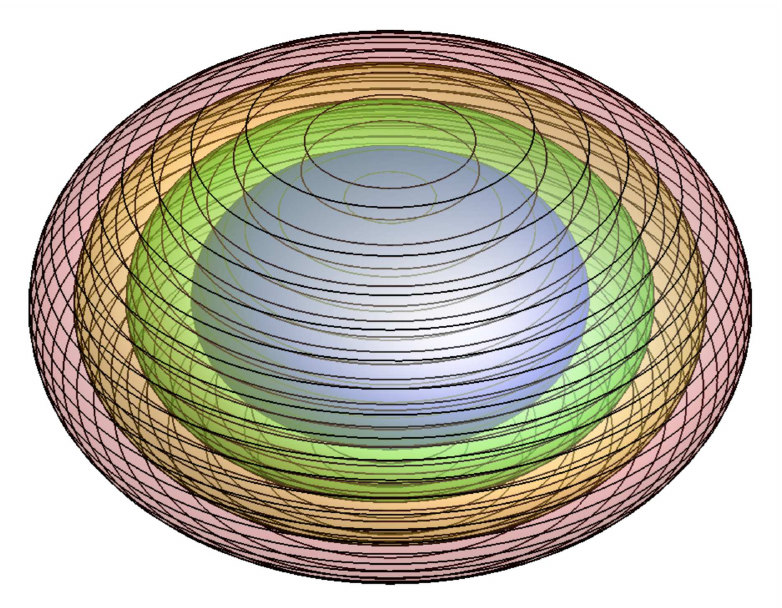}
		}
	\end{center}
	\caption{Three-dimensional plots of the condensate density in the Thomas-Fermi approximation. From left to right, there is a spherical BEC, a prolate spheroidal BEC and an oblate spheroidal BEC. Different shaded areas illustrate the fact that the density continuously varies, being greatest in the centre, and with surfaces of constant density being similar-shaped spheroidal surfaces.}
	\label{fig:TFDensities}
\end{figure*}

In terms of experimental parameters, $R$ is given by

\begin{equation}\label{eq:SphericalR}
R= ( 15 N a_s R_0^4)^{1/5},
\end{equation}

where $R_0$ is defined in the previous section. The density of the BEC sphere is illustrated in Figure \ref{fig:TFSphereDensity} where contours represent surfaces of constant density, which are spherical surfaces. In contrast to the Gaussian approximation, the Bose gas will clearly be of a larger size in this strongly, repulsive interaction regime.

The  density function for a spheroidal BEC in the TF approximation can be found by inserting the potential \eref{eq:SpheroidalTrap} into the Gross-Pitaevskii equation \eref{eq:GP}, and dropping the kinetic term,  to obtain:

\begin{equation} \label{eq:ProlateDensity}
\rho^{spheroid}_0(\bs{r})= \frac{5}{2} \rho^{spheroid}_{0} (1-r^2_{\rho} /a^2 - z^2/c^2),
\end{equation}
where  $\rho^{spheroid}_{0} := M / ((4/3) \pi a^2 c)$. This density distribution is illustrated in Figures \ref{fig:TFProlateDensity} and \ref{fig:TFOblateDensity} for a prolate ($a<c$) and oblate ($a>c$) spheroid where contours are surfaces of constant density, which are similar-shaped spheroidal surfaces. In terms of experimental parameters, the equatorial and polar radii are:
\begin{eqnarray}
a = (15 N a_s a_0^4 \lambda_{\omega})^{1/5},\\
c = a / \lambda_{\omega},
\end{eqnarray}
where $a_0$ is defined in the previous section for the Gaussian approximation.  For the TF approximation to be very good, we require that \cite{BECReview}: 
\begin{equation} \label{eq:TF}
N \gg \frac{s_0}{a_s}.
\end{equation}
Furthermore, the TF approximation is less accurate near the boundaries of the condensate. Here the density abruptly vanishes in the TF approximation, but in reality there is a more gradual decrease such that the condensate wavefunction will tend to, but never actually reach, zero \cite{YellowBook}.

\begin{figure*}[t!]
	\begin{center}
		\subfigure{%
			\put(110,150){(a)}
			\label{fig:TFBECVsUniformSphere}
			\includegraphics[width=0.465\textwidth]{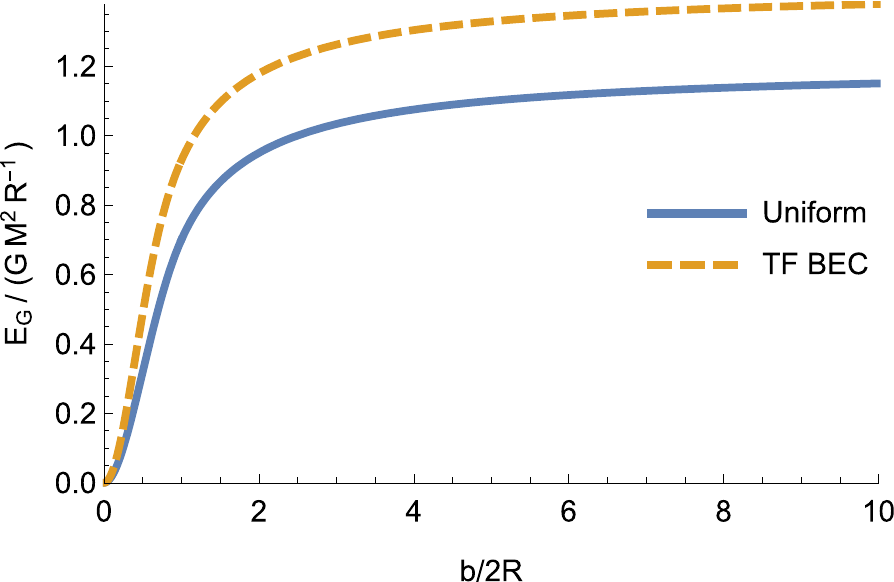}            
		}%
		\hspace{0.1cm}
		\subfigure{%
			\put(110,150){(b)}
			\label{fig:TFBECVsGaussBEC}
			\includegraphics[width=0.465\textwidth]{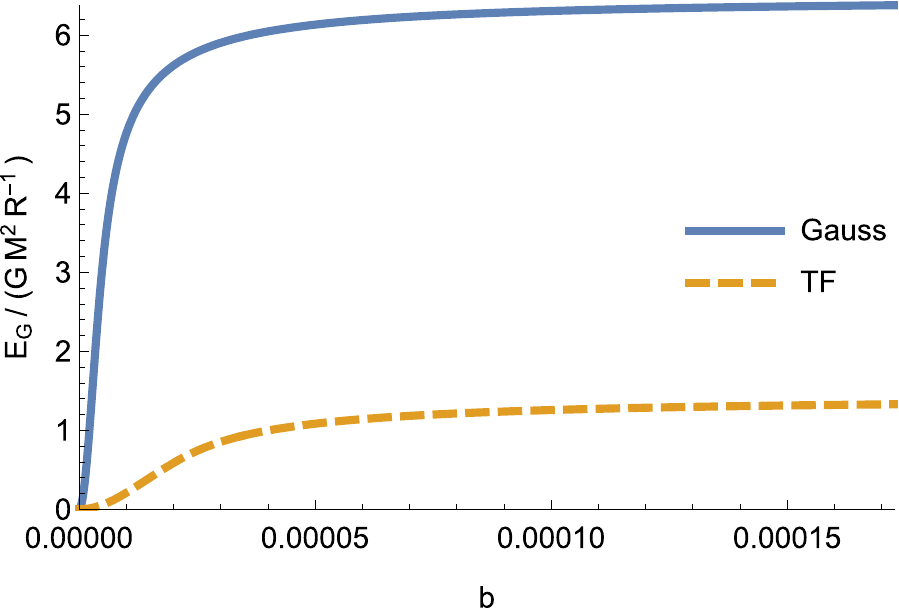}
		}\\ 
	\end{center}
	\vspace{-0.5cm}
	\caption{On the left is the gravitational self-energy of the difference between  displaced spherical BECs (in the TF approximation) and displaced uniform spheres, $E_G/(G M^2 R^{-1})$, against $b/(2R)$ where $R$ is the radius of the spheres, $M$ is their mass and $b$ is the distance between the centres of the sphere states. On the right is $E_G/(G M^2 R^{-1})$ against $b$ of spherical $^{133}\mathrm{Cs}$ BECs in the TF  and Gaussian  approximations with  $10^6$ atoms, the same trapping frequency $\omega_0 = 100\unit{Hz}$, and with the standard scattering length in the former regime, but with zero scattering in the latter so that we have an ideal BEC in that case. }
	\label{fig:BECVsUniformSphere}
\end{figure*}

\subsubsection{\small{Self-energy  of the difference between spherical BECs}} \label{sec:EGSphericalBECs} \hfill \break

\vspace{-4mm}

Now that we have mass distributions for BEC spheres and spheroids, we can determine the value of $E_G$ for the different shapes and density functions using \eref{eq:EG1} or \eref{eq:EG2}.  An approach to this is discussed in   \ref{app:BECSphere}-\ref{app:BECGaussianSphere} where we also calculate the gravitational potential of these objects. For the spherical BEC in the Gaussian approximation (see  \ref{app:GaussBECSphere}), we find

\begin{equation}
E_G= 
\frac{G M^2}{R'_0} \Big(  \sqrt{\frac{2}{\pi}} - \frac{1}{2 \lambda_0} \mathrm{erf}(\sqrt{2} \lambda_0)\Big),
\end{equation}
where we have  defined $\lambda_0 := b / (2 R'_0)$ with $R'_0$ given by \eref{eq:Rprimed}.   In contrast, in the TF  regime, we obtain (see  \ref{app:BECSphere}):

\begin{equation}
E_G = \cases{
	\frac{10 G M^2}{7 R} \Big(  2 \lambda^2 - \frac{21}{5} \lambda^4 + \frac{7}{2}  \lambda^5  - \frac{3}{4} \lambda^7 +  \frac{1}{10} \lambda^9 \Big)  &if  $0 \leq \lambda \leq 1$, \\
	\frac{10 G M^2}{7 R}   \Big( 1 - \frac{7}{20 \lambda}\Big) &if  $\lambda \geq 1$,}
\end{equation}
where $\lambda := b / (2R)$ and $R$ is given by \eref{eq:SphericalR}.

These self-energy differences are illustrated in Figure \ref{fig:BECVsUniformSphere}. For the same total mass and volume (and so average density),  $E_G$ of a spherical BEC in the TF regime is always greater than that of a uniform one. This is exemplified by Figure \ref{fig:TFBECVsUniformSphere} and is due to the fact that the density is more constrained towards the centre. The fact that $E_G$ is different  despite the potential outside a non-uniform sphere being the same as a uniform sphere, could provide a further possibility for distinguishing the state reduction process considered here to other models.

In Figure \ref{fig:TFBECVsGaussBEC}, we plot  $E_G$ of a spherical BEC in the TF regime against the Gaussian regime for a $^{133}\mathrm{Cs}$ BEC with $10^6$ atoms, the same trapping frequency $\omega_0 = 100\unit{Hz}$, and with the standard scattering length in the former regime, but with zero scattering in the latter so that we have an ideal BEC. For all values of $b$, the Gaussian $E_G$ is always greater than the TF case. This is principally due to $R_0$ being much smaller than $R$ in this case, with the gap increasing as $N$  increases.  Therefore, with attractive interactions, we would expect $E_G$ to increase further under the condition that all other BEC parameters, apart from the scattering length, stay the same.

\subsubsection{\small{Self-energy of the difference between spheroidal BECs}} \label{sec:EGBECSpheroids} \hfill \break

\vspace{-4mm}

\begin{figure*}[t!]
	\begin{center}
		\hspace{-0.65cm}
		\subfigure{%
			\put(120,140){(a)}
			\label{fig:EGBECSphereAllSpheroidse05}
			\includegraphics[trim={1cm 0 0 0},clip,width=0.55\textwidth]{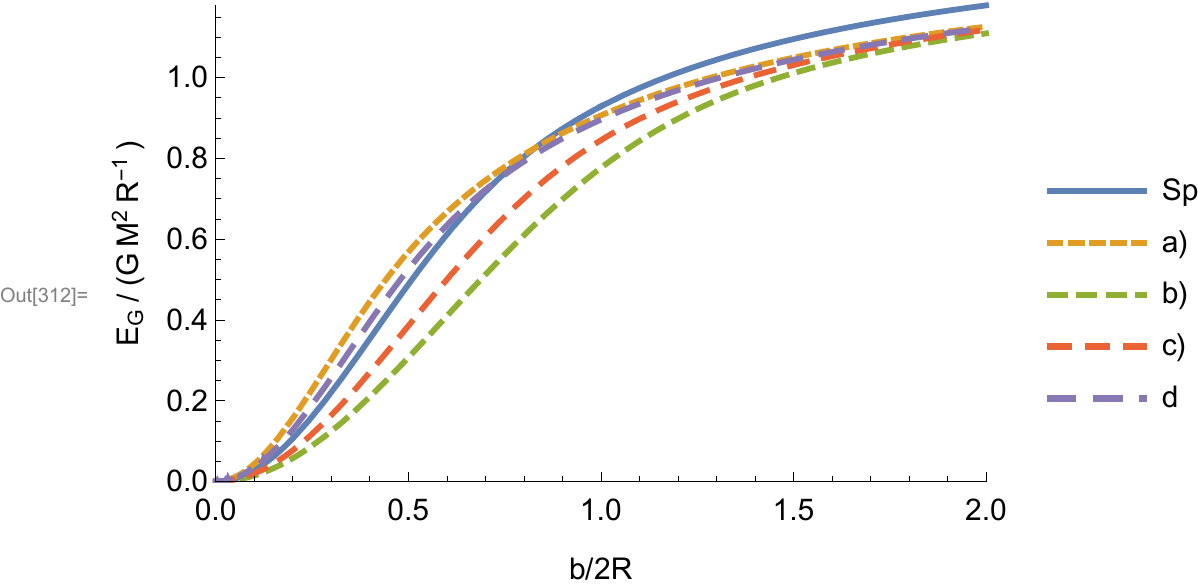}            
		}%
		\subfigure{%
			\put(120,140){(b)}
			\label{fig:EGBECSphereAllSpheroidse001}
			\includegraphics[width=0.45\textwidth]{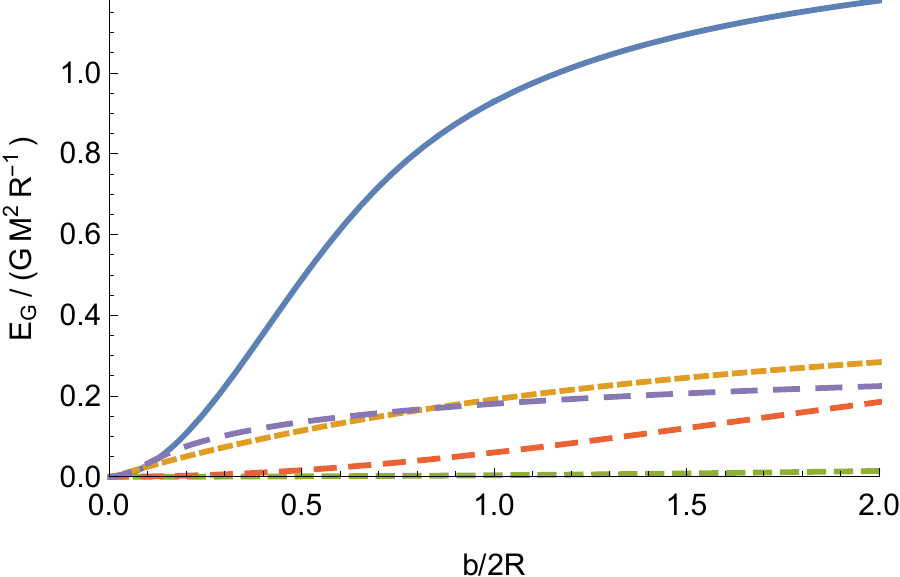}
		}\\ 
	\end{center}
	\vspace{-0.5cm}
	\caption{Both plots are of the gravitational self-energy of the difference between  displaced  spherical and spheroidal BEC mass distributions (in the TF regime), $E_G$, against $b/(2R)$, where $R$ is the radius of the spheres, $M$ is their mass and $b$ is the distance between the centres of the states. All mass distributions have the same total mass and volume. The solid line is for the spherical case, and the various dashed and dotted lines are for the a), b), c) and d) spheroidal configurations illustrated in Figure \ref{fig:Configurations}. The left plot is for $\epsilon = 0.5$ (ellipticity $e=0.87$), and the right plot is for  $\epsilon = 0.01$ and (ellipticity $e=0.99995$).}
	\label{fig:EGBECSphereAllSpheroids}
\end{figure*}  

The generic value of $E_G$ for spheroidal BECs is more complicated (see \ref{app:BECSpheroid} and \ref{app:BECGaussianSphere}) and here we just provide the expression for configuration b) in Figure \ref{fig:Configurations} (the symmetry-axis displaced prolate) for the TF regime in the limit of high ellipticity  (i.e.\ to first order in $\epsilon$, which is defined by \eref{eq:epsilon}):

\begin{equation} \label{eq:EGBECProlate}
E_G = \cases{
	\frac{10 G M^2}{7 a} \Big( \mathtt{A}/24 - \mathtt{B} \ln \epsilon \Big)  &if  $0 \leq \lambda \leq 1$, \\
	\frac{10 G M^2}{7 a}  \Big( \mathtt{C}/1536 -  \ln \epsilon  \Big)  &if  $\lambda \geq 1$,}
\end{equation}

where $\lambda := b / (2c)$ and
\begin{eqnarray}
\mathtt{A}&:=144 \lambda^2(\ln 2 - 1) - 168 \lambda^4(3\ln 2 - 4)-378\lambda^5+ 133 \lambda^7 -25 \lambda^9,\\  
\mathtt{B} &:= 10 (6\lambda^2  - 21 \lambda^4 + 21 \lambda^6 - 6 \lambda^7 + \lambda^9),\\ \nonumber
\mathtt{C}&:= 6 \gamma \coth^{-1} (1-2\lambda) -  6\kappa  \coth^{-1} (1+2\lambda)\\\nonumber &~~~~~~~~~~~~~~-768(1-6\lambda^2 + 21 \lambda^4) \coth^{-1} \lambda \\\nonumber&~~~~~~~~~~~~~~-64(12 \lambda^7-66 \lambda^5-284 \lambda^3 +81 \lambda - 24 \ln 2) \\
&~~~~~~~~~~~~~~-315\lambda(16\lambda^2-3) \ln \frac{\lambda^2-1}{\lambda^2}
\end{eqnarray}

with
\begin{eqnarray}\label{key}
\gamma &:= (1-2 \lambda)^3 \Big[128 + \lambda\Big( 2 \lambda[207+4\lambda (2 \lambda(\lambda+3)(2 \lambda-3))]+453\Big)\Big], \\
\kappa &:= (1+2 \lambda)^3 \Big[128 + \lambda\Big(  2 \lambda[207+4\lambda (2 \lambda(\lambda-3)(2 \lambda+3) )]-453\Big)\Big].
\end{eqnarray}

Assuming the TF regime, the value of $E_G$ for the four configurations a), b), c) and d) (see Figure \ref{fig:Configurations}) of BEC spheroids is plotted in Figures \ref{fig:EGBECSphereAllSpheroidse05} and \ref{fig:EGBECSphereAllSpheroidse001} against the BEC sphere case for $\epsilon =0.5$ and $\epsilon =0.01$. As in the uniform case, the value of $E_G$ for configurations b) and c) is always less than that of a BEC sphere, whereas, the other spheroidal configurations can have larger $E_G$ values at certain displacements and ellipticity values. However, again, the sphere always has the greatest $E_G$ at infinity - the values of $E_G$ in the BEC TF case compared to the uniform case \eref{eq:EGSphereInfty}-\eref{eq:EGOblateInfty}  are just $25/21 \approx 1.2$ larger for each object.

\begin{figure*}[t!]
	\begin{center}
		\subfigure{%
			\put(67,140){(a)}
			\label{fig:EGBECOblateVsSphere}
			\includegraphics[width=0.325\textwidth]{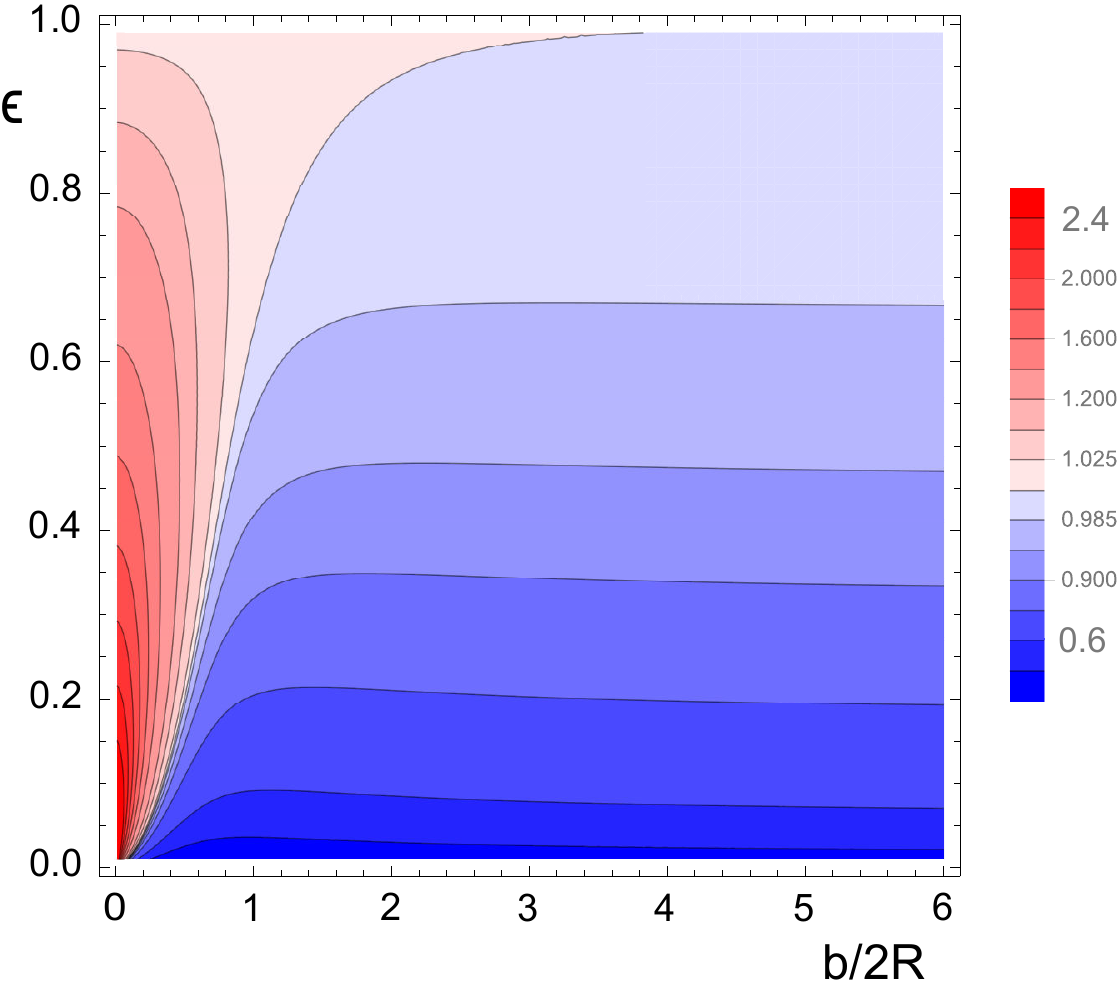}            
		}%
		\subfigure{%
			\put(67,140){(b)}
			\label{fig:EGBECParProlateVsSphere}
			\includegraphics[width=0.325\textwidth]{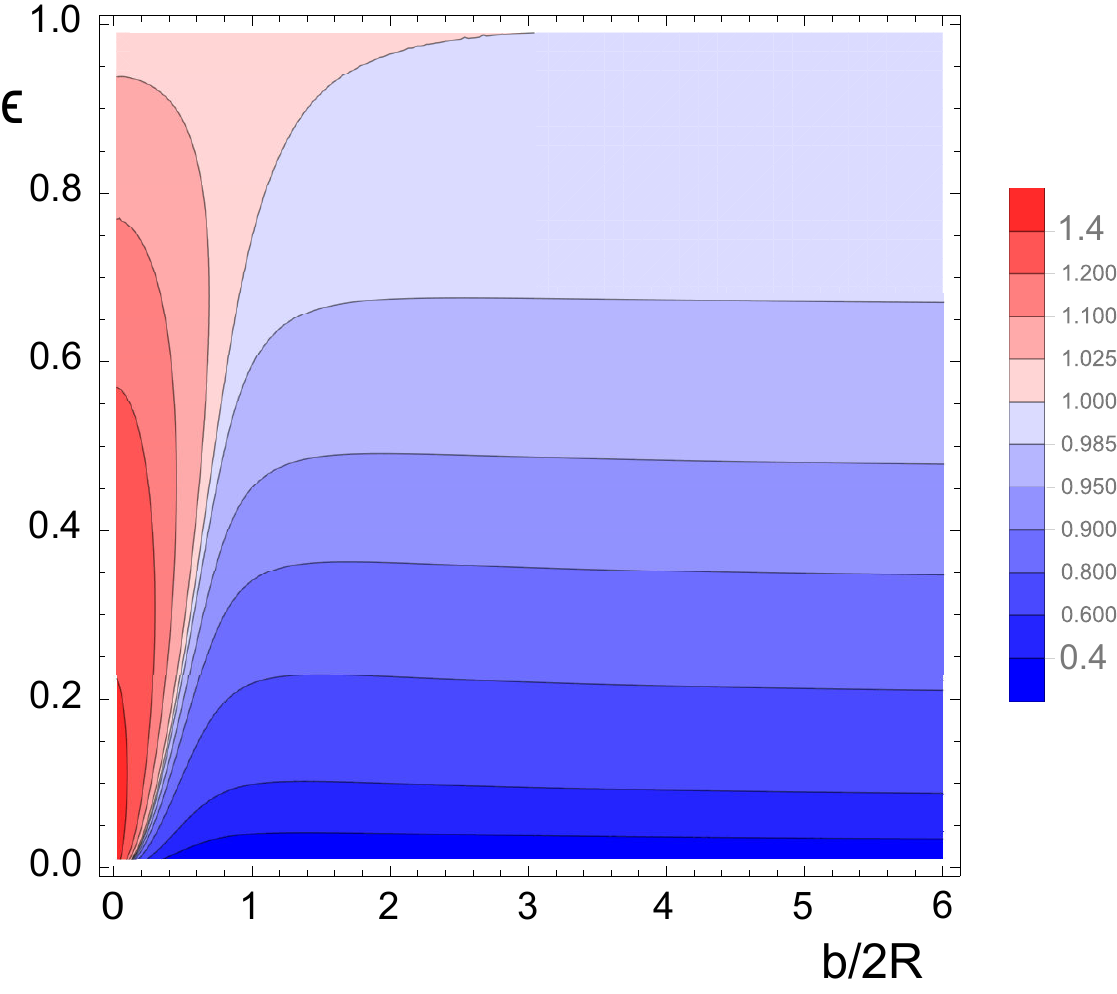}
		}%
		\subfigure{%
			\put(67,140){(c)}
			\label{fig:EGBECOblateVsParProlate}
			\includegraphics[width=0.325\textwidth]{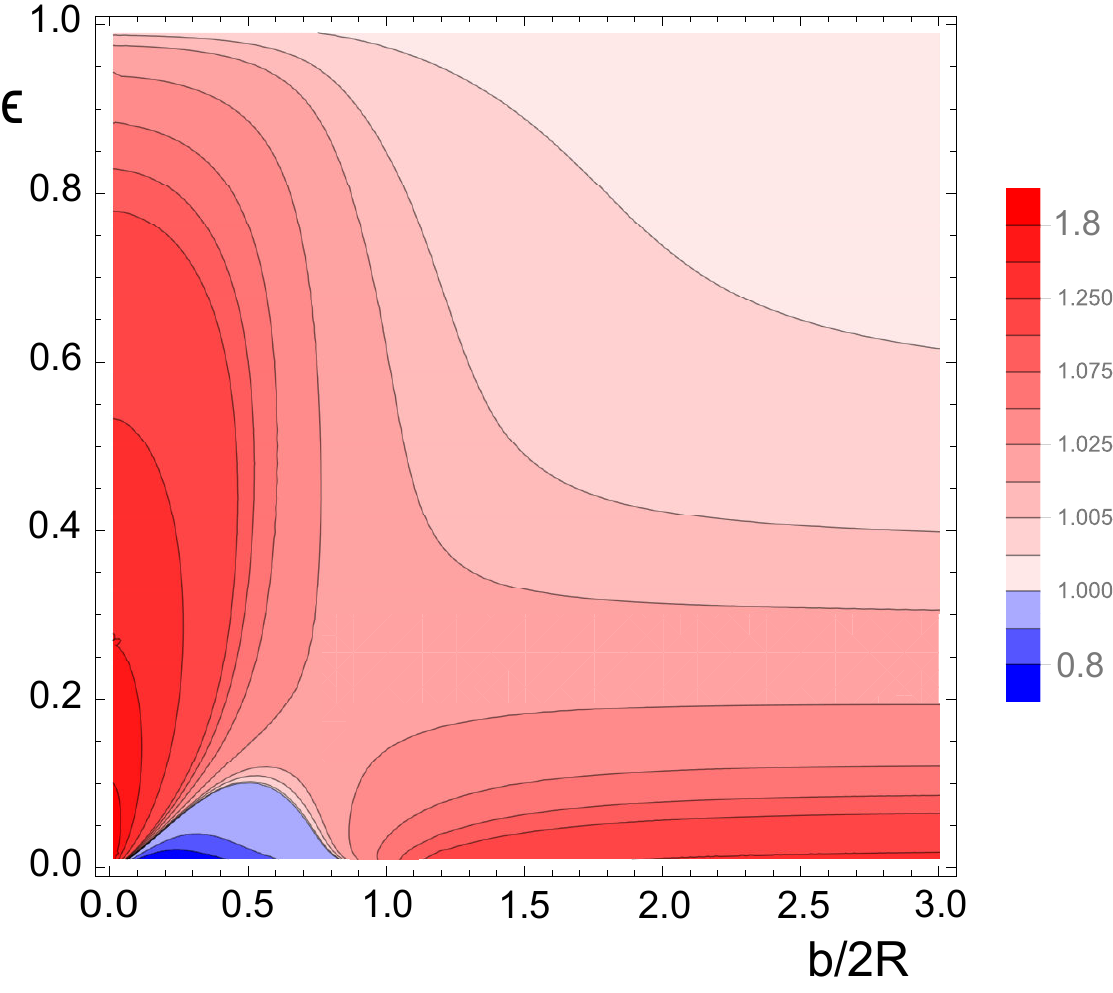}
		}\\ 
	\end{center}
	\vspace{-0.5cm}
	\caption{On the left is a contour plot of the gravitational self-energy of the difference between  displaced  oblate BECs  (displaced along the symmetry axes) over the that of displaced BEC spheres,  $E^{a)}_G/E^{sphere}_G$, in the TF regime. The x-axis is the distance $b$ between the centres of the states divided by two times the radius $R$, and the y-axis is the value of $\epsilon$ for the spheroid. The middle plot is as the left but with equatorial displaced prolate BECs rather than oblate ones i.e.\ $E^{c)}_G/E^{sphere}_G$, and the right plot is of $E^{a)}_G/E^{c)}_G$ for BECs in the TF regime.} 
	\label{fig:EGOblateVsSpheroid}
\end{figure*}

In Figures \ref{fig:EGBECOblateVsSphere}, \ref{fig:EGBECParProlateVsSphere} and \ref{fig:EGBECOblateVsParProlate} we compare  $E_G$ of the BEC sphere with a spheroidal BEC in configuration a), the BEC sphere with a spheroidal BEC in configuration d), and the spheroidal configuration a) with d), for all BECs in the TF regime and assuming the same volume and density for the different objects. These are very similar  to the uniform cases \ref{fig:EGOblateVsSphere}-\ref{fig:EGParProlateVsSphere} and illustrate again that it may be preferable to use spheroidal rather than spherical objects for testing GQSR.

\begin{figure*}[t!]
	\begin{center}
		\includegraphics[width=0.55\textwidth]{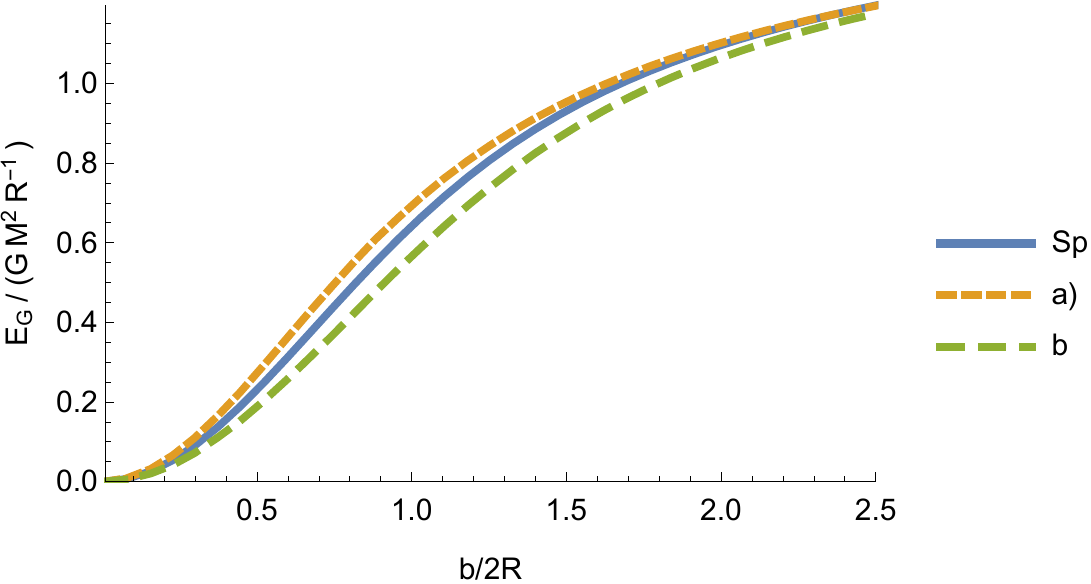}
	\end{center}
	\vspace{-0.5cm}
	\caption{The gravitational self-energy of the difference between  displaced  spherical and spheroidal BEC mass distributions, $E_G$, in the Gaussian regime, against $\lambda=b/(2R)$ where $R$ is the radius of the sphere, $M$ is the mass and $b$ is the distance between the centres of the states. All mass distributions have the same total mass and volume, and the spheroidal BECs have $\epsilon=0.75$. The solid blue line is for the sphere, and the two dashed lines are for the spheroidal configurations a) and b).} \label{fig:BECGaussian}
\end{figure*}

In Figure \ref{fig:BECGaussian}, we also plot spherical and spheroidal configurations a) and b) for BECs in the Gaussian approximation with $\epsilon = 0.75$ ($e \approx 0.7$) and displacement $b$ from zero to $10 R$. As in the TF approximation, the oblate case can have a value of $E_G$ that is greater than the spherical case. Note that for high values of ellipticity, it is possible to enter a quasi-one or two dimensional regime where the quantum and thermal motion can be frozen in two or one dimensions (see e.g.\ \cite{PitaevskiiBook}). This is to be distinguished from the case when the BEC looks lower dimensional from only a geometrical point of view but locally has a three-dimensional character. In certain configurations, it can be a good approximation to neglect the spatial dependence of the density in one or two dimensions, potentially simplifying the calculation of $E_G$ for such BEC states.

\subsubsection{\small{Self-energy difference in BEC experiments}} \hfill \break

\vspace{-4mm}
Now that we have calculated $E_G$ for mass distributions that can be generated by BEC experiments, let us consider what sort of experimental parameters might be required to test the gravitationally-induced state reduction model.  
Taking a spherical BEC for simplicity, when the separation of the two BEC states is of the order of their (effective) diameter, the value of $E_G$ is of the order (assuming $\gamma = 1 / (8 \pi)$ in \eref{eq:EG1}):
\begin{equation} \label{eq:EGBECTouching}
E_G \sim \frac{ G m^2 N^2}{ R}. 
\end{equation}
For example, in the TF approximation, when  two spherical BEC states are touching, the value of $E_G$ is found to be:
\begin{equation}
E_G = \frac{13 G m^2 N^2}{14 R}. 
\end{equation}

Using this expression for $E_G$, for a $^{133}\mathrm{Cs}$ BEC of radius $1 \, \mathrm{\mu m}$, we would need around $4 \times 10^9$ atoms in each state for a collapse lifetime of around $2\,\mathrm{s}$. In \eref{eq:EGBECTouching}, there is a stronger dependence on the number of atoms $N$ than on the radius $R$, and so $N = 4 \times 10^{10}$ and $R= 0.1\unit{mm}$ would also cause the same collapse rate, while potentially being more experimentally feasible due to the reduced density  (see Section \ref{sec:EnvDecoherence}). On the other hand, if, for example, $\gamma = 8 \pi$ were found to be more appropriate  in \eref{eq:EG1}, then a collapse time of $2\unit{s}$ would occur when $N \approx 10^{9}$ and $R= 0.1\unit{mm}$ or $N \approx 10^{8}$ and $R= 1\unit{\mu m}$. Allowing for smaller timescales than $2\unit{s}$ would also improve the required values for $N$ and $R$.

Although such numbers of atoms have not been achieved yet for a $^{133}\unit{Cs}$ BEC experiment, over  $10^9$ atoms were reported for a hydrogen BEC in 1998 \cite{HydrogenBEC}, and over $10^8$  atoms  for a $^{23}\unit{Na}$ BEC in 2006 \cite{LargeNaBEC} (also see \cite{Naik2005} for a  $^{23}\unit{Na}$ BEC of over $10^7$ atoms in 2004). These were single-well rather than double-well BECs, and so not large macroscopic superposition states. However, in Sections \ref{sec:DoubleWellBEC} and \ref{sec:EnvDecoherence} we discuss how large macroscopic states, such as NOON states, or approximations to these, could be generated in double-well BECs, and what sort of experimental parameters would be required in order for GQSR to be seen in the presence of environmental decoherence.

\subsection{Generating macroscopic  superposition states with double-well BECs} \label{sec:DoubleWellBEC}

A double-well BEC can, in principle, be used to create a macroscopic superposition state (see, for example, \cite{CatStatesBECs1998,MacroscopicSuperpositionsRuostekoski,MacroscopicStatesBEC1999,MacroscopicStatesBECs2001,CatStatesRuostekoski,CatsBECsPhaseImprint,CatStatesPhase2003,CatsAttractiveBECs,MacroscopicSuperpositionsDunningham,CatStatesBECs,NOONRepulsive}). The full Hamiltonian of this system is:

\begin{equation} \label{eq:DWHamiltonian}
\hat{H} = \int d^3 \bs{r} \hat{\Psi}\da(\bs{r}) \hat{H}_{DW} \hat{\Psi}(\bs{r}) + \frac{1}{2} g \int d^3 \bs{r}  \hat{\Psi}\da(\bs{r}) \hat{\Psi}\da(\bs{r}) \hat{\Psi}(\bs{r}) \hat{\Psi}(\bs{r}),
\end{equation}

where
\begin{eqnarray}\label{key}
&[\hat{\Psi}(\bs{r}) ,\hat{\Psi}\da(\bs{r}')] = \delta (\bs{r} - \bs{r}');\\
&\hat{H}_{DW} := \Big[-\frac{\hbar^2}{2m} \nabla^2 + V_{DW} (\bs{r}) \Big];
\end{eqnarray}

$V_{DW}$ is the particular double-well potential, which we take to be symmetric; and we have assumed that the inter-atomic interaction can be well-approximated by two-body $s$-wave scattering. 

Assuming that the energy barrier between the two wells is large enough, we make the ansatz that the BEC can be described as consisting of atoms that occupy a condensed state $\ket{\psi_L}$ of the left well, or a condensed state $\ket{\psi_R}$ of the right well, which are taken to be approximately orthogonal, $\langle \psi_L| \psi_R \rangle \approx 0$.  That is, we assume that $\hat{\Psi}$ can be approximated by:

\begin{equation} \label{eq:TwoModeAnsatz}
\hat{\Psi}(\bs{r},t) = \psi_L(\bs{r},t) \hat{a}_L (t) + \psi_R(\bs{r},t) \hat{a}_R (t),
\end{equation} 
where $\hat{a}_L$ and $\hat{a}_R$ are the annihilation operators for the states $\ket{\psi}_L$ and $\ket{\psi}_R$, which have localized wavefunctions $\psi_L$ and $\psi_R$. These obey the usual bosonic commutation rules:
\begin{equation}
[\hat{a}_i, \hat{a}\da_j] = \delta_{ij},\hspace{2cm}
[\hat{a}_i, \hat{a}_j] = 0,
\end{equation}
where $i,j=L,R$. The wavefunctions  $\psi_L$ and $\psi_R$ are assumed to have negligible overlap such that they are approximately orthogonal:

\begin{equation}
\int d\bs{r} \psi_i^{\ast}  (\bs{r},t) \psi_j (\bs{r},t)  \approx \delta_{ij},
\end{equation}
and the operator for the total number of particles, which is  conserved, is given by

\begin{equation}
\hat{N} := \int d \bs{r} \hat{\Psi} \hat{\Psi}\da  = \hat{a}_L\da \hat{a}_L + \hat{a}_R\da \hat{a}_R.
\end{equation}

In the non-linear tight-binding approximation \cite{NonLinearTBApprox},  an adiabatic approximation is applied where $\psi_L$ and $\psi_R$ are real and their spatial profiles adapt adiabatically to the instantaneous number of particles. In this tight-binding approximation, the wavefunctions   depend implicitly on time $t$ through the number of particles in each well $N_i = \braket{\hat{a}_i\da \hat{a}_i}$, i.e.\ $\psi_i(\bs{r},t) = \psi_i (\bs{r},N_i(t))$. In our large separation approximation, and assuming macroscopic occupation of the two states, the wavefunctions $\psi_L$ and $\psi_R$ (when multiplied by $\sqrt{N_L}$ and $\sqrt{N_R}$) obey, to a good approximation,  solutions of the Gross-Pitaevskii equation \eref{eq:GP} with potential $V_{DW}$ \cite{NonLinearTBApprox} (for, alternatively, a full variational approach, see \cite{DALTON2011668}).\footnote{In Section \ref{sec:BECSpheroids}, we used single-well potentials  to determine the density and shape of the two superposed states. This is a  good approximation at larger separations, but at smaller separations, as the states start to overlap further, the full character of the double-well potential will become more important, modifying the density, and the two-mode approximation discussed here will loose its validity. However, when comparing to environmental decoherence in Section \ref{sec:EnvDecoherence} we work with a rate of state reduction that is most appropriate  when the states are not overlapping.}

Plugging our ansatz \eref{eq:TwoModeAnsatz} into the Hamiltonian \eref{eq:DWHamiltonian}, we obtain

\begin{equation} \label{eq:HTwoMode}
\hat{H} = \hat{H}_1  + \hat{H}_2,
\end{equation}

where
\begin{eqnarray}
\hat{H}_1 &:= \xi_L \hat{a}_L\da \hat{a}_L +   \xi_R \hat{a}_R\da \hat{a}_R + J_{LR}  (\hat{a}_L\da \hat{a}_R + \hat{a}_R\da \hat{a}_L),\\\nonumber
\hat{H}_2 &:= U_L \hat{a}_L\datwo \hat{a}_L^2 + U_R  \hat{a}_R\datwo \hat{a}_R^2 + 4 U_{LRLR}  \hat{a}_L\da \hat{a}_R\da \hat{a}_L \hat{a}_R \\\nonumber
&~~~~~~~+ 2 U_{LLLR} (\hat{a}\datwo_L \hat{a}_L \hat{a}_R + ~h.c.)+  2 U_{RRRL} (\hat{a}\datwo_R  \hat{a}_R \hat{a}_L + ~h.c.) \\
&~~~~~~~+ U_{LLRR} (\hat{a}\datwo_L \hat{a}\datwo_R + ~h.c.),
\end{eqnarray}

with 
\begin{eqnarray}
\xi_L := \int d^3 \bs{r} \psi_L \hat{H}_{DW} \psi_L,~ \xi_R := \int d^3 \bs{r} \psi_R \hat{H}_{DW} \psi_R, \\
J_{LR} := \int d^3 \bs{r} \psi_L \hat{H}_{DW} \psi_R,~U_L := g\int  d^3 \bs{r} \psi_L^4, ~U_R := g\int  d^3 \bs{r} \psi_R^4,\\
U_{LRLR} := g\int  d^3 \bs{r} \psi_L^2 \psi_R^2,~U_{LLRR} := g\int  d^3 \bs{r} \psi_L^{ 2} \psi_R^2,\\
U_{LLLR} := g\int  d^3 \bs{r} \psi_L^2 \psi_L \psi_R,~U_{RRRL} := g\int  d^3 \bs{r} \psi_R^2 \psi_R \psi_L,
\end{eqnarray}

and we are taking $\psi_L$ and $\psi_R$ to be real. The Hamiltonian \eref{eq:HTwoMode} can be shown to contain an analytic solution \cite{AnalyticalBECs1}. It can also be approximated by an extended two-mode Bose-Hubbard model in the non-linear tight-binding approximation \cite{NonLinearTBApprox}. In the case that the spatial profile of $\psi_L$ and $\psi_R$ is approximately independent of the number of particles in each well (the standard tight-binding approximation), it can be further approximated with the two-mode version of the Bose-Hubbard model \cite{NonLinearTBApprox,BoseHubbard,BoseHubbardLattice,BoseHubbardBEC1,PitaevskiiBook}:

\begin{equation} \label{eq:BoseHubbard}
\hat{H} =  E_{LR} (\hat{a}_L\da \hat{a}_R + \hat{a}_R\da \hat{a}_L) + \frac{1}{2}  U ( \hat{a}_L\datwo \hat{a}^2_L + \hat{a}_R\datwo \hat{a}^2_R),
\end{equation}

where $U = U_L$; we have assumed $U_L=U_R$; we have removed terms proportional to the number operator $\hat{N}$ since this commutes with $\hat{H}$; and we have neglected any atomic collisions in the overlapping region of the two modes. Here the $J_{LR}$ terms are responsible for quantum tunnelling between the two wells, and the $U$ terms are the atom-atom interactions within each well.

There have been several proposals for generating  a macroscopic superposition state (Schr\"{o}dinger cat state) in a double-well BEC. For example, in \cite{CatStatesBECs1998,CatsAttractiveBECs} it is considered that, starting from a repulsive BEC, if the interaction strength $g$ is varied adiabatically to a negative value (using a Feshbach resonance), then a cat state can be prepared. This occurs because a NOON state is the ground state of the two-mode Bose-Hubbard model \eref{eq:BoseHubbard} with strong attractive interactions. In \cite{CatStatesBECs}, it is shown that the ground state becomes degenerate with the first excited state in this case, such that there needs to be an exponentially long time to create an exact NOON state. However, in \cite{DemonstratingNOON} it was found that, for realistic parameters and time-scales, an approximate NOON state can be generated with a smooth change in the scattering length. An alternative to this method is to use a Feshbach resonance  to make a sudden change in the scattering length \cite{CatStatesBECs}. For example, a repulsive BEC could be prepared in a single-well and then the tunnelling barrier is raised adiabatically to divide the well into two equal parts (forming a so-called `coherent' state \cite{PitaevskiiBook} when neglecting interactions), then a Feshbach resonance is used to suddenly switch $g$ from a positive to a negative value such that the state dynamically evolves to a large macroscopic superposition state.

Another possibility would be to set the scattering length to zero and drive the system to an upper excited state, then slowly increase the interactions (keeping them repulsive) while, at the same time, decreasing the inter-well tunnelling to zero \cite{NOONRepulsive}. This method is possible since a NOON state is the upper energy state of the repulsive Bose-Hubbard model, and has the advantage that the BEC does not need to move to an attractive state, which can become unstable \cite{BECReview}. Rather than modifying the scattering length, a cat state could also be generated by manipulating the BEC with an external laser \cite{MacroscopicSuperpositionsRuostekoski,CatStatesRuostekoski,CatsBECsPhaseImprint}. For example, in \cite{CatsBECsPhaseImprint}, it is suggested that a far off-resonance laser could be used to imprint a $\pi$-phase on one of the wells such that the quantum wavepacket bifurcates.  The tunnelling barrier is then raised to halt the evolution and fix the cat state.

Once a macroscopic superposition state, such as a NOON state, has been prepared, we need to make sure that we can experimentally distinguish it from a classical statistical mixture. For a double-well BEC, one possibility is to look for a a non-zero Nth-order correlation  $\braket{\hat{a}_L\daN \hat{a}_R^N}$ \cite{DemonstratingNOON,QuantifyingNOONCoherence}.  For an exact NOON state, $\ket{\mathrm{NOON}}$, where we have a superposition of $N$ particles in the left-hand state $\ket{\psi_L}$ and $N$ particles in the right-hand state $\ket{\psi_R}$, which we write as $(\ket{\mathrm{N0}} + \ket{\mathrm{0N}})/\sqrt{2}$, we have

\begin{equation} \label{eq:NOON0}
\bra{\mathrm{NOON}} \hat{a}_L\daN \hat{a}_R^N \ket{\mathrm{NOON}} = \frac{N!}{2},
\end{equation}

whereas, for a statistical mixture, we have zero. Experimental methods for measuring $\braket{\hat{a}_L\daN \hat{a}_R^N}$ in double-well BECs can be found in \cite{DemonstratingNOON,QuantifyingNOONCoherence}.

As well as being able to distinguish a NOON state from a statistical mixture, we also need need to make sure that we can experimentally distinguish the GQSR process from environmental decoherence.  That is, we would ideally like the objective collapse rate to be greater than the rate of environmental decoherence. Given an initial NOON state, $\ket{\mathrm{NOON}}$, we can use \eref{eq:Ps}-\eref{eq:Pd} to write down the density operator for the state under the process of GQSR:
\begin{eqnarray}
\hat{\rho} (t) &= \frac{1}{2} e^{- E_G t / \hbar} \Big[ \ket{\mathrm{N0}} + \ket{\mathrm{0N}}\Big]\Big[ \bra{\mathrm{N0}} + \bra{\mathrm{0N}}\Big]\\ &\hspace{1cm}+ \frac{1}{2} \Big(1-e^{- E_G t / \hbar}\Big) \Big[ \ket{\mathrm{N0}} \bra{\mathrm{N0}} + \ket{\mathrm{0N}}\bra{\mathrm{0N}}\Big].
\end{eqnarray}

In terms of the annihilation operators of the left and right states, $\hat{a}_L$ and $\hat{a}_R$, we have 

\begin{equation}
\ket{\mathrm{NOON}}= \frac{1}{\sqrt{2 N!}} (\hat{a}_L\daN + \hat{a}_R\daN) \ket{0},
\end{equation}

such  that the $N$-particle correlation $\braket{\hat{a}_L\daN \hat{a}_R^N}$ evolves in time as 
\begin{equation}\label{eq:NOONCollapseRate}
\braket{\hat{a}_L\daN \hat{a}_R^N}(t) = e^{-E_G t /\hbar} \braket{\hat{a}_L\daN \hat{a}_R^N}_0,
\end{equation}

where $\braket{\hat{a}_L\daN \hat{a}_R^N}_0$ is given by \eref{eq:NOON0}. We now compare this evolution of the $N$-particle correlation to that imposed by various environmental decoherence channels in double-well BECs.

\subsection{Environmental decoherence} \label{sec:EnvDecoherence}

There are several channels of  environmental decoherence in BEC systems. Here we concentrate on the prominent ones due to three-body recombination, interactions with the thermal cloud, and  interactions with foreign atoms. We also briefly discuss noise  due to the trapping potential.

\subsubsection{\small{Three-body recombination}} \label{sec:ThreeBody} \hfill \break

\vspace{-4mm}

Three-body recombination is the process where three atoms in the condensate collide to form a molecule (atom-atom bound state) and a single atom, which can both then escape the trap. This process often limits the lifetime and size of condensates. In \cite{3BodyDecoherence}, a master equation was derived for three-body loss in the Born-Markov approximation and for a BEC with repulsive interactions.  Since this is a three-body problem, this master equation is of the following  form for a double-well BEC in the two-mode approximation \cite{CSLBECs}:

\begin{equation}\label{key}
\frac{d \hat{\rho} (t)}{dt} = - \frac{i}{\hbar} [ \hat{H}, \hat{\rho}(t)] + \gamma_3 \sum_{k = L,R}[ \hat{a}_k^3 \hat{\rho}(t) \hat{a}\dathree_k - \frac{1}{2} \{ \hat{a}_k\dathree \hat{a}^3_k, \hat{\rho}(t)\}],  
\end{equation}

where
\begin{eqnarray}
\gamma_3 &:= \frac{K_3}{72} \int d^3 \bs{x} |\psi_{L,R}(\bs{r})|^6\\
&= \frac{K_3}{72} n^2,
\end{eqnarray}

with $n$  the condensate number density and $K_3$  the recombination event rate, which can be approximated as \cite{ThreeBodyRate}:

\begin{equation}\label{key}
K_3 = 23 \frac{\hbar}{m} a_s^4.
\end{equation} 

The N-particle correlation for a NOON state under this master equation is then \cite{CSLBECs}:

\begin{equation}
\braket{\hat{a}\daN_L \hat{a}^N_R} (t) = e^{- \gamma_3 N t} \braket{\hat{a}\daN_L \hat{a}^N_R}_0.
\end{equation} 

Comparing to the gravitationally-induced collapse rate  for a NOON state \eref{eq:NOONCollapseRate}, we   require that

\begin{equation}
E_G / \hbar \gg \gamma_3 N.
\end{equation}

Taking $E_G$ to be of the form \eref{eq:EGBECTouching} for simplicity,  we  need

\begin{equation}
\frac{ G m^2 N}{ R} \gg  \frac{23 }{72 m} a_s^4 \hbar^2 n^2.
\end{equation}

Assuming, for example, a $^{133}\unit{Cs}$ BEC with $N \sim 4 \times 10^9$ and $R \sim 10 \, \mu m$ (such that $\tau \sim 10\,s$), then to obtain a three-body recombination rate that is ten times slower in the TF regime, we would need to utilize a Feshbach resonance in order to reduce the scattering length  by approximately four orders of magnitude (and we take the trapping frequency to be around $300\unit{Hz}$). Increasing the number of atoms to $4 \times 10^{10}$ instead, then a radius $R \sim 0.1\unit{mm}$ (trap frequency  $10\unit{Hz}$) and a reduction in the scattering length by three orders of magnitude would be enough.   Assuming instead a Gaussian approximation, then in order to operate in this regime, the trapping frequency and/or  scattering length need to be reduced further, which only lowers the decoherence rate. Note that, if  it were found to be more appropriate to take $\gamma = 8 \pi$ rather than $\gamma = 1 / (8 \pi)$ in \eref{eq:EG1}, then this would increase $E_G$ by almost three orders of magnitude,  significantly improving the experimental feasibility. For example, in this case we could have around $6 \times 10^8$ atoms and $R = 0.1\unit{mm}$, with the interaction strength reduced by two orders of magnitude.

In several of the proposals to create a NOON state that were discussed in Section \ref{sec:DoubleWellBEC}, the (attractive or repulsive) interaction strength is modified. For example, in \cite{NOONRepulsive}, the repulsive interaction strength is  increased while the inter-well tunnelling is reduced to zero.  In this case, once the NOON state is prepared (or a good approximation to one) the interaction strength would  then likely have to be reduced in order to prolong the coherence of, at least an approximation to, the state in light of three-body interactions. Alternatively, other methods could be employed to inhibit three-body decay, such as using an external laser    \cite{InhibitingThreeBody,InhibitingThreeBodyZeno} or lowering the effective dimensionality of the BEC \cite{ThreeBodyOneD,ThreeBodyTwoD,ThreeBodyTwoD2015}. Here we have  assumed a three dimensional BEC throughout. However, although condensation cannot occur in one or two dimensional uniform systems, with a harmonic trap it is possible to have condensation in an ideal Bose gas in two dimensions,  and macroscopic occupation of the lowest energy state in one dimension at finite temperatures \cite{BECReview}. These lower dimensional systems can be  achieved when one or two of the harmonic trapping frequencies are much higher than the  others, i.e.\ in the limit of a very flat oblate spheroid or  thin prolate spheroid. Unlike in three dimensions, in a Bose gas of one or two dimensions, the three-body decay can become temperature dependent and vanishing at absolute zero. Therefore, reducing the effective dimensionality of the trap, and operating at low temperatures  may be another possibility for inhibiting decoherence due to three-body decay. 

As stated in Section \ref{sec:GaussianDensity}, for a BEC with attractive interactions, the condensate becomes unstable if the number of atoms exceeds a critical value $N_c$, which for a spherical trap at zero temperature is given by \eref{eq:CritNum}.  Therefore, if a NOON state is formed with an attractive BEC, the number of atoms $N$ needs to be lower than   $N_c$ \cite{PitaevskiiBook}. One possibility is to increase $N_c$ by lowering the scattering length. However, an exact NOON state from this method is only obtained in the limit of infinite attractive interactions \cite{DemonstratingNOON}. Therefore, a lower $a_s$ would likely lead to an approximation to a NOON state, for which we would have to calculate the rate of GQSR, and will be the concern of future work. It may instead be preferable to utilize one of the methods outlined in Section \ref{sec:EnvDecoherence} that generates a NOON state with a repulsive BEC.

\subsubsection{\small{Thermal cloud interactions}} \label{sec:Thermal} \hfill \break

\vspace{-4mm}

Interactions between the condensate atoms and atoms in the thermal cloud (the noncondensed atoms due to a finite temperature) will also lead to decoherence of a NOON state \cite{CatsThermal,MacroscopicStatesBECs2001}.  These interactions can be of three types: single particle loss $C + NC \rightarrow NC + NC$, two particle loss $C + C \rightarrow NC + NC$, and scattering $C + NC \rightarrow C + NC$ (together with the opposite processes) \cite{ThermalInteractionsThesis}. In \cite{CatsThermal}, assuming a Born-Markov  and standard tight-binding approximation,  a quantum master equation was derived for the scattering process, where the thermal cloud environment can learn the quantum state of the condensate system.  This is of the form:

\begin{equation}\label{eq:ThermalMasterEq}
\frac{d \hat{\rho} (t)}{dt} = - \frac{i}{\hbar} [ \hat{H}, \hat{\rho}(t)] - \gamma_t [ \hat{a}\da_L \hat{a}_L - \hat{a}\da_R \hat{a}_R, [\hat{a}\da_L \hat{a}_L - \hat{a}\da_R \hat{a}_R, \hat{\rho}(t)]],
\end{equation}

where 
\begin{equation}
\gamma_t :=64 \pi^4 a^2_s n_{th} v_t,
\end{equation}

with $v_t := \sqrt{2 k_B T / m}$ being the thermal velocity of the atoms in the thermal cloud;  $T$  the temperature; and $n_{th}$ the thermal cloud number density, which can be approximated by:
\begin{eqnarray}
n_{th} &= \frac{e^{-\mu /k_B T}}{V_{th}} \Big(\frac{k_B T}{\hbar \omega}\Big)^3,\\
V_{th} &:= \frac{4}{3} \pi R^3_{th},\\
R_{th} &:= \sqrt{\frac{2 k_B T}{m \omega^2}} ,
\end{eqnarray}

where $\omega := (\omega_x \omega_y \omega_z)^{1/3}$ and $\omega_{x,y,z}$ are the various harmonic trapping frequencies. In \cite{QuantumKineticTheoryIII,QuantumKineticTheoryIV}, a master equation was derived for   the scattering of thermal particles off a single condensate within the TF regime.  The rate $\gamma_t$ in this case becomes \cite{QuantumKineticTheoryIV}:
\begin{eqnarray}
\gamma_t &= \frac{4 k_B T \mu_{TF}^4}{9 \pi^4 \hbar^5 \omega^4 N^2} e^{\mu/k_B T},
\end{eqnarray}

where  $\mu$ is the chemical potential of the non-condensed cloud, we have assume a spherical trapping potential, and 

\begin{equation}
\mu_{TF} = \frac{1}{2} \hbar \omega \Big( 15 N a / R_0\Big)^{2/5}
\end{equation} 
is the chemical potential of a spherical BEC in the TF regime.

The N-particle correlation for a NOON state under the master equation \eref{eq:ThermalMasterEq} is:

\begin{equation}
\braket{\hat{a}\daN_L \hat{a}^N_R} (t) = e^{- \gamma_t N^2 t} \braket{\hat{a}\daN_L \hat{a}^N_R}_0.
\end{equation} 

Taking $E_G$ to be of the form \eref{eq:EGBECTouching}, we require

\begin{equation}
\frac{ G m^2 }{ R \hbar} \gg  \gamma_t.
\end{equation}

Assuming a Gaussian $^{133}\unit{Cs}$  BEC with the interaction strength reduced by six orders of magnitude and $4 \times 10^9$ atoms, as considered as a possibility in the previous section, we would need to increase the trapping potential so that the radius is of order $1\unit{\mu m}$, and operate at a temperature $T \lesssim 1 \unit{nK}$.   A temperature of $0.5\unit{nK}$ has been achieved  for a low-density $^{23}\unit{Na}$ BEC in  a single-well potential \cite{05nKBEC}. If, on the other hand, we want to work in the TF regime, then more challenging experimental parameters appear to be necessary. For example, environmental decoherence would be five times slower than collapse when there is  
around $4 \times 10^{11}$ atoms in a condensate of radius of $0.1\unit{mm}$, the interaction strength is reduced by a further   two orders of magnitude as compared to the TF regime considered in the previous section, and the temperature is  $T \lesssim 0.1 \unit{nK}$. Therefore, as suggested in the previous section, if a NOON state is prepared by  changing the interaction strength then, to prolong the lifetime of the state, it would be preferable to subsequently significantly reduce the interaction strength so that we are working with an approximately ideal gas. 

The temperature bound and/or the condensate radius can be increased in  the Gaussian approximation by further lowering the interaction strength and, at the same time, either keeping the total atom number the same or increasing it.  Also, trap engineering and symmetrization of the environment would help  \cite{CatsThermal}.  However,   it is possible that a Born-Markov approximation is not appropriate for the description of thermal cloud decoherence in this case, such that the estimates provided here would be inaccurate \cite{CatStatesBECs}. Furthermore, as discussed in the previous section, it is possible that these values could be improved if we took  $\gamma = 8 \pi$ rather than  $\gamma =1/ (8 \pi)$ in \eref{eq:EG1}. For example, in this case it would be possible to lower the total atom number to $10^8$ while keeping the rest of the parameters the same.

\subsubsection{\small{Foreign atom interactions}} \label{sec:Vacuum} \hfill \break

\vspace{-4mm}

Decoherence can also occur due to interactions with background gas particles at room temperature.\footnote{Here we assume the background gases operate at room temperature, but it is also possible that the vacuum chamber could be  cryogenically cooled.}. These foreign particles collide with the condensate atoms and can either cause them to leave the trap entirely or heat up \cite{VacuumRateAndHeating}. Assuming that all collisions cause atoms to leave the condensate, a master equation for this process was derived in \cite{VacuumMaterEq} assuming a Born-Markov approximation.  Since this is a one-body process, this master equation is of the form:

\begin{equation}\label{eq:vacuumMasterEq}
\frac{d \hat{\rho} (t)}{dt} = - \frac{i}{\hbar} [ \hat{H}, \hat{\rho}(t)] + \gamma_f \sum_{k = L,R}[ \hat{a}_k \hat{\rho}(t) \hat{a}\da_k - \frac{1}{2} \{ \hat{a}_k\da \hat{a}_k, \hat{\rho}(t)\}].
\end{equation}

A rough estimate of the rate $\gamma_f$ can be calculated assuming only $s$-wave scattering \cite{VacuumMaterEq}:
\begin{equation}
\gamma_t \sim \frac{1}{\sqrt{6}} \sigma (u_f) n_f u_f,
\end{equation}

where $n_f$ is the number density of foreign atoms, $u_f $ is their average velocity, and $\sigma (u_f)$ is the cross-section for the process. Using kinetic theory, we can approximate  these quantities by \cite{VacuumRateAndHeating}:

\begin{equation}
u_f = \sqrt{2 k_B T / m_f},\\
n_f = P / (k_B T),\\
\sigma_f = 7.57 (1.033)^2 \Big(\frac{C_6}{\hbar u_f}\Big)^{2/5},
\end{equation}

where $P$ is the pressure of the vacuum chamber and $C_6$ is the Van der Waals constants from the Van der Waals potential $V(r) = - C_6 / r^6$.  Various interaction cross-sections have been calculated for these processes in \cite{VacuumRateAndHeating} assuming $T$ is room temperature.

The N-particle correlation for a NOON state under the master equation \eref{eq:vacuumMasterEq} is then:

\begin{equation}
\braket{\hat{a}\daN_L \hat{a}^N_R} (t) = e^{- \gamma_f N t} \braket{\hat{a}\daN_L \hat{a}^N_R}_0,
\end{equation} 

so that we require
\begin{equation}
\frac{ G m^2 N }{ R \hbar} \gg  \gamma_f.
\end{equation}

Note that the environmental decoherence rate here is equal to the atomic loss rate. This is because, for an exact NOON state, the loss of one atom means that the density operator is now a mixture of the states $\ket{(\mathrm{N}-1) 0}$ and $\ket{0 (\mathrm{N}-1)}$, neither of which is itself a NOON state \cite{CatStatesBECs}. Therefore, one scattering event is enough to collapse the NOON state into all-left or all-right states. However, in practice it is unlikely that an exact NOON state will be formed, and instead a more general macroscopic superposition state, such as $\ket{\psi} = (\ket{\mathrm{N}/10,9\mathrm{N}/10} + \ket{9\mathrm{N}/10,\mathrm{N}/10})/\sqrt{2}$, would be more probable. Indeed, these type of states would be formed in the process where the scattering length is suddenly changed to a negative value \cite{CatStatesBECs}. Single-atom losses for these states would still result in similarly `good' macroscopic superposition states, such that the effect of scattering a foreign atom may not have a significant detrimental effect  \cite{CatStatesBECs}. Of course we also need to determine how the GQSR rate might change for the approximate NOON states, and this will be the concern of future work.

\subsubsection{\small{Decoherence from the trapping potential}} \label{sec:TrapNoise} \hfill \break

\vspace{-4mm}

Optical, magnetic or opto-magnetic traps can be used for the implementation of the double-well potential. These electromagnetic traps can also cause decoherence of a NOON state. For example, in an optical trap, decoherence of NOON states can come from spontaneously scattered/diffracted photons \cite{DecoherencePhotonEmission,CatStatesBECs}, and phase noise  of the laser beam \cite{PhaseLaserNoiseBEC}. However, to generate the required large numbers of atoms, it is likely that a pure magnetic trap should be used, such as that in \cite{BFieldDoubleWellTrap,SchummThesis}.  Surprisingly, decoherence of macroscopic superposition states due to fluctuations of a  magnetic field has been found to be  independent of the total particle number \cite{MagNoiseInDoubleWell},  improving the feasibility of generating such states.

\setcounter{footnote}{0}

\section{Conclusions} \label{sec:Conclusions}

We have investigated testing a unified theory of GR  and QT with a Bose-Einstein condensate. In particular, we have considered testing a proposal for a unified theory that is based on the `gravitizing quantum theory' approach rather than the conventional `quantizing gravity' approach. In Section \ref{sec:PenroseCollapse}, we examined how, if we attempt to make QT consistent with the equivalence principle of GR, then a possible resolution is to consider making modifications to QT that would lead to a violation of the superposition principle of QT where the degree of violation is dependent on the gravitational interaction and configuration of the system. Since this increases for more massive systems, the proposal can provide an objective state reduction that is consistent with current experiments,  thus resolving the measurement problem of QT, which would, on other hand, be expected to persist for the `quantizing gravity' approach and conventional quantum gravity theories. QT is predicted to breakdown  when the mass of a quantum system is near the Planck mass scale, allowing for experimental tests that are far more achievable than those generally required for distinguishing conventional quantum gravity theories, where the relevant effects are anticipated near the Planck length scale.

In the proposal considered here for a unified theory of GR and QT,  quantum superposition states are expected to decay to localized states with an average lifetime that is (in the Newtonian limit) reciprocally related to the self-energy of the difference between the mass distributions of the localized states, $E_G$ \cite{Penrose1996}, which is dependent on the mass and configuration of the system. This has been generically considered for displaced, uniform, spherical mass distributions. However, BECs tend to have non-uniform mass distributions, and so we have extended this to the quadratic and Gaussian density distributions that are usually found in BEC experiments, but which may also be applicable to other systems. Since they are often  generated in BEC experiments, we have also considered non-uniform spheroidal mass distributions, as well as uniform ones that would be approximated in nano/micro-object experiments, finding that the average lifetime of state reduction can be reduced compared to the spherical case. Due to the particular dependence that the gravitationally-induced quantum state reduction (GQSR) considered here has on the geometry of the superposed objects, this analysis could also be used to distinguish the GQSR from other, and potentially non-gravitational, collapse models, such as the continuous spontaneous
localization (CSL) model \cite{CSL}.

To probe the GQSR, we have considered a BEC in a double-well trap that is placed in a macroscopic superposition state of two locations.
Assuming that the state reduction is a Poisson process similar to particle decay, we have compared the rate of wavefunction collapse against prominent channels of environmental decoherence in BEC systems. For the rate of decoherence to be significantly less than the rate of collapse, we estimate that the BEC should have greater than $10^8$ or $10^9$ atoms, depending on the choice of a free parameter in the GQSR proposal, and that the scattering length is reduced using an external magnetic field while maintaining a macroscopic superposition state. Being able to control the atom-atom interactions  provides a unique 
asset to BEC tests.

We have concentrated on exact NOON states for estimating collapse and decoherence rates. However, as with experimental proposals based on nano/micro-objects, these states would be challenging to create and approximations to these  states are more likely to be generated in  experiments. Although estimating environmental decoherence for approximate NOON states is a relatively simple task, the GQSR needs to be extended to be able to handle these states.
One possibility is to follow the approach of Di\'{o}si \cite{Diosi1989}, but there may be other, more general, alternatives, which will be the concern of future work.  We have also concentrated on only three-dimensional BECs, but prolate and oblate BECs with high ellipticity could move into a quasi-one and -two dimensional regime, potentially reducing environmental decoherence processes such as that from three-body recombination \cite{ThreeBodyOneD,ThreeBodyTwoD,ThreeBodyTwoD2015}. In this case, environmental decoherence could be reduced relative to spherical BECs, whereas the collapse rate would be increased,  improving the feasibility of experimental tests. 

If signals of this proposal were not observed in experiments, then, depending on the achievable experimental parameters, this could place severe constraints on the model (for example, the value of $\gamma$), and potentially rule it out. It would also likely place constraints on other models of objective state reduction, such as CSL \cite{CSL}. However, if signals were observed, then we would have the first evidence of how GR and QT must combine to form a consistent, unified theory. Furthermore, it would  explain the mysterious measurement process in QT and provide it with a well-defined classical limit.

\ack
We thank participants of the Gravity in the lab 2018 workshop in Benasque, Spain for useful discussions and comments. In particular, we would like to thank  Philippe Bouyer, Devang Naik, Hendrik Ulbricht, Marko Toro\v{s}, Daniel Goldwater and Michael Tolan. RP  would  like to  thank John Moussouris for a personal endowment  through the University of Oxford. RH and IF would like to acknowledge the support of the Penrose Institute.

\newpage 
\appendix

\setcounter{footnote}{0}

\section{$E_G$ for uniform, spherical mass distributions} \label{app:UniformSphere}

Here we calculate the self-energy of the difference between uniform, displaced spherical mass distributions. Taking the radius of the sphere states to be $R$ and their mass $M$, then their density functions are defined by
\begin{equation}
\rho(\bs{r})  = \cases{
	\rho_0 := M / (\frac{4}{3} \pi R^3) &if $\bs{r}$ is inside sphere, \\
	0  &otherwise.}
\end{equation}
In terms of the step function $\theta(x)$, we can write the density functions as
\begin{eqnarray}\label{key}
\rho(\bs{r}) &= \rho_0 \theta(R-r_{\rho}) \theta(R^2-r^2_{\rho}-z^2),\\
\rho'(\bs{r}) &= \rho_0 \theta(R-r_{\rho}) \theta(R^2-r^2_{\rho}-(z-b)^2), 
\end{eqnarray}

where $r_{\rho}$ is the cylindrical radial coordinate, and we have taken the  sphere states to be displaced along the $z$ coordinate by a distance $b$, with the $\rho(\bs{r})$ sphere state being at the origin of our coordinate system.

These density functions can be plugged into \eref{eq:EG2}. However, we find it simpler to work with \eref{eq:EG1} to calculate $E_G$ for this situation. In this case we need the gravitational potential inside and outside a sphere:
\begin{equation}\label{eq:PhiInsphere}
\phi = \cases{
	\phi_{in} := -\frac{G M}{R} \Big(\frac{3}{2} - \frac{r^2}{2 R^2}\Big)  &if  $r \leq R$, \\ \label{eq:PhiOutsphere}
	\phi_{out} :=  -\frac{G M}{r} &if  $r \geq R$,}
\end{equation}
where now we are working in spherical coordinates $(r, \theta, \psi)$. Taking $\gamma = 1 / (8 \pi)$ then, due to the symmetry of the problem, we can use
\begin{eqnarray}\label{key}
E_G &= \frac{1}{2} \int (\phi - \phi')(\rho' - \rho) d^3 \bs{r}\\ \label{eq:EGSphereSimple}
&= \int \phi (\rho' - \rho) d^3 \bs{r}.
\end{eqnarray}

We first consider the term $\phi \rho'$, which is related to the gravitational interaction energy \cite{Penrose1996}. When $b > 2 R$ then, following Gauss's law, the gravitational interaction energy is simply $- G M^2 / b$. We can  calculate this by  choosing the origin of our coordinate system to be at the centre of the $\rho(\bs{r})$ sphere and  integrate its potential over the density of the other sphere using surfaces of constant radial coordinate $r$:
\begin{equation}
\int \phi \rho' d^3 \bs{r} = - \rho_0 \int^{2 \pi}_0 \int^{\cos^{-1} (z_r/r)}_0  \int^{b+R}_{b-R} \phi_{out} r^2 \sin \theta dr d \theta d \psi = -\frac{G M^2}{b}.
\end{equation} 
Using the same method for $R \leq b \leq 2 R$, we find:
\begin{eqnarray}\nonumber
\int \phi \rho' d^3 \bs{r} &= - \rho_0 \int^{2 \pi}_0 \int^{\cos^{-1} (z_r/r)}_0 \Big[ \int^{R}_{b-R} \phi_{in} \\&~~~~~~~~~~~~~~~~~~~+ \int^{b+R}_{R} \phi_{out}\Big]
r^2 \sin \theta dr d \theta d \psi\\
&= \frac{G M^2}{R} \Big(-\frac{6}{5} + 2 \lambda^2 - \frac{3}{2} \lambda^3 + \frac{1}{5} \lambda^5\Big).
\end{eqnarray}

Finally, for $b \leq R$, we have:
\begin{eqnarray}\nonumber
\int \phi \rho' d^3 \bs{r} &= - \rho_0 \int^{2 \pi}_0 \Big[\Big( \int^{\pi}_0 \int^{R-b}_{0} + \int^{\cos^{-1} (z_r/r)}_0  \int^{R}_{R-b}\Big)  \phi_{in} \\&~~~~~~~~~~+ \int^{\cos^{-1} (z_r/r)}_0  \int^{R}_{R+b} \phi_{out} \Big]
r^2 \sin \theta dr d \theta d\psi\\
&= \frac{G M^2}{R} \Big(-\frac{6}{5} + 2 \lambda^2 - \frac{3}{2} \lambda^3 + \frac{1}{5} \lambda^5\Big),
\end{eqnarray}  

which, as expected is the same as the previous result. Now we consider the term $\phi \rho$ in \eref{eq:EGSphereSimple}. We can simply extract this from the above result when $b=0$ such that:
\begin{equation}
\int \phi \rho d^3 \bs{r}  = -\frac{6 G M^2}{5 R},
\end{equation}
which is simply twice the gravitational  self-energy of a sphere. Putting this altogether, we then obtain:
\begin{equation}
E_G = \cases{
	\frac{G M^2}{R} \Big(  2 \lambda^2 - \frac{3}{2} \lambda^3 + \frac{1}{5} \lambda^5\Big) &if  $0 \leq \lambda \leq 1$, \\
	\frac{G M^2}{R} \Big( \frac{6}{5} -  \frac{1}{2 \lambda}\Big) &if  $\lambda \geq 1$.}
\end{equation}

\section{$E_G$ for BEC, spherical mass distributions in the Thomas-Fermi approximation}
\label{app:BECSphere}

Here we calculate the self-energy of the difference between   displaced spherical BEC mass distributions in the Thomas-Fermi approximation. Taking the radius of the sphere states to be $R$ and their total mass $M$, then their density functions are defined by
\begin{eqnarray}\label{key}
\rho(\bs{r}) = \rho_0 (\bs{r}) \theta(R-r_{\rho}) \theta(R^2-r^2_{\rho}-z^2),\\
\rho'(\bs{r}) = \rho'_0 (\bs{r}) \theta(R-r_{\rho}) \theta(R^2-r^2_{\rho}-(z-b)^2), 
\end{eqnarray}

where
\begin{eqnarray}\label{key}
\rho_0(\bs{r}) &:= \frac{5}{2} \rho_0 (1 - (r_{\rho}^2 + z^2)/R^2) = \frac{5}{2} \rho_0 (1 -r^2/R^2)  ,\\
\rho'_0(\bs{r}) &:= \frac{5}{2} \rho_0 (1 - (r_{\rho}^2 + [z-b]^2)/R^2) = \frac{5}{2} \rho_0 (1 -r'^2/R^2),
\end{eqnarray}

with $r'^2 = r^2 + b^2 - 2 r b \cos \theta$ and $\rho_0 := M/((4/3) \pi R^3)$ as before. We again use \eref{eq:EG1} to calculate $E_G$ for this situation. In this case we need the gravitational potential inside and outside the sphere. From Gauss's law, the outside potential is of course the same as in the uniform situation. To find the inner potential, we can also apply  Gauss's law:
\begin{equation}
\oint \bs{g}.\bs{dS} = - 4 \pi G M_r,
\end{equation}
where we choose a spherical surface of constant radius $r$ within the sphere such that $M_r$ is the total mass within this spherical surface and is given by
\begin{eqnarray}
M_r &= 4 \pi \int_0^r \rho(\bs{r}') r^{'2} dr',\\
&=  \frac{M}{2 R^5} (5 R^2 r^3 - 3 r^5),
\end{eqnarray}

where $M$ is the total mass of the spherical BEC. Therefore, the field inside the sphere is given by:
\begin{equation}
\bs{g} = - \frac{G M}{2 R^5 r^2} (5 R^2 r^3 - 3r^5),
\end{equation}
and the potential then can be found through:
\begin{eqnarray}
\phi_{in} (r) &= - \int^R_{\infty} \frac{-G M}{r^2} dr - \int_R^r  \frac{-G M}{2 R^5 r^2} (5 R^2 r^3 - 3r^5)\\ \label{eq:phiBECinside}
&= - \frac{G M}{8 R^5} (15 R^4 - 10R^2 r^2 + 3 r^4).
\end{eqnarray}

The gravitational potential of a spherical BEC in the Thomas-Fermi approximation is then
\begin{equation}\label{key}
\phi = \cases{
	\phi_{in} := -\frac{G M}{8 R} \Big(15 - 10 r^2/R^2 + 3 r^4/R^4\Big)  &if  $r \leq R$, \\
	\phi_{out} :=  -\frac{G M}{r} &if $r \geq R$.}
\end{equation}
The rest of the calculation now proceeds similar to the uniform case. We first consider the term $\phi \rho'$ in \eref{eq:EGSphereSimple}. When $b > 2 R$ we find, due to Gauss's law again, that this is simply $- G M^2 / b$ as before. For $0 \leq b \leq 2 R$ we choose the origin of our coordinate system to be at the centre of the $\rho(\bs{r})$ sphere state and again integrate its potential over the density of the other sphere state using surfaces of constant radial coordinate $r$:
\begin{eqnarray}\nonumber
\int \phi \rho' d^3 \bs{r} &= -  \int^{2 \pi}_0 \int^{\cos^{-1} (z_r/r)}_0 \Big[ \int^{R}_{b-R} \phi_{in} \\&~~~~~~~~~~~~~~~~~~~+ \int^{b+R}_{R} \phi_{out}\Big]
\rho'_0(\bs{r}) r^2 \sin \theta dr d \theta  d\psi \\
&= \frac{G M^2}{R} \Big(-\frac{10}{7} + \frac{20}{7}  \lambda^2 - 6 \lambda^4 + 5 \lambda^6 - \frac{15}{14} \lambda^7 + \frac{1}{7} \lambda^9\Big).
\end{eqnarray} 
For the term $\phi \rho$ in \eref{eq:EGSphereSimple} we can again simply extract this from the above result when $b=0$:
\begin{equation}
\int \phi \rho = -\frac{10 G M^2}{7 R}.
\end{equation}
Putting this altogether, we  obtain:
\begin{equation}
E_G= \cases{
	\frac{G M^2}{R} \Big(  \frac{20}{7} \lambda^2 - 6 \lambda^4 + 5 \lambda^5  - \frac{15}{14} \lambda^7 +  \frac{1}{7} \lambda^9 \Big) &if $0 \leq \lambda \leq 1$, \\
	\frac{G M^2}{R}   \Big( \frac{10}{7} - \frac{1}{2 \lambda}\Big) &if  $\lambda \geq 1$.}
\end{equation}

\newpage

\section{$E_G$ for BEC, spherical mass distributions in the Gaussian approximation}
\label{app:GaussBECSphere}

Here we calculate the self-energy of the difference between   displaced spherical BEC mass distributions in the Gaussian approximation. Taking the sphere to have total mass $M$, then the density functions are defined by
\begin{eqnarray}\label{eq:SphereGauss}
\rho(\bs{r}) = \frac{4}{3 \sqrt{\pi}} \rho_0 e^{-r^2/R_0^{'2}},\\
\rho'(\bs{r}) = \frac{4}{3 \sqrt{\pi}} \rho_0 e^{-r'^2/R_0^{'2}},
\end{eqnarray}

where $\rho_0 := M/((4/3) \pi R_0^{'3})$ and $R'_0$ is given by \eref{eq:Rprimed} and can be taken as a measure for the size of the condensate \cite{BECReview}. However, we do not take this as a discontinuous cut in the density and instead keep the the wavefunction of the condensate has having infinite extent. Following the procedure outlined in the previous section using Gauss's law (or using the method outlined in  \ref{app:BECGaussianSphere}), the potential of a Gaussian sphere is found to be
\begin{equation} \label{eq:phiGauss}
\phi(r) = - \frac{G M}{r} \mathrm{erf} (r/R_0'),
\end{equation}
where $\mathrm{erf}(x)$ is the error function. This tends to $-2 G M / ( \sqrt{\pi} R'_0)$ in the limit that $r \rightarrow 0$.  The $\phi \rho$ term of  \eref{eq:EGSphereSimple} is then found to be
\begin{equation}
\int d^3 \bs{r} \phi \rho = \int^{2\pi}_0 \int^{\pi}_0 \int^{\infty}_0 \rho(\bs{r}) \phi(r) = - \frac{ G M^2}{R_0'} \sqrt{\frac{2}{\pi}}, 
\end{equation}
and the $\phi \rho'$ term is
\begin{equation}\label{key}
\int d^3 \bs{r} \phi \rho' = \int^{2\pi}_0 \int^{\pi}_0 \int^{\infty}_0 \rho'(\bs{r}) \phi(r) = - \frac{ G M^2}{b} \mathrm{erf} \Big(\frac{b}{\sqrt{2} R'_0}\Big),
\end{equation}
where we have used the identity \cite{ErrorFunctionIdentities}:
\begin{equation}\label{key}
\int^{\infty}_0 \Big[e^{-(x-A)^2} - e^{-(x+A)^2}\Big] \mathrm{erf}(x) dx \equiv \sqrt{\pi}  \mathrm{erf}(A/\sqrt{2}).
\end{equation}
Putting this altogether, we  obtain:
\begin{equation}
E_G= 
\frac{G M^2}{R_0'} \Big(  \sqrt{\frac{2}{\pi}} - \frac{1}{2 \lambda} \mathrm{erf}(\sqrt{2} \lambda)\Big)
\end{equation}
where we have defined $\lambda := b / (2 R_0')$.

\section{$E_G$ for uniform, spheroidal mass distributions} \label{app:UniformSpheroid}

Here we consider the self-energy of the difference between uniform, displaced spheroidal mass distributions. Following the previous sections, we work with \eref{eq:EG1} to calculate $E_G$. In this case we need the gravitational potential inside and outside the spheroid.  This is simplest in spheroidal coordinates: in prolate spheroidal coordinates, the gravitational potential of prolate spheroid is \cite{SpheroidGravPotExt,SpheroidGravPotInside}:
\begin{eqnarray}\label{key}
\phi = \cases{
	\phi^{prolate}_{in} := -\frac{G M}{l} \Big(Q_0(\xi_0) G_1(\xi,\eta) + \frac{\xi(\xi^2-1)}{\xi_0[\xi_0^2-1]}  G_2(\xi,\eta) \\~~~~~~~~~~~~~~~~~~~~~~~~~~+ \frac{\xi_0^2 - \xi^2}{\xi_0 [\xi_0^2 -1]} G_3(\xi,\eta) \Big) &\hspace{-2cm}if  $\xi \leq \xi_0$ \\
	\phi^{prolate}_{out} :=  -\frac{G M}{ l} \Big(Q_0 (\xi) - Q_2 (\xi) P_2 (\eta) \Big) &\hspace{-2cm}if  $\xi \geq \xi_0$,}
\end{eqnarray}
where 
\begin{eqnarray}
G_1(\xi,\eta):=1 - P_2(\xi) P_2(\eta),\\
G_2(\xi,\eta):=\frac{3}{2} P_2 (\eta)  \xi,\\
G_3(\xi,\eta):=\frac{1}{2}  + P_2(\xi) P_2(\eta),
\end{eqnarray}
$P_n(x)$ and $Q_n(x)$ are Legendre  polynomials of the first and second kind:\footnote{Often $Q_0(x)$ is alternatively defined as 
	\begin{equation}\label{key}
	Q_0(x) := \frac{1}{2} \ln \Big(\frac{1+x}{1-x}\Big) = \tan^{-1} x.
	\end{equation}}
\begin{eqnarray}
P_2(x) &= \frac{1}{2} (3 x^2 - 1),\\
Q_0(x) &= (1/2) \ln\Big(\frac{x + 1}{x - 1}\Big)= \tan^{-1} x + \frac{1}{2} i \pi,\\
Q_2(x) &= P_2(x) Q_0(x) - \frac{3}{2} x;
\end{eqnarray}
$l = \sqrt{c^2-a^2}$ is the focal distance with $a$ and $c$  the  equatorial and polar radii respectively (which are respectively the semi-minor and semi-major axes for the prolate but semi-major and semi-minor for the oblate case); $(\xi,\eta,\psi)$ are prolate spheroidal coordinates with
\begin{eqnarray}
x/l &= \sqrt{\xi^2-1} \sin \nu  \cos \psi,\\
y/l &= \sqrt{\xi^2-1} \sin \nu  \sin \psi,\\
z/l &= \xi \cos \nu,
\end{eqnarray}
using $\eta = \cos \nu$; and $\xi_0 := c / l$ is the value of $\xi$ at the surface of the prolate spheroid. For the potential of the oblate spheroid, just replace $\xi $ with $i \xi$, $l$ by $-i l$ and $\xi_0$ with $i \xi_0$ \cite{SpheroidGravPotExt,SpheroidGravPotInside}. 

We first consider the term $\phi \rho$ in \eref{eq:EG1} for a prolate spheroid:
\begin{eqnarray}\label{key}
\int d^3 \bs{r} \phi \rho &= \rho^{spheroid}_0 \int^{2 \pi}_0 \int^{1}_{-1} \int^{\xi_0}_1 l^3 (\xi^2-\eta^2) \phi^{prolate}_{in} d \xi d \eta d \psi\\ &= \frac{6 G M^2}{5 l} \coth^{-1} \xi_0 , \\
&= \frac{6 G M^2}{5 l} \tanh^{-1} e,
\end{eqnarray}
where  $e$ is the ellipticity  ($e:= c/l$ for the prolate case) and $\rho^{spheroid}_0 = M / ((4/3) \pi a^2 c)$ is the density of a uniform spheroid. For an oblate spheroid, we just need to replace $\xi_0$ with $i \xi_0$ to obtain
\begin{equation}\label{key}
\frac{6 G M^2}{5 l} \cot^{-1} \xi_0  = \frac{6 G M^2}{5 l} \sin^{-1} e.
\end{equation}
where $e:=a/l$ is now the ellipticity of an oblate spheroid with $l=\sqrt{a^2-c^2}$  its focal distance.

We now consider the term $\phi \rho'$ in \eref{eq:EG1}. For the prolate spheroid in configuration b) in Figure \ref{fig:Configurations}, we choose to integrate over surfaces of constant $\xi$. When $b \geq 2 c$, we use
\begin{eqnarray}
\int \phi \rho' d^3 \bs{r} &=  \rho^{spheroid}_0 \int^{2 \pi}_0 \int^1_{\eta_{\xi}}  \int^{(b+c)/l}_{(b-c)/l} 
l^3 (\xi^2-\eta^2) \phi^{prolate}_{out} d \xi d \eta d \psi,
\end{eqnarray}  

where $\eta_{\xi}$ is the $\eta$-coordinate where the  $\xi$-surface meets with the surface of the $\rho'$ spheroid. When $0 \leq b \leq 2 c$, we use
\begin{eqnarray}
\int \phi \rho' d^3 \bs{r} &= \rho^{spheroid}_0 \int^{2 \pi}_0 \int^1_{\eta_{\xi}} \Big[ \int^{c/l}_{(b-c)/l} \phi^{prolate}_{in} \\&\hspace{1cm}+ \int^{(b+c)/l}_{c/l} \phi^{prolate}_{out}\Big]
l^3 (\xi^2-\eta^2) d \xi d \eta d \psi.
\end{eqnarray}  

The result for $E_G$ to first order in $\epsilon$, where $a =  \epsilon c$, for a prolate spheroid is provided in Section \ref{sec:UniformSpheroid}. 

Another option, which is more suited to an oblate spheroidal coordinate system, is to integrate over surfaces of constant $\eta$.    Choosing now to work instead with an oblate spheroid  coordinate system then, when $b \geq 2 c$, we use

\begin{equation}
\int \phi \rho' d^3 \bs{r} =  \rho^{spheroid}_0 \int^{2 \pi}_0 \int^1_{\eta_{max}}  \int^{\xi_2}_{\xi_1} 
l^3 (\xi^2+\eta^2) \phi^{oblate}_{out} d \xi d \eta d \psi,
\end{equation}  

where $\xi_1$ and $\xi_2$ are the two values of $\xi$ where the constant $\eta$-surface crosses the $\rho'$ spheroid state, and $\eta_{max}$ is the value of $\eta$ where there is only one $\xi$ solution i.e.\ $\xi_1=\xi_2$. When $0 \leq b \leq 2 c$, we use
\begin{eqnarray}
\int \phi \rho' d^3 \bs{r} &= \rho^{spheroid}_0 \int^{2 \pi}_0 \Big[ \int^1_{\eta_{int}}  \Big( \int^{\xi_0}_{\xi_1} \phi^{oblate}_{in} + \int^{\xi_2}_{\xi_0} \phi^{oblate}_{out}\Big) \\&\hspace{1cm}+ \int^{\eta_{int}}_{\eta_{max}} \int^{\xi_2}_{\xi_1} \phi^{oblate}_{out}\Big]
l^3 (\xi^2+\eta^2) d \xi d \eta d \psi,
\end{eqnarray}

where $\eta_{int} = b/(2l \xi_0)$ is the $\eta$-coordinate where the two spheroids meet.  Once the result for $E_G$ is obtained, the prolate spheroid case can be found via $\xi_0 \rightarrow i \xi_0$ as above.  When $\epsilon \ll 1$, where now $c :=  \epsilon a$, $E_G$ for an oblate spheroid displaced along its symmetry axis can be approximated by \eref{eq:EGUniformProlate} in Section \ref{sec:UniformSpheroid}.

Unlike in the spherical case, equipotential surfaces are not similar-shaped spheroids (or confocal spheroids), emphasizing that Gauss's law is not as useful for these objects. Therefore, integrating over constant $\xi$ or $\eta$ surfaces is not as simple. An alternative is to use cylindrical coordinates where the prolate spheroid potential inside and outside the spheroid is given by:
\begin{eqnarray}\label{eq:PhiProlateInCyl}
\phi^{prolate}_{in} &:= 
-\frac{3 G M }{4 l^3} \Big[(2l^2 + D) \mathrm{csch}^{-1} \Big( \frac{a}{l}\Big) -  \frac{l ( c^2 r^2 - 2 a^2 z^2)}{a^2 c} 	\Big], \\ \label{eq:PhiProlateOutCyl}
\phi^{prolate}_{out} &:= -\frac{3 G M }{4 l^3} \Big[ (2 l^2 + D) \sinh^{-1} \Big(\frac{\sqrt{2} l}{ B_p}\Big) - 
\frac{\sqrt{2} l ( A_p D+l^2 C)}{ E_p B_p^2 } \Big],
\end{eqnarray}

and the oblate potential is found by taking $l \rightarrow -i l$, to obtain:
\begin{eqnarray}\label{eq:PhiOblateInCyl}
\phi^{oblate}_{in} &:= 
-\frac{3 G M }{4 l^3} \Big[(2l^2 - D) \csc^{-1} \Big(\frac{a}{l}\Big) +  \frac{l ( c^2 r^2 - 2 a^2 z^2)}{a^2 c} 	\Big], \\ \label{eq:PhiOblateOutCyl}
\phi^{oblate}_{out} &:= -\frac{3 G M}{4 l^3} \Big[ (2 l^2 - D) \sin^{-1} \Big(\frac{\sqrt{2} l}{B_o}\Big) + 
\frac{\sqrt{2} l ( A_o D-l^2 C)}{ E_o B^2_o} \Big],
\end{eqnarray}
where
\begin{eqnarray}
A_p &:= r^2 + z^2 +\sqrt{l^4 + 2l^2 (r^2-z^2) + (z^2+r^2)^2},\\
A_o &:=r^2 + z^2 +\sqrt{z^4+ 2 z^2(r^2+l^2)  + (l^2-r^2)^2},\\
B_p&:=\sqrt{A_p-l^2},~~E_p :=\sqrt{A_p+l^2}, \\
B_o&:=\sqrt{A_o+l^2},~~E_o := \sqrt{A_o-l^2},\\
C &:=r^2+2z^2,\\
D &:=r^2-2z^2.
\end{eqnarray}
To second order in ellipticity, the potentials, in spherical coordinates,  become:
\begin{eqnarray}\label{key}
\phi = \cases{
	\phi_{in} := \phi_{in}^{sphere} \pm \frac{G M}{5 R^3} P_2 (\cos \theta) e^2  &if  $r \leq r(\theta)$, \\
	\phi_{out} :=  \phi_{out}^{sphere} \pm \frac{G M R^2}{5 r^3} P_2 (\cos \theta) e^2  &if  $r \geq r(\theta)$,}
\end{eqnarray}

where $+$ is for the prolate case, $-$ is for the oblate case, $\phi^{sphere}_{in/out}$ is given by \eref{eq:PhiInsphere}-\eref{eq:PhiOutsphere},  $r(\theta) := c [1- e^2 \cos^2 \theta]^{-1/2}$ (with the respective definitions of ellipticity for the two spheroidal cases), and we have taken both spheroids to have the same volume as a sphere with radius $R$.

Using the full potentials in cylindrical coordinates, for the spheroidal cases in configuration a) and b) in Figure \ref{fig:Configurations}, the $\phi \rho$ term is then:
\begin{eqnarray}
\int \phi \rho d^3 \bs{r} &= \rho^{spheroid}_{0} \int^{2 \pi}_0 \int^{c}_{-c} \int^{a \sqrt{1 - z^2/c^2}}_0 \phi_{in}^{spheroid} r dr dz d \psi. 
\end{eqnarray}

where $\phi_{in}^{spheroid}$ is given by either \eref{eq:PhiProlateInCyl} or \eref{eq:PhiOblateInCyl}. 
For $b \geq 2 c$, the $\phi \rho'$ term is:

\begin{equation}\label{key}
\int \phi \rho' d^3 \bs{r} =  \rho^{spheroid}_{0} \int^{2 \pi}_0 \int^{b+c}_{b-c} \int^{a \sqrt{1 - (z-b)^2/c^2}}_0 \phi_{out}^{spheroid} r dr dz d \psi,
\end{equation}
where $\phi_{out}^{spheroid}$ is given by either \eref{eq:PhiProlateOutCyl} or \eref{eq:PhiOblateOutCyl}; and for $0 \leq b \leq 2c$, we can use
\begin{eqnarray}
&\rho^{spheroid}_{0} \int^{2 \pi}_0 \Big[\Big(\int^{b/2}_{b-c} \int^{a \sqrt{1 - (z-b)^2/c^2}}_0
+\int^{c}_{b/2} \int^{a \sqrt{1 - z^2/c^2}}_0\Big)\phi_{in}^{spheroid}\\&+ \Big(\int^{c}_{b/2}
\int^{a \sqrt{1-(z-b)^2/c^2}}_{a \sqrt{1 - z^2/c^2}}
+\int^{b+c}_{c} \int^{a \sqrt{1 - (z-b)^2/c^2}}_0\Big)\phi_{out}^{spheroid}\Big] r dr dz d \psi. 
\end{eqnarray}

For the oblate and prolate configurations c) and d) in Figure \ref{fig:Configurations}, the above procedure is just slightly modified.

\section{$E_G$ for BEC, spheroidal mass distributions  in the Thomas-Fermi limit} \label{app:BECSpheroid}
Within the Thomas-Fermi approximation, the density of spheroidal BECs is given by (see \eref{eq:ProlateDensity}):
\begin{eqnarray} 
\rho^{oblate}_0(\bs{r})&= \frac{5}{2} \rho^{spheroid}_{0} (1-r^2_{\rho} /a^2 - z^2/c^2)\\
&= \frac{5}{2} \rho^{spheroid}_{0} \frac{\xi_0^2-\xi^2}{\xi_0^2[\xi_0^2+1]} [\xi_0^2+ \eta^2]\\ \label{eq:rhoOblate}
&= \frac{5 \epsilon^2}{2}  \rho^{spheroid}_{0}  \Big(1 - e'^2[\xi^2 + \eta^2(1-e'^2 \xi^2)]\Big),\\ \label{eq:rhoProlate}
\rho^{prolate}_0(\bs{r}) &= \frac{5}{2\epsilon^2} \rho^{spheroid}_{0} \Big(1 - e^2[\xi^2 + \eta^2(1-e^2 \xi^2)]\Big)
\end{eqnarray}
for the respective coordinate systems, where $e':= l/c$ is the second ellipticity for an oblate spheroid.

We now find the gravitational potential of these spheroidal BECs by summing the individual potentials from point-like sources of mass $d m = \rho(\bs{r}') d^3 \bs{r}'$ where $d^3 \bs{r}'$ is the volume element of the spheroid, and $\rho(\bs{r})$ is its density function, i.e.\ we use \eref{eq:phi}:

\begin{equation}
\phi(\bs{x}) \equiv - G \int \frac{\rho(\bs{r'})}{r} d^3 \bs{r'},
\end{equation}

with $r := |\bs{r} - \bs{r'}|$ the distance from the point source to the point of interest.  Working in prolate spheroidal coordinates we then have

\begin{equation} \label{eq:dphi}
\phi = - G \int \frac{1}{r} \rho(\xi',\eta') l^3 [\xi'^2-\eta'^2] d\xi' d \eta' d \psi'. 
\end{equation}

Following \cite{SpheroidGravPotExt,SpheroidGravPotInside}, in prolate spheroidal coordinates, the ratio of $l$ to $r$ can be expressed as:
\begin{eqnarray}\label{eq:loverr}
\frac{l}{r} &= \sum^{\infty}_{n=0} (2n +1) P_n(\eta) P_n(\eta') Q_n(\xi)P_n(\xi') \\&\hspace{1cm}+ 2 \sum^{\infty}_{n=1} (2n+1) \sum^n_{m=1} (-1)^m \Big(\frac{(n-m)!}{(n+m)!}\Big)^2\\&\hspace{1cm}
\times P^{m}_n (\eta) P^m_n (\eta') Q^m_n (\xi) P^m_n(\xi') \cos [m (\psi - \psi')],
\end{eqnarray}

for $\xi > \xi'$, and
\begin{eqnarray}\label{eq:loverr}
\frac{l}{r} = \sum^{\infty}_{n=0} (2n +1) P_n(\eta) P_n(\eta') P_n(\xi)Q_n(\xi') + f(\cos [m(\psi - \psi')]),
\end{eqnarray}

for $\xi < \xi'$, where $P^m_n(x)$ and $Q^m_n(x)$ are the associated Legendre polynomials of the first and second kind and $f$ is an unimportant function of $\cos [m(\psi - \psi')]$ since, when inserting \eref{eq:loverr} into \eref{eq:dphi}, this term, and the second term in \eref{eq:loverr}, vanish once we integrate over $\psi$ \cite{SpheroidGravPotExt,SpheroidGravPotInside}.  For the prolate spheroid, we then end up with:
\begin{eqnarray}
\phi_{out} &= - 2 \pi l^2 G \sum^{\infty}_{n=0} (2n + 1) Q_n (\xi) P_n(\eta) \\&\hspace{1cm}\times \int^1_{-1} \int^{\xi_0}_1  \rho(\xi',\eta') P_n (\xi') P_n (\eta') [\xi'^2-\eta'^2] d\xi' d \eta',\\
\phi_{in} &=  - 2 \pi l^2 G \sum^{\infty}_{n=0} (2n + 1) \Big[ Q_n (\xi) P_n(\eta) \int^1_{-1} \int^{\xi}_1  \rho(\xi',\eta') P_n (\xi') P_n (\eta') \\&\hspace{0.7cm}+ P_n (\xi) P_n(\eta) \int^1_{-1} \int^{\xi_0}_{\xi}  \rho(\xi',\eta') Q_n (\xi') P_n (\eta')\Big] [\xi'^2-\eta'^2] d\xi' d \eta'.
\end{eqnarray}

Using \eref{eq:rhoProlate} for $\rho(\xi',\eta')$, we then find that:
\begin{eqnarray}\label{key}
\phi^{prolate}_{out} &=-\frac{G M}{ 112 l} \Big(16  Q_0(\xi_0) F_1(\xi,\eta) + \frac{\xi^2 - \xi_0^2}{ \xi_0^3 [ \xi_0^2 - 1]^2} F_2(\xi,\eta) \\&\hspace{1cm}+ \frac{\xi^2(\xi^2-1)}{ \xi_0^3 [ \xi_0^2 - 1]^2} F_3(\xi,\eta)\Big),\\
\phi^{prolate}_{out} &=-\frac{G M}{ 7 l} \Big(7 Q_0 (\xi)
- 10 Q_2 (\xi) P_2 (\eta) + 3 Q_4(\xi) P_4(\eta) \Big),
\end{eqnarray}

where
\begin{eqnarray}
F_1(\xi,\eta) &:= 7 - 10 P_2(\xi)  P_2(\eta) + 3 P_4 (\xi) P_4 (\eta),\\
F_2(\xi,\eta) &:= 14[5 \xi_0^2 - 7 \xi_0^4 + \xi^2 (3 \xi_0^2 -1)] \\&+ 40 P_2(\xi) P_2(\eta) [2 \xi_0^2  - 4 \xi_0^4  + \xi^2(3 \xi_0^2-4) + 3 \xi^4 ] \\& - 8 P_4(\xi) P_4(\eta) [10 \xi_0^2 - 6 \xi_0^4 - \xi^2(13 + 6\xi_0^2) + 15 \xi^4],\\
F_4(\xi,\eta) &:= 60 P_2(\eta) [\xi_0^4 - 6 \xi_0^2 + 2 \xi^2 - 3 \xi^4] \\&+ 5 P_4 (\eta)(21 \xi^2 -11) [3 \xi_0^2 - \xi^2(1 + 7 \xi_0^2) + 5 \xi^4],\\
P_4(x) &:= \frac{1}{8} (3 - 30 x^2 + 35 x^4),\\
Q_4(x) &:= \frac{1}{48} \xi (110 -210 \xi^2) + P_4(\xi) Q_0(\xi),
\end{eqnarray}
and  we have used the orthogonality relationship of Legendre polynomials of the first kind:

\begin{equation} \label{eq:LegendreOrthog}
\int^1_{-1} P_n(x) P_m(x) dx = \frac{2}{2n +1} \delta_{mn}.
\end{equation}

In contrast to the uniform case, we now have  Legendre polynomials of the fourth degree. Also note that, unlike in the spherical case, the potential outside the BEC spheroid is different to the uniform spheroid. To obtain the oblate potentials in oblate coordinates, we just make the changes $\xi \rightarrow i \xi$, $\xi_0 \rightarrow i \xi_0$ and $l \rightarrow -il$.

It is also possible to find the potentials in  cylindrical coordinates by taking the inverse transformations:
\begin{eqnarray}\label{key}
\xi &= \frac{1}{2l} \Big[\sqrt{r^2 + (z+l)^2} + \sqrt{r^2 + (z-l)^2}\Big],\\
\eta &= \frac{1}{2l} \Big[\sqrt{r^2 + (z+l)^2} - \sqrt{r^2 + (z-l)^2}\Big]
\end{eqnarray} 

for prolate spheroidal coordinates, and (taking $l \rightarrow - il$):
\begin{eqnarray}\label{key}
\xi &= \frac{1}{2l} \Big[\sqrt{r^2 + (z-il)^2} + \sqrt{r^2 + (z+il)^2}\Big] \equiv\frac{1}{\sqrt{2} l} E_o,\\
\eta &= \frac{i}{2l} \Big[\sqrt{r^2 + (z-il)^2} - \sqrt{r^2 + (z+il)^2}\Big] \equiv \frac{1}{\sqrt{2} l} F_o,
\end{eqnarray} 

for oblate spheroidal coordinates, where $E_o$ is defined in the previous section, and

\begin{equation}\label{key}
F_o := \sqrt{ l^2 - r^2 - z^2 + \sqrt{(l^2-r^2)^2 + 2(l^2+r^2)z^2 + z^4}}.
\end{equation}

In the appropriate limit, i.e.\ $a \rightarrow c \rightarrow R$ (and so $l \rightarrow 0$), it is possible to show that these potentials become the spherical BEC potentials provided in \ref{app:BECSphere}. To calculate $E_G$ for the different spheroidal configurations, the procedure in the previous section can be followed with the uniform potentials and density functions replaced with those above. Alternatively, it is possible to integrate over spheroidal surfaces of constant density using the area element $l^2 [ (\xi^2 - \eta(\xi)^2)(1 - \eta(\xi)^2 + (\xi^2-1)\eta'(\xi)^2)]^{1/2}$ for prolate coordinates where $\eta' :=  d \eta(\xi) / d \xi$.

\section{BEC spheroidal mass distributions in the Gaussian limit} \label{app:BECGaussianSphere}

Here we calculate the gravitational potential due to a spheroidal BEC in the Gaussian limit for small ellipticity values. We work in spherical coordinates to easily compare  to the spherical BEC case.  In general, the potential can be calculated from \eref{eq:dphi}, which, in spherical coordinates is

\begin{equation}\label{key}
\phi(r,\theta,\psi) = - G \int^{2 \pi}_0 \int^\pi_0 \int_0^{\infty} \frac{\rho(r',\theta',\psi')}{|\bs{r}-\bs{r}'|} r'^2 \sin \theta' dr' d\theta' d\psi'.
\end{equation}

Since the spheroidal density does not depend on $\psi$, we can set $\psi=0$ such that 

\begin{equation}\label{key}
|\bs{r}-\bs{r}'|=\sqrt{r^2 + r'^2 - 2 r r' (\sin \theta \sin \theta' \cos \psi' + \cos \theta \cos \theta')}.
\end{equation}

We can then expand the above in terms of Legendre polynomials of the first kind:

\begin{equation}\label{key}
\frac{1}{2 \pi} \int^{2 \pi}_0 \frac{1}{|\bs{r}-\bs{r}'|} d \psi' = \cases{ \frac{1}{r} \sum_{n=0,\infty} \Big(\frac{r'}{r}\Big)^n P_n (\cos \theta)  P_n (\cos \theta') &if $r > r'$,\\ 
	\frac{1}{r'} \sum_{n=0,\infty} \Big(\frac{r}{r'}\Big)^n P_n (\cos \theta)  P_n (\cos \theta') &if $r < r'$,}
\end{equation}

so that \eref{eq:dphi} becomes
\begin{eqnarray}\label{eq:phiSphericalCoords}
\phi(r,\theta) &= - 2 \pi G P_n (\cos \theta) \int^\pi_0 \Big[ \frac{1}{r^{n+1}} \int^r_0 r'^{n+2}  \\&\hspace{1cm}+ r^n \int^\infty_r r'^{1-n} \Big] \rho(r',\theta') P_n(\cos \theta') \sin \theta' dr' d\theta'.
\end{eqnarray}

The density of the spheroidal BEC in the Gaussian limit is given by  \eref{eq:OblateDensityGauss}.  Assuming the spheroid to have the same volume as a sphere with radius $R$, then in the limit of small ellipticity, the density function becomes
\begin{eqnarray}\label{key}
\rho (r,\theta) &=  \rho^{Gauss}_{sphere} \Big[1\pm \frac{2 r^2 e^2}{3 R^2}  P_2(\cos \theta) \\&\hspace{1cm}- \frac{r^2 e^4}{135 R^2} \Big(35 -  \frac{14 r^2}{R^2} - 10  (\pm 7\beta +  \frac{2 r^2}{R^2})P_2(\cos \theta) \\&\hspace{1cm}- \frac{36 r^2}{R^2}  P_4(\cos \theta) \Big)\Big],
\end{eqnarray} 

where $\rho^{Gauss}_{sphere}$ is given by \eref{eq:SphereGauss}; $+$ is for the prolate case, which has $\beta=1$; and $-$ is for the oblate case, which has  $\beta=2$. Inserting the density expression into \eref{eq:phiSphericalCoords}, and using the orthogonality relationship \eref{eq:LegendreOrthog}, we find 
\begin{eqnarray}\label{key}
\phi(r,\theta) &= \phi_{sphere}^{Gauss} \pm \frac{G M e^2}{6  R r^3} P_2(\cos \theta) \Big[e(r) X_2(r) -21 R^5 \mathrm{erf} \Big(\frac{r}{R}\Big) \Big] \\&\hspace{1cm}+ \frac{G M e^4}{R^3} \Big(\frac{1}{45} e(r) (r^2 + R^2) \\&\hspace{1cm}\pm \frac{1}{63 \beta r^3} \Big[ e(r) X_4 (r) - 21 R^5 \mathrm{erf}\Big(\frac{r}{R}\Big)  \Big] P_2(\cos \theta) \\&\hspace{1cm}+ \frac{1}{140 r^5} \Big[ e(r) Y_4(r) - 105 R^7 \mathrm{erf} \Big(\frac{r}{R}\Big)\Big]P_4(\cos \theta) \Big),
\end{eqnarray}

where
\begin{eqnarray}\label{key}
e(r) &:= \frac{2}{\sqrt{\pi}} e^{-r^2/R^2},\\
X_2(r) &:= 2r^3 + 3 r R^2,\\
X_4(r) &:= \pm 2 \beta r^5 + 14 r^3 R^2 + 21 r R^4,\\
Y_4(r) &:=  r(  8 r^6 + 28 r^4 R^2 + 70 r^2 R^4 + 105 R^6),
\end{eqnarray}

$\phi_{sphere}^{Gauss}$ is given by \eref{eq:phiGauss}; and, again, $+$ is for the prolate case, which has $\beta=1$; and $-$ is for the oblate case, which has  $\beta=2$.

\newpage
\bibliography{references}

\begin{thebibliography}{115}%
\makeatletter
\providecommand \@ifxundefined [1]{%
 \@ifx{#1\undefined}
}%
\providecommand \@ifnum [1]{%
 \ifnum #1\expandafter \@firstoftwo
 \else \expandafter \@secondoftwo
 \fi
}%
\providecommand \@ifx [1]{%
 \ifx #1\expandafter \@firstoftwo
 \else \expandafter \@secondoftwo
 \fi
}%
\providecommand \natexlab [1]{#1}%
\providecommand \enquote  [1]{``#1''}%
\providecommand \bibnamefont  [1]{#1}%
\providecommand \bibfnamefont [1]{#1}%
\providecommand \citenamefont [1]{#1}%
\providecommand \href@noop [0]{\@secondoftwo}%
\providecommand \href [0]{\begingroup \@sanitize@url \@href}%
\providecommand \@href[1]{\@@startlink{#1}\@@href}%
\providecommand \@@href[1]{\endgroup#1\@@endlink}%
\providecommand \@sanitize@url [0]{\catcode `\\12\catcode `\$12\catcode
  `\&12\catcode `\#12\catcode `\^12\catcode `\_12\catcode `\%12\relax}%
\providecommand \@@startlink[1]{}%
\providecommand \@@endlink[0]{}%
\providecommand \url  [0]{\begingroup\@sanitize@url \@url }%
\providecommand \@url [1]{\endgroup\@href {#1}{\urlprefix }}%
\providecommand \urlprefix  [0]{URL }%
\providecommand \Eprint [0]{\href }%
\providecommand \doibase [0]{http://dx.doi.org/}%
\providecommand \selectlanguage [0]{\@gobble}%
\providecommand \bibinfo  [0]{\@secondoftwo}%
\providecommand \bibfield  [0]{\@secondoftwo}%
\providecommand \translation [1]{[#1]}%
\providecommand \BibitemOpen [0]{}%
\providecommand \bibitemStop [0]{}%
\providecommand \bibitemNoStop [0]{.\EOS\space}%
\providecommand \EOS [0]{\spacefactor3000\relax}%
\providecommand \BibitemShut  [1]{\csname bibitem#1\endcsname}%
\let\auto@bib@innerbib\@empty
\bibitem [{\citenamefont {Penrose}(1965)}]{Penrose1965}%
  \BibitemOpen
  \bibfield  {author} {\bibinfo {author} {\bibfnamefont {R.}~\bibnamefont
  {Penrose}},\ }\href {\doibase 10.1098/rspa.1965.0058} {\bibfield  {journal}
  {\bibinfo  {journal} {Proceedings of the Royal Society of London. Series A,
  Mathematical and Physical Sciences}\ }\textbf {\bibinfo {volume} {284}},\
  \bibinfo {pages} {159} (\bibinfo {year} {1965})}\BibitemShut {NoStop}%
\bibitem [{\citenamefont {{Penrose}}(1998)}]{Penrose1993}%
  \BibitemOpen
  \bibfield  {author} {\bibinfo {author} {\bibfnamefont {R.}~\bibnamefont
  {{Penrose}}},\ }in\ \href@noop {} {\emph {\bibinfo {booktitle} {{General
  Relativity and Gravitation 13. Part 1: Plenary Lectures 1992. Proceedings of
  the Thirteenth International Conference on General Relativity and Gravitation
  held at Cordoba, Argentina, 28 June - 4 July 1992}}}},\ \bibinfo {editor}
  {edited by\ \bibinfo {editor} {\bibfnamefont {R.~J.}\ \bibnamefont
  {{Gleiser}}}, \bibinfo {editor} {\bibfnamefont {K.~N.}\ \bibnamefont
  {{Kozameh}}}, \ and\ \bibinfo {editor} {\bibfnamefont {O.~M.}\ \bibnamefont
  {{Moreschi}}}}\ (\bibinfo  {publisher} {Institute of Physics Publishing},\
  \bibinfo {year} {1998})\ pp.\ \bibinfo {pages} {179--189}\BibitemShut
  {NoStop}%
\bibitem [{\citenamefont {{Penrose}}(1996)}]{Penrose1996}%
  \BibitemOpen
  \bibfield  {author} {\bibinfo {author} {\bibfnamefont {R.}~\bibnamefont
  {{Penrose}}},\ }\href {\doibase 10.1007/BF02105068} {\bibfield  {journal}
  {\bibinfo  {journal} {Gen. Relat. Gravit.}\ }\textbf {\bibinfo {volume}
  {28}},\ \bibinfo {pages} {581} (\bibinfo {year} {1996})}\BibitemShut
  {NoStop}%
\bibitem [{\citenamefont {{Penrose}}(2004)}]{RoadToReality}%
  \BibitemOpen
  \bibfield  {author} {\bibinfo {author} {\bibfnamefont {R.}~\bibnamefont
  {{Penrose}}},\ }\href@noop {} {\emph {\bibinfo {title} {The road to reality :
  a complete guide to the laws of the universe}}}\ (\bibinfo  {publisher}
  {Jonathan Cape},\ \bibinfo {year} {2004})\BibitemShut {NoStop}%
\bibitem [{\citenamefont {Penrose}(2009)}]{Penrose2009}%
  \BibitemOpen
  \bibfield  {author} {\bibinfo {author} {\bibfnamefont {R.}~\bibnamefont
  {Penrose}},\ }\href {http://stacks.iop.org/1742-6596/174/i=1/a=012001}
  {\bibfield  {journal} {\bibinfo  {journal} {Journal of Physics: Conference
  Series}\ }\textbf {\bibinfo {volume} {174}},\ \bibinfo {pages} {012001}
  (\bibinfo {year} {2009})}\BibitemShut {NoStop}%
\bibitem [{\citenamefont {Penrose}(2014)}]{Penrose2014}%
  \BibitemOpen
  \bibfield  {author} {\bibinfo {author} {\bibfnamefont {R.}~\bibnamefont
  {Penrose}},\ }\bibfield  {booktitle} {\emph {\bibinfo {booktitle}
  {{Proceedings, Horizons of Quantum Physics: Taipei, Taiwan, October 14-18,
  2012}}},\ }\href {\doibase 10.1007/s10701-013-9770-0} {\bibfield  {journal}
  {\bibinfo  {journal} {Found. Phys.}\ }\textbf {\bibinfo {volume} {44}},\
  \bibinfo {pages} {557} (\bibinfo {year} {2014})}\BibitemShut {NoStop}%
\bibitem [{\citenamefont {Everett}(1957)}]{ManyWorlds}%
  \BibitemOpen
  \bibfield  {author} {\bibinfo {author} {\bibfnamefont {H.}~\bibnamefont
  {Everett}},\ }\href {\doibase 10.1103/RevModPhys.29.454} {\bibfield
  {journal} {\bibinfo  {journal} {Rev. Mod. Phys.}\ }\textbf {\bibinfo {volume}
  {29}},\ \bibinfo {pages} {454} (\bibinfo {year} {1957})}\BibitemShut
  {NoStop}%
\bibitem [{\citenamefont {Rovelli}(1996)}]{RQM}%
  \BibitemOpen
  \bibfield  {author} {\bibinfo {author} {\bibfnamefont {C.}~\bibnamefont
  {Rovelli}},\ }\href {\doibase 10.1007/BF02302261} {\bibfield  {journal}
  {\bibinfo  {journal} {International Journal of Theoretical Physics}\ }\textbf
  {\bibinfo {volume} {35}},\ \bibinfo {pages} {1637} (\bibinfo {year}
  {1996})}\BibitemShut {NoStop}%
\bibitem [{\citenamefont
  {{K{\'a}rolyh{\'a}zy}}(1966)}]{karolyhazy1966gravitation}%
  \BibitemOpen
  \bibfield  {author} {\bibinfo {author} {\bibfnamefont {F.}~\bibnamefont
  {{K{\'a}rolyh{\'a}zy}}},\ }\href {\doibase 10.1007/BF02717926} {\bibfield
  {journal} {\bibinfo  {journal} {Nuovo Cimento A (1965-1970)}\ }\textbf
  {\bibinfo {volume} {42}},\ \bibinfo {pages} {390} (\bibinfo {year}
  {1966})}\BibitemShut {NoStop}%
\bibitem [{\citenamefont {{K{\'a}rolyh{\'a}zy}}(1974)}]{karolyhazy1974}%
  \BibitemOpen
  \bibfield  {author} {\bibinfo {author} {\bibfnamefont {F.}~\bibnamefont
  {{K{\'a}rolyh{\'a}zy}}},\ }\href {http://cds.cern.ch/record/430144}
  {\bibfield  {journal} {\bibinfo  {journal} {Magyar Fizikai Folyoirat}\
  }\textbf {\bibinfo {volume} {22}},\ \bibinfo {pages} {23} (\bibinfo {year}
  {1974})}\BibitemShut {NoStop}%
\bibitem [{\citenamefont {{K{\'a}rolyh{\'a}zy}}, \citenamefont {{Frenkel}},\
  and\ \citenamefont {{Luk{\'a}cs}}(1986)}]{karolyhazy1986}%
  \BibitemOpen
  \bibfield  {author} {\bibinfo {author} {\bibfnamefont {F.}~\bibnamefont
  {{K{\'a}rolyh{\'a}zy}}}, \bibinfo {author} {\bibfnamefont {A.}~\bibnamefont
  {{Frenkel}}}, \ and\ \bibinfo {author} {\bibfnamefont {B.}~\bibnamefont
  {{Luk{\'a}cs}}},\ }in\ \href@noop {} {\emph {\bibinfo {booktitle} {Quantum
  Concepts in Space and Time}}},\ \bibinfo {editor} {edited by\ \bibinfo
  {editor} {\bibfnamefont {R.}~\bibnamefont {{Penrose}}}\ and\ \bibinfo
  {editor} {\bibfnamefont {C.~J.}\ \bibnamefont {{Isham}}}}\ (\bibinfo
  {publisher} {Oxford University Press},\ \bibinfo {year} {1986})\ pp.\
  \bibinfo {pages} {109--128}\BibitemShut {NoStop}%
\bibitem [{\citenamefont {{Kibble}}(1981)}]{Kibble1980}%
  \BibitemOpen
  \bibfield  {author} {\bibinfo {author} {\bibfnamefont {T.~W.~B.}\
  \bibnamefont {{Kibble}}},\ }in\ \href@noop {} {\emph {\bibinfo {booktitle}
  {{Quantum Gravity 2: a Second Oxford Symposium}}}},\ \bibinfo {editor}
  {edited by\ \bibinfo {editor} {\bibfnamefont {C.~J.}\ \bibnamefont
  {{Isham}}}, \bibinfo {editor} {\bibfnamefont {R.}~\bibnamefont {{Penrose}}},
  \ and\ \bibinfo {editor} {\bibfnamefont {D.~W.}\ \bibnamefont {{Sciama}}}}\
  (\bibinfo  {publisher} {Oxford University Press},\ \bibinfo {year} {1981})\
  pp.\ \bibinfo {pages} {63--80}\BibitemShut {NoStop}%
\bibitem [{\citenamefont {Di\'{o}si}(1984)}]{DIOSI1984199}%
  \BibitemOpen
  \bibfield  {author} {\bibinfo {author} {\bibfnamefont {L.}~\bibnamefont
  {Di\'{o}si}},\ }\href {\doibase 10.1016/0375-9601(84)90397-9} {\bibfield
  {journal} {\bibinfo  {journal} {Physics Letters A}\ }\textbf {\bibinfo
  {volume} {105}},\ \bibinfo {pages} {199 } (\bibinfo {year}
  {1984})}\BibitemShut {NoStop}%
\bibitem [{\citenamefont {Di\'{o}si}(1987)}]{DIOSI1987377}%
  \BibitemOpen
  \bibfield  {author} {\bibinfo {author} {\bibfnamefont {L.}~\bibnamefont
  {Di\'{o}si}},\ }\href {\doibase https://doi.org/10.1016/0375-9601(87)90681-5}
  {\bibfield  {journal} {\bibinfo  {journal} {Physics Letters A}\ }\textbf
  {\bibinfo {volume} {120}},\ \bibinfo {pages} {377 } (\bibinfo {year}
  {1987})}\BibitemShut {NoStop}%
\bibitem [{\citenamefont {Di\'{o}si}\ and\ \citenamefont
  {Luk\'{a}cs}(1987)}]{DisoiLukas87}%
  \BibitemOpen
  \bibfield  {author} {\bibinfo {author} {\bibfnamefont {L.}~\bibnamefont
  {Di\'{o}si}}\ and\ \bibinfo {author} {\bibfnamefont {B.}~\bibnamefont
  {Luk\'{a}cs}},\ }\href {\doibase 10.1002/andp.19874990703} {\bibfield
  {journal} {\bibinfo  {journal} {Annalen der Physik}\ }\textbf {\bibinfo
  {volume} {499}},\ \bibinfo {pages} {488} (\bibinfo {year}
  {1987})}\BibitemShut {NoStop}%
\bibitem [{\citenamefont {Ghirardi}, \citenamefont {Grassi},\ and\
  \citenamefont {Rimini}(1990)}]{GRWGravity}%
  \BibitemOpen
  \bibfield  {author} {\bibinfo {author} {\bibfnamefont {G.}~\bibnamefont
  {Ghirardi}}, \bibinfo {author} {\bibfnamefont {R.}~\bibnamefont {Grassi}}, \
  and\ \bibinfo {author} {\bibfnamefont {A.}~\bibnamefont {Rimini}},\ }\href
  {\doibase 10.1103/PhysRevA.42.1057} {\bibfield  {journal} {\bibinfo
  {journal} {Phys. Rev. A}\ }\textbf {\bibinfo {volume} {42}},\ \bibinfo
  {pages} {1057} (\bibinfo {year} {1990})}\BibitemShut {NoStop}%
\bibitem [{\citenamefont {Di\'osi}(1989)}]{Diosi1989}%
  \BibitemOpen
  \bibfield  {author} {\bibinfo {author} {\bibfnamefont {L.}~\bibnamefont
  {Di\'osi}},\ }\href {\doibase 10.1103/PhysRevA.40.1165} {\bibfield  {journal}
  {\bibinfo  {journal} {Phys. Rev. A}\ }\textbf {\bibinfo {volume} {40}},\
  \bibinfo {pages} {1165} (\bibinfo {year} {1989})}\BibitemShut {NoStop}%
\bibitem [{\citenamefont {Di\'osi}(1990)}]{Diosi1990Relativistic}%
  \BibitemOpen
  \bibfield  {author} {\bibinfo {author} {\bibfnamefont {L.}~\bibnamefont
  {Di\'osi}},\ }\href {\doibase 10.1103/PhysRevA.42.5086} {\bibfield  {journal}
  {\bibinfo  {journal} {Phys. Rev. A}\ }\textbf {\bibinfo {volume} {42}},\
  \bibinfo {pages} {5086} (\bibinfo {year} {1990})}\BibitemShut {NoStop}%
\bibitem [{\citenamefont {{Percival}}(1995)}]{Percival503}%
  \BibitemOpen
  \bibfield  {author} {\bibinfo {author} {\bibfnamefont {I.~C.}\ \bibnamefont
  {{Percival}}},\ }\href {\doibase 10.1098/rspa.1995.0139} {\bibfield
  {journal} {\bibinfo  {journal} {Proceedings of the Royal Society of London A:
  Mathematical, Physical and Engineering Sciences}\ }\textbf {\bibinfo {volume}
  {451}},\ \bibinfo {pages} {503} (\bibinfo {year} {1995})}\BibitemShut
  {NoStop}%
\bibitem [{\citenamefont {{Feynman}}(1957)}]{FeynmanQG}%
  \BibitemOpen
  \bibfield  {author} {\bibinfo {author} {\bibfnamefont {R.}~\bibnamefont
  {{Feynman}}},\ }in\ \href
  {http://www.edition-open-sources.org/sources/5/index.html} {\emph {\bibinfo
  {booktitle} {{Chapel Hill Conference Proceedings}}}},\ \bibinfo {editor}
  {edited by\ \bibinfo {editor} {\bibfnamefont {C.~M.}\ \bibnamefont {DeWitt}}\
  and\ \bibinfo {editor} {\bibfnamefont {D.}~\bibnamefont {Rickles}}}\
  (\bibinfo  {publisher} {Edition Open Access},\ \bibinfo {year} {1957})\ pp.\
  \bibinfo {pages} {250--256}\BibitemShut {NoStop}%
\bibitem [{\citenamefont {Carney}, \citenamefont {Stamp},\ and\ \citenamefont
  {Taylor}()}]{DanCarneyReview}%
  \BibitemOpen
  \bibfield  {author} {\bibinfo {author} {\bibfnamefont {D.}~\bibnamefont
  {Carney}}, \bibinfo {author} {\bibfnamefont {P.~C.~E.}\ \bibnamefont
  {Stamp}}, \ and\ \bibinfo {author} {\bibfnamefont {J.~M.}\ \bibnamefont
  {Taylor}},\ }\href@noop {} {\bibinfo  {journal} {Classical and Quantum
  Gravity}\ }\BibitemShut {NoStop}%
\bibitem [{\citenamefont {Christodoulou}\ and\ \citenamefont
  {Rovelli}(2019)}]{RovelliQGExp}%
  \BibitemOpen
\bibfield  {journal} {  }\bibfield  {author} {\bibinfo {author} {\bibfnamefont
  {M.}~\bibnamefont {Christodoulou}}\ and\ \bibinfo {author} {\bibfnamefont
  {C.}~\bibnamefont {Rovelli}},\ }\href {\doibase
  https://doi.org/10.1016/j.physletb.2019.03.015} {\bibfield  {journal}
  {\bibinfo  {journal} {Physics Letters B}\ }\textbf {\bibinfo {volume}
  {792}},\ \bibinfo {pages} {64 } (\bibinfo {year} {2019})}\BibitemShut
  {NoStop}%
\bibitem [{\citenamefont {Bose}\ \emph {et~al.}(2017)\citenamefont {Bose},
  \citenamefont {Mazumdar}, \citenamefont {Morley}, \citenamefont {Ulbricht},
  \citenamefont {Toro\ifmmode~\check{s}\else \v{s}\fi{}}, \citenamefont
  {Paternostro}, \citenamefont {Geraci}, \citenamefont {Barker}, \citenamefont
  {Kim},\ and\ \citenamefont {Milburn}}]{BoseQGExp}%
  \BibitemOpen
  \bibfield  {author} {\bibinfo {author} {\bibfnamefont {S.}~\bibnamefont
  {Bose}}, \bibinfo {author} {\bibfnamefont {A.}~\bibnamefont {Mazumdar}},
  \bibinfo {author} {\bibfnamefont {G.~W.}\ \bibnamefont {Morley}}, \bibinfo
  {author} {\bibfnamefont {H.}~\bibnamefont {Ulbricht}}, \bibinfo {author}
  {\bibfnamefont {M.}~\bibnamefont {Toro\ifmmode~\check{s}\else \v{s}\fi{}}},
  \bibinfo {author} {\bibfnamefont {M.}~\bibnamefont {Paternostro}}, \bibinfo
  {author} {\bibfnamefont {A.~A.}\ \bibnamefont {Geraci}}, \bibinfo {author}
  {\bibfnamefont {P.~F.}\ \bibnamefont {Barker}}, \bibinfo {author}
  {\bibfnamefont {M.~S.}\ \bibnamefont {Kim}}, \ and\ \bibinfo {author}
  {\bibfnamefont {G.}~\bibnamefont {Milburn}},\ }\href {\doibase
  10.1103/PhysRevLett.119.240401} {\bibfield  {journal} {\bibinfo  {journal}
  {Phys. Rev. Lett.}\ }\textbf {\bibinfo {volume} {119}},\ \bibinfo {pages}
  {240401} (\bibinfo {year} {2017})}\BibitemShut {NoStop}%
\bibitem [{\citenamefont {Marletto}\ and\ \citenamefont
  {Vedral}(2017)}]{VedralQGExp}%
  \BibitemOpen
  \bibfield  {author} {\bibinfo {author} {\bibfnamefont {C.}~\bibnamefont
  {Marletto}}\ and\ \bibinfo {author} {\bibfnamefont {V.}~\bibnamefont
  {Vedral}},\ }\href {\doibase 10.1103/PhysRevLett.119.240402} {\bibfield
  {journal} {\bibinfo  {journal} {Phys. Rev. Lett.}\ }\textbf {\bibinfo
  {volume} {119}},\ \bibinfo {pages} {240402} (\bibinfo {year}
  {2017})}\BibitemShut {NoStop}%
\bibitem [{\citenamefont {Casimir}\ and\ \citenamefont
  {Polder}(1948)}]{CPInteraction}%
  \BibitemOpen
  \bibfield  {author} {\bibinfo {author} {\bibfnamefont {H.~B.~G.}\
  \bibnamefont {Casimir}}\ and\ \bibinfo {author} {\bibfnamefont
  {D.}~\bibnamefont {Polder}},\ }\href {\doibase 10.1103/PhysRev.73.360}
  {\bibfield  {journal} {\bibinfo  {journal} {Phys. Rev.}\ }\textbf {\bibinfo
  {volume} {73}},\ \bibinfo {pages} {360} (\bibinfo {year} {1948})}\BibitemShut
  {NoStop}%
\bibitem [{\citenamefont {Marletto}\ and\ \citenamefont
  {Vedral}(2018)}]{VlatkoClassical}%
  \BibitemOpen
  \bibfield  {author} {\bibinfo {author} {\bibfnamefont {C.}~\bibnamefont
  {Marletto}}\ and\ \bibinfo {author} {\bibfnamefont {V.}~\bibnamefont
  {Vedral}},\ }\href {\doibase 10.1103/PhysRevD.98.046001} {\bibfield
  {journal} {\bibinfo  {journal} {Phys. Rev. D}\ }\textbf {\bibinfo {volume}
  {98}},\ \bibinfo {pages} {046001} (\bibinfo {year} {2018})}\BibitemShut
  {NoStop}%
\bibitem [{\citenamefont {{Christodoulou}}\ and\ \citenamefont
  {{Rovelli}}(2018)}]{RovelliDiscreteTime}%
  \BibitemOpen
  \bibfield  {author} {\bibinfo {author} {\bibfnamefont {M.}~\bibnamefont
  {{Christodoulou}}}\ and\ \bibinfo {author} {\bibfnamefont {C.}~\bibnamefont
  {{Rovelli}}},\ }\href@noop {} {\bibfield  {journal} {\bibinfo  {journal}
  {arXiv e-prints}\ } (\bibinfo {year} {2018})},\ \Eprint
  {http://arxiv.org/abs/1812.01542} {arXiv:1812.01542} \BibitemShut {NoStop}%
\bibitem [{\citenamefont {Penrose}()}]{Penrose2000}%
  \BibitemOpen
  \bibfield  {author} {\bibinfo {author} {\bibfnamefont {R.}~\bibnamefont
  {Penrose}},\ }\enquote {\bibinfo {title} {Wavefunction collapse as a real
  gravitational effect},}\ in\ \href {\doibase 10.1142/9781848160224_0013}
  {\emph {\bibinfo {booktitle} {Mathematical Physics 2000}}},\ \bibinfo
  {editor} {edited by\ \bibinfo {editor} {\bibfnamefont {A.}~\bibnamefont
  {{Fokas}}}, \bibinfo {editor} {\bibfnamefont {T.~W.~B.}\ \bibnamefont
  {{Kibble}}}, \bibinfo {editor} {\bibfnamefont {A.}~\bibnamefont
  {{Grigouriou}}}, \ and\ \bibinfo {editor} {\bibfnamefont {B.}~\bibnamefont
  {{Zegarlinski}}}}\ (\bibinfo  {publisher} {Imperial College Press})\ pp.\
  \bibinfo {pages} {266--282}\BibitemShut {NoStop}%
\bibitem [{\citenamefont {Marshall}\ \emph {et~al.}(2003)\citenamefont
  {Marshall}, \citenamefont {Simon}, \citenamefont {Penrose},\ and\
  \citenamefont {Bouwmeester}}]{MirrorSuperposition}%
  \BibitemOpen
  \bibfield  {author} {\bibinfo {author} {\bibfnamefont {W.}~\bibnamefont
  {Marshall}}, \bibinfo {author} {\bibfnamefont {C.}~\bibnamefont {Simon}},
  \bibinfo {author} {\bibfnamefont {R.}~\bibnamefont {Penrose}}, \ and\
  \bibinfo {author} {\bibfnamefont {D.}~\bibnamefont {Bouwmeester}},\ }\href
  {\doibase 10.1103/PhysRevLett.91.130401} {\bibfield  {journal} {\bibinfo
  {journal} {Phys. Rev. Lett.}\ }\textbf {\bibinfo {volume} {91}},\ \bibinfo
  {pages} {130401} (\bibinfo {year} {2003})}\BibitemShut {NoStop}%
\bibitem [{\citenamefont {Weaver}\ \emph {et~al.}(2016)\citenamefont {Weaver},
  \citenamefont {Pepper}, \citenamefont {Luna}, \citenamefont {Buters},
  \citenamefont {Eerkens}, \citenamefont {Welker}, \citenamefont {Perock},
  \citenamefont {Heeck}, \citenamefont {de~Man},\ and\ \citenamefont
  {Bouwmeester}}]{TrampolineResonators}%
  \BibitemOpen
  \bibfield  {author} {\bibinfo {author} {\bibfnamefont {M.~J.}\ \bibnamefont
  {Weaver}}, \bibinfo {author} {\bibfnamefont {B.}~\bibnamefont {Pepper}},
  \bibinfo {author} {\bibfnamefont {F.}~\bibnamefont {Luna}}, \bibinfo {author}
  {\bibfnamefont {F.~M.}\ \bibnamefont {Buters}}, \bibinfo {author}
  {\bibfnamefont {H.~J.}\ \bibnamefont {Eerkens}}, \bibinfo {author}
  {\bibfnamefont {G.}~\bibnamefont {Welker}}, \bibinfo {author} {\bibfnamefont
  {B.}~\bibnamefont {Perock}}, \bibinfo {author} {\bibfnamefont
  {K.}~\bibnamefont {Heeck}}, \bibinfo {author} {\bibfnamefont
  {S.}~\bibnamefont {de~Man}}, \ and\ \bibinfo {author} {\bibfnamefont
  {D.}~\bibnamefont {Bouwmeester}},\ }\href {\doibase 10.1063/1.4939828}
  {\bibfield  {journal} {\bibinfo  {journal} {Applied Physics Letters}\
  }\textbf {\bibinfo {volume} {108}},\ \bibinfo {pages} {033501} (\bibinfo
  {year} {2016})}\BibitemShut {NoStop}%
\bibitem [{\citenamefont {Eerkens}\ \emph {et~al.}(2015)\citenamefont
  {Eerkens}, \citenamefont {Buters}, \citenamefont {Weaver}, \citenamefont
  {Pepper}, \citenamefont {Welker}, \citenamefont {Heeck}, \citenamefont
  {Sonin}, \citenamefont {de~Man},\ and\ \citenamefont
  {Bouwmeester}}]{SideBandCooling}%
  \BibitemOpen
  \bibfield  {author} {\bibinfo {author} {\bibfnamefont {H.~J.}\ \bibnamefont
  {Eerkens}}, \bibinfo {author} {\bibfnamefont {F.~M.}\ \bibnamefont {Buters}},
  \bibinfo {author} {\bibfnamefont {M.~J.}\ \bibnamefont {Weaver}}, \bibinfo
  {author} {\bibfnamefont {B.}~\bibnamefont {Pepper}}, \bibinfo {author}
  {\bibfnamefont {G.}~\bibnamefont {Welker}}, \bibinfo {author} {\bibfnamefont
  {K.}~\bibnamefont {Heeck}}, \bibinfo {author} {\bibfnamefont
  {P.}~\bibnamefont {Sonin}}, \bibinfo {author} {\bibfnamefont
  {S.}~\bibnamefont {de~Man}}, \ and\ \bibinfo {author} {\bibfnamefont
  {D.}~\bibnamefont {Bouwmeester}},\ }\href {\doibase 10.1364/OE.23.008014}
  {\bibfield  {journal} {\bibinfo  {journal} {Opt. Express}\ }\textbf {\bibinfo
  {volume} {23}},\ \bibinfo {pages} {8014} (\bibinfo {year}
  {2015})}\BibitemShut {NoStop}%
\bibitem [{\citenamefont {Adler}(2007)}]{Adler2007}%
  \BibitemOpen
  \bibfield  {author} {\bibinfo {author} {\bibfnamefont {S.~L.}\ \bibnamefont
  {Adler}},\ }\href {http://stacks.iop.org/1751-8121/40/i=4/a=011} {\bibfield
  {journal} {\bibinfo  {journal} {Journal of Physics A: Mathematical and
  Theoretical}\ }\textbf {\bibinfo {volume} {40}},\ \bibinfo {pages} {755}
  (\bibinfo {year} {2007})}\BibitemShut {NoStop}%
\bibitem [{\citenamefont {Pino}\ \emph {et~al.}(2018)\citenamefont {Pino},
  \citenamefont {Prat-Camps}, \citenamefont {Sinha}, \citenamefont
  {Venkatesh},\ and\ \citenamefont
  {Romero-Isart}}]{SuperconductingMicrosphere}%
  \BibitemOpen
  \bibfield  {author} {\bibinfo {author} {\bibfnamefont {H.}~\bibnamefont
  {Pino}}, \bibinfo {author} {\bibfnamefont {J.}~\bibnamefont {Prat-Camps}},
  \bibinfo {author} {\bibfnamefont {K.}~\bibnamefont {Sinha}}, \bibinfo
  {author} {\bibfnamefont {B.~P.}\ \bibnamefont {Venkatesh}}, \ and\ \bibinfo
  {author} {\bibfnamefont {O.}~\bibnamefont {Romero-Isart}},\ }\href
  {http://stacks.iop.org/2058-9565/3/i=2/a=025001} {\bibfield  {journal}
  {\bibinfo  {journal} {Quantum Science and Technology}\ }\textbf {\bibinfo
  {volume} {3}},\ \bibinfo {pages} {025001} (\bibinfo {year}
  {2018})}\BibitemShut {NoStop}%
\bibitem [{\citenamefont {Fuentes}\ and\ \citenamefont
  {Penrose}(2018)}]{FuentesPenrose2018}%
  \BibitemOpen
  \bibfield  {author} {\bibinfo {author} {\bibfnamefont {I.}~\bibnamefont
  {Fuentes}}\ and\ \bibinfo {author} {\bibfnamefont {R.}~\bibnamefont
  {Penrose}},\ }\enquote {\bibinfo {title} {Quantum state reduction via
  gravity, and possible tests using {B}ose–{E}instein condensates},}\ in\
  \href {\doibase 10.1017/9781316995457.012} {\emph {\bibinfo {booktitle}
  {Collapse of the Wave Function: Models, Ontology, Origin, and
  Implications}}}\ (\bibinfo  {publisher} {Cambridge University Press},\
  \bibinfo {year} {2018})\ p.\ \bibinfo {pages} {187–206}\BibitemShut
  {NoStop}%
\bibitem [{\citenamefont {{Haine}}(2018)}]{Haine}%
  \BibitemOpen
  \bibfield  {author} {\bibinfo {author} {\bibfnamefont {S.~A.}\ \bibnamefont
  {{Haine}}},\ }\href@noop {} {\bibfield  {journal} {\bibinfo  {journal} {arXiv
  e-prints}\ } (\bibinfo {year} {2018})},\ \Eprint
  {http://arxiv.org/abs/1810.10202} {arXiv:1810.10202 [quant-ph]} \BibitemShut
  {NoStop}%
\bibitem [{\citenamefont {Kovachy}\ \emph {et~al.}(2015)\citenamefont
  {Kovachy}, \citenamefont {Asenbaum}, \citenamefont {Overstreet},
  \citenamefont {Donnelly}, \citenamefont {Dickerson}, \citenamefont
  {Sugarbaker}, \citenamefont {Hogan},\ and\ \citenamefont
  {Kasevich}}]{KasevichHalfMetre}%
  \BibitemOpen
  \bibfield  {author} {\bibinfo {author} {\bibfnamefont {T.}~\bibnamefont
  {Kovachy}}, \bibinfo {author} {\bibfnamefont {P.}~\bibnamefont {Asenbaum}},
  \bibinfo {author} {\bibfnamefont {C.}~\bibnamefont {Overstreet}}, \bibinfo
  {author} {\bibfnamefont {C.}~\bibnamefont {Donnelly}}, \bibinfo {author}
  {\bibfnamefont {S.}~\bibnamefont {Dickerson}}, \bibinfo {author}
  {\bibfnamefont {A.}~\bibnamefont {Sugarbaker}}, \bibinfo {author}
  {\bibfnamefont {J.}~\bibnamefont {Hogan}}, \ and\ \bibinfo {author}
  {\bibfnamefont {M.}~\bibnamefont {Kasevich}},\ }\href {\doibase
  10.1038/nature16155} {\bibfield  {journal} {\bibinfo  {journal} {Nature}\
  }\textbf {\bibinfo {volume} {528}},\ \bibinfo {pages} {530} (\bibinfo {year}
  {2015})}\BibitemShut {NoStop}%
\bibitem [{\citenamefont {Gaunt}\ \emph {et~al.}(2013)\citenamefont {Gaunt},
  \citenamefont {Schmidutz}, \citenamefont {Gotlibovych}, \citenamefont
  {Smith},\ and\ \citenamefont {Hadzibabic}}]{UniformBECTrap}%
  \BibitemOpen
  \bibfield  {author} {\bibinfo {author} {\bibfnamefont {A.~L.}\ \bibnamefont
  {Gaunt}}, \bibinfo {author} {\bibfnamefont {T.~F.}\ \bibnamefont
  {Schmidutz}}, \bibinfo {author} {\bibfnamefont {I.}~\bibnamefont
  {Gotlibovych}}, \bibinfo {author} {\bibfnamefont {R.~P.}\ \bibnamefont
  {Smith}}, \ and\ \bibinfo {author} {\bibfnamefont {Z.}~\bibnamefont
  {Hadzibabic}},\ }\href {\doibase 10.1103/PhysRevLett.110.200406} {\bibfield
  {journal} {\bibinfo  {journal} {Phys. Rev. Lett.}\ }\textbf {\bibinfo
  {volume} {110}},\ \bibinfo {pages} {200406} (\bibinfo {year}
  {2013})}\BibitemShut {NoStop}%
\bibitem [{\citenamefont {Cirac}\ \emph {et~al.}(1998)\citenamefont {Cirac},
  \citenamefont {Lewenstein}, \citenamefont {M\o{}lmer},\ and\ \citenamefont
  {Zoller}}]{CatStatesBECs1998}%
  \BibitemOpen
  \bibfield  {author} {\bibinfo {author} {\bibfnamefont {J.~I.}\ \bibnamefont
  {Cirac}}, \bibinfo {author} {\bibfnamefont {M.}~\bibnamefont {Lewenstein}},
  \bibinfo {author} {\bibfnamefont {K.}~\bibnamefont {M\o{}lmer}}, \ and\
  \bibinfo {author} {\bibfnamefont {P.}~\bibnamefont {Zoller}},\ }\href
  {\doibase 10.1103/PhysRevA.57.1208} {\bibfield  {journal} {\bibinfo
  {journal} {Phys. Rev. A}\ }\textbf {\bibinfo {volume} {57}},\ \bibinfo
  {pages} {1208} (\bibinfo {year} {1998})}\BibitemShut {NoStop}%
\bibitem [{\citenamefont {Ho}\ and\ \citenamefont
  {Ciobanu}(2004)}]{CatsAttractiveBECs}%
  \BibitemOpen
  \bibfield  {author} {\bibinfo {author} {\bibfnamefont {T.-L.}\ \bibnamefont
  {Ho}}\ and\ \bibinfo {author} {\bibfnamefont {C.~V.}\ \bibnamefont
  {Ciobanu}},\ }\href {\doibase 10.1023/B:JOLT.0000024552.87247.eb} {\bibfield
  {journal} {\bibinfo  {journal} {Journal of Low Temperature Physics}\ }\textbf
  {\bibinfo {volume} {135}},\ \bibinfo {pages} {257} (\bibinfo {year}
  {2004})}\BibitemShut {NoStop}%
\bibitem [{\citenamefont {Huang}\ and\ \citenamefont
  {Moore}(2006)}]{CatStatesBECs}%
  \BibitemOpen
  \bibfield  {author} {\bibinfo {author} {\bibfnamefont {Y.~P.}\ \bibnamefont
  {Huang}}\ and\ \bibinfo {author} {\bibfnamefont {M.~G.}\ \bibnamefont
  {Moore}},\ }\href {\doibase 10.1103/PhysRevA.73.023606} {\bibfield  {journal}
  {\bibinfo  {journal} {Phys. Rev. A}\ }\textbf {\bibinfo {volume} {73}},\
  \bibinfo {pages} {023606} (\bibinfo {year} {2006})}\BibitemShut {NoStop}%
\bibitem [{\citenamefont {Bychek}, \citenamefont {Maksimov},\ and\
  \citenamefont {Kolovsky}(2018)}]{NOONRepulsive}%
  \BibitemOpen
  \bibfield  {author} {\bibinfo {author} {\bibfnamefont {A.~A.}\ \bibnamefont
  {Bychek}}, \bibinfo {author} {\bibfnamefont {D.~N.}\ \bibnamefont
  {Maksimov}}, \ and\ \bibinfo {author} {\bibfnamefont {A.~R.}\ \bibnamefont
  {Kolovsky}},\ }\href {\doibase 10.1103/PhysRevA.97.063624} {\bibfield
  {journal} {\bibinfo  {journal} {Phys. Rev. A}\ }\textbf {\bibinfo {volume}
  {97}},\ \bibinfo {pages} {063624} (\bibinfo {year} {2018})}\BibitemShut
  {NoStop}%
\bibitem [{\citenamefont {{Quandt-Wiese}}(2017)}]{EGOtherShapes}%
  \BibitemOpen
  \bibfield  {author} {\bibinfo {author} {\bibfnamefont {G.}~\bibnamefont
  {{Quandt-Wiese}}},\ }\href@noop {} {\bibfield  {journal} {\bibinfo  {journal}
  {ArXiv e-prints}\ } (\bibinfo {year} {2017})},\ \Eprint
  {http://arxiv.org/abs/1701.00353} {arXiv:1701.00353 [quant-ph]} \BibitemShut
  {NoStop}%
\bibitem [{\citenamefont {Mead}(1964)}]{PhysRev.135.B849}%
  \BibitemOpen
  \bibfield  {author} {\bibinfo {author} {\bibfnamefont {C.~A.}\ \bibnamefont
  {Mead}},\ }\href {\doibase 10.1103/PhysRev.135.B849} {\bibfield  {journal}
  {\bibinfo  {journal} {Phys. Rev.}\ }\textbf {\bibinfo {volume} {135}},\
  \bibinfo {pages} {B849} (\bibinfo {year} {1964})}\BibitemShut {NoStop}%
\bibitem [{\citenamefont {Tawfik}\ and\ \citenamefont {Diab}(2014)}]{GUP}%
  \BibitemOpen
  \bibfield  {author} {\bibinfo {author} {\bibfnamefont {A.}~\bibnamefont
  {Tawfik}}\ and\ \bibinfo {author} {\bibfnamefont {A.}~\bibnamefont {Diab}},\
  }\href {\doibase 10.1142/S0218271814300250} {\bibfield  {journal} {\bibinfo
  {journal} {International Journal of Modern Physics D}\ }\textbf {\bibinfo
  {volume} {23}},\ \bibinfo {pages} {1430025} (\bibinfo {year} {2014})},\
  \Eprint {http://arxiv.org/abs/https://doi.org/10.1142/S0218271814300250}
  {https://doi.org/10.1142/S0218271814300250} \BibitemShut {NoStop}%
\bibitem [{\citenamefont {Greenberger}\ and\ \citenamefont
  {Overhauser}(1979)}]{CoherenceNeutron}%
  \BibitemOpen
  \bibfield  {author} {\bibinfo {author} {\bibfnamefont {D.~M.}\ \bibnamefont
  {Greenberger}}\ and\ \bibinfo {author} {\bibfnamefont {A.~W.}\ \bibnamefont
  {Overhauser}},\ }\href {\doibase 10.1103/RevModPhys.51.43} {\bibfield
  {journal} {\bibinfo  {journal} {Rev. Mod. Phys.}\ }\textbf {\bibinfo {volume}
  {51}},\ \bibinfo {pages} {43} (\bibinfo {year} {1979})}\BibitemShut {NoStop}%
\bibitem [{\citenamefont {Beyer}\ and\ \citenamefont
  {Nitsch}(1986)}]{NonRelCOW}%
  \BibitemOpen
  \bibfield  {author} {\bibinfo {author} {\bibfnamefont {H.}~\bibnamefont
  {Beyer}}\ and\ \bibinfo {author} {\bibfnamefont {J.}~\bibnamefont {Nitsch}},\
  }\href {\doibase https://doi.org/10.1016/0370-2693(86)91579-0} {\bibfield
  {journal} {\bibinfo  {journal} {Physics Letters B}\ }\textbf {\bibinfo
  {volume} {182}},\ \bibinfo {pages} {211 } (\bibinfo {year}
  {1986})}\BibitemShut {NoStop}%
\bibitem [{\citenamefont {{Rosu}}(1999)}]{Rosu1999}%
  \BibitemOpen
  \bibfield  {author} {\bibinfo {author} {\bibfnamefont {H.~C.}\ \bibnamefont
  {{Rosu}}},\ }\href@noop {} {\bibfield  {journal} {\bibinfo  {journal}
  {Gravitation and Cosmology}\ }\textbf {\bibinfo {volume} {5}},\ \bibinfo
  {pages} {81} (\bibinfo {year} {1999})}\BibitemShut {NoStop}%
\bibitem [{\citenamefont {Colella}, \citenamefont {Overhauser},\ and\
  \citenamefont {Werner}(1975)}]{COW}%
  \BibitemOpen
  \bibfield  {author} {\bibinfo {author} {\bibfnamefont {R.}~\bibnamefont
  {Colella}}, \bibinfo {author} {\bibfnamefont {A.~W.}\ \bibnamefont
  {Overhauser}}, \ and\ \bibinfo {author} {\bibfnamefont {S.~A.}\ \bibnamefont
  {Werner}},\ }\href {\doibase 10.1103/PhysRevLett.34.1472} {\bibfield
  {journal} {\bibinfo  {journal} {Phys. Rev. Lett.}\ }\textbf {\bibinfo
  {volume} {34}},\ \bibinfo {pages} {1472} (\bibinfo {year}
  {1975})}\BibitemShut {NoStop}%
\bibitem [{\citenamefont {Colella}\ and\ \citenamefont
  {Overhauser}(1980)}]{AfterCOW}%
  \BibitemOpen
  \bibfield  {author} {\bibinfo {author} {\bibfnamefont {R.}~\bibnamefont
  {Colella}}\ and\ \bibinfo {author} {\bibfnamefont {A.~W.}\ \bibnamefont
  {Overhauser}},\ }\href {http://www.jstor.org/stable/27849721} {\bibfield
  {journal} {\bibinfo  {journal} {American Scientist}\ }\textbf {\bibinfo
  {volume} {68}},\ \bibinfo {pages} {70} (\bibinfo {year} {1980})}\BibitemShut
  {NoStop}%
\bibitem [{\citenamefont {Werner}(1994)}]{COWPhase}%
  \BibitemOpen
  \bibfield  {author} {\bibinfo {author} {\bibfnamefont {S.~A.}\ \bibnamefont
  {Werner}},\ }\href {http://stacks.iop.org/0264-9381/11/i=6A/a=016} {\bibfield
   {journal} {\bibinfo  {journal} {Classical and Quantum Gravity}\ }\textbf
  {\bibinfo {volume} {11}},\ \bibinfo {pages} {A207} (\bibinfo {year}
  {1994})}\BibitemShut {NoStop}%
\bibitem [{\citenamefont {Rauch}\ and\ \citenamefont
  {Werner}(2015)}]{NeutronInterferometryBook}%
  \BibitemOpen
  \bibfield  {author} {\bibinfo {author} {\bibfnamefont {H.}~\bibnamefont
  {Rauch}}\ and\ \bibinfo {author} {\bibfnamefont {S.~A.}\ \bibnamefont
  {Werner}},\ }\href {\doibase 10.1080/00107514.2016.1156770} {\emph {\bibinfo
  {title} {Neutron interferometry: lessons in experimental quantum mechanics,
  wave-particle duality, and entanglement}}},\ \bibinfo {edition} {2nd}\ ed.,\
  Vol.~\bibinfo {volume} {57}\ (\bibinfo  {publisher} {Oxford University
  Press},\ \bibinfo {year} {2015})\ pp.\ \bibinfo {pages}
  {284--285}\BibitemShut {NoStop}%
\bibitem [{\citenamefont {Fulling}(1973)}]{UnruhEffectFulling}%
  \BibitemOpen
  \bibfield  {author} {\bibinfo {author} {\bibfnamefont {S.~A.}\ \bibnamefont
  {Fulling}},\ }\href {\doibase 10.1103/PhysRevD.7.2850} {\bibfield  {journal}
  {\bibinfo  {journal} {Phys. Rev. D}\ }\textbf {\bibinfo {volume} {7}},\
  \bibinfo {pages} {2850} (\bibinfo {year} {1973})}\BibitemShut {NoStop}%
\bibitem [{\citenamefont {Davies}(1975)}]{UnruhEffectDavies}%
  \BibitemOpen
  \bibfield  {author} {\bibinfo {author} {\bibfnamefont {P.~C.~W.}\
  \bibnamefont {Davies}},\ }\href {http://stacks.iop.org/0305-4470/8/i=4/a=022}
  {\bibfield  {journal} {\bibinfo  {journal} {Journal of Physics A:
  Mathematical and General}\ }\textbf {\bibinfo {volume} {8}},\ \bibinfo
  {pages} {609} (\bibinfo {year} {1975})}\BibitemShut {NoStop}%
\bibitem [{\citenamefont {Unruh}(1976)}]{UnruhEffect}%
  \BibitemOpen
  \bibfield  {author} {\bibinfo {author} {\bibfnamefont {W.~G.}\ \bibnamefont
  {Unruh}},\ }\href {\doibase 10.1103/PhysRevD.14.870} {\bibfield  {journal}
  {\bibinfo  {journal} {Phys. Rev. D}\ }\textbf {\bibinfo {volume} {14}},\
  \bibinfo {pages} {870} (\bibinfo {year} {1976})}\BibitemShut {NoStop}%
\bibitem [{\citenamefont {{Oosterkamp}}\ and\ \citenamefont
  {{Zaanen}}(2014)}]{PenroseCollapseClock}%
  \BibitemOpen
  \bibfield  {author} {\bibinfo {author} {\bibfnamefont {T.~H.}\ \bibnamefont
  {{Oosterkamp}}}\ and\ \bibinfo {author} {\bibfnamefont {J.}~\bibnamefont
  {{Zaanen}}},\ }\href@noop {} {\bibfield  {journal} {\bibinfo  {journal}
  {ArXiv e-prints}\ } (\bibinfo {year} {2014})},\ \Eprint
  {http://arxiv.org/abs/1401.0176} {arXiv:1401.0176 [quant-ph]} \BibitemShut
  {NoStop}%
\bibitem [{\citenamefont {Penrose}(1998)}]{penrose1998quantum}%
  \BibitemOpen
  \bibfield  {author} {\bibinfo {author} {\bibfnamefont {R.}~\bibnamefont
  {Penrose}},\ }\href@noop {} {\bibfield  {journal} {\bibinfo  {journal}
  {Philosophical Transactions of the Royal Society of London. Series A:
  Mathematical, Physical and Engineering Sciences}\ }\textbf {\bibinfo {volume}
  {356}},\ \bibinfo {pages} {1927} (\bibinfo {year} {1998})}\BibitemShut
  {NoStop}%
\bibitem [{\citenamefont {Moroz}, \citenamefont {Penrose},\ and\ \citenamefont
  {Tod}(1998)}]{Moroz_1998}%
  \BibitemOpen
  \bibfield  {author} {\bibinfo {author} {\bibfnamefont {I.~M.}\ \bibnamefont
  {Moroz}}, \bibinfo {author} {\bibfnamefont {R.}~\bibnamefont {Penrose}}, \
  and\ \bibinfo {author} {\bibfnamefont {P.}~\bibnamefont {Tod}},\ }\href
  {\doibase 10.1088/0264-9381/15/9/019} {\bibfield  {journal} {\bibinfo
  {journal} {Classical and Quantum Gravity}\ }\textbf {\bibinfo {volume}
  {15}},\ \bibinfo {pages} {2733} (\bibinfo {year} {1998})}\BibitemShut
  {NoStop}%
\bibitem [{\citenamefont {Scala}\ \emph {et~al.}(2013)\citenamefont {Scala},
  \citenamefont {Kim}, \citenamefont {Morley}, \citenamefont {Barker},\ and\
  \citenamefont {Bose}}]{BoseMatterWave}%
  \BibitemOpen
  \bibfield  {author} {\bibinfo {author} {\bibfnamefont {M.}~\bibnamefont
  {Scala}}, \bibinfo {author} {\bibfnamefont {M.~S.}\ \bibnamefont {Kim}},
  \bibinfo {author} {\bibfnamefont {G.~W.}\ \bibnamefont {Morley}}, \bibinfo
  {author} {\bibfnamefont {P.~F.}\ \bibnamefont {Barker}}, \ and\ \bibinfo
  {author} {\bibfnamefont {S.}~\bibnamefont {Bose}},\ }\href {\doibase
  10.1103/PhysRevLett.111.180403} {\bibfield  {journal} {\bibinfo  {journal}
  {Phys. Rev. Lett.}\ }\textbf {\bibinfo {volume} {111}},\ \bibinfo {pages}
  {180403} (\bibinfo {year} {2013})}\BibitemShut {NoStop}%
\bibitem [{\citenamefont {Pitaevskii}()}]{Pitaevskii1961}%
  \BibitemOpen
  \bibfield  {author} {\bibinfo {author} {\bibfnamefont {L.~P.}\ \bibnamefont
  {Pitaevskii}},\ }\href@noop {} {\bibinfo  {journal} {Soviet Physics
  JETP-USSR}\ ,\ \bibinfo {pages} {451}}\BibitemShut {NoStop}%
\bibitem [{\citenamefont {Gross}(1961)}]{Gross61}%
  \BibitemOpen
\bibfield  {journal} {  }\bibfield  {author} {\bibinfo {author} {\bibfnamefont
  {E.}~\bibnamefont {Gross}},\ }\href {\doibase 10.1007/BF02731494} {\bibfield
  {journal} {\bibinfo  {journal} {Il Nuovo Cimento (1955-1965)}\ }\textbf
  {\bibinfo {volume} {20}},\ \bibinfo {pages} {454} (\bibinfo {year}
  {1961})}\BibitemShut {NoStop}%
\bibitem [{\citenamefont {P\'erez-Garc\'{\i}a}\ \emph
  {et~al.}(1996)\citenamefont {P\'erez-Garc\'{\i}a}, \citenamefont {Michinel},
  \citenamefont {Cirac}, \citenamefont {Lewenstein},\ and\ \citenamefont
  {Zoller}}]{GaussianAnsatzAttractive}%
  \BibitemOpen
  \bibfield  {author} {\bibinfo {author} {\bibfnamefont {V.~M.}\ \bibnamefont
  {P\'erez-Garc\'{\i}a}}, \bibinfo {author} {\bibfnamefont {H.}~\bibnamefont
  {Michinel}}, \bibinfo {author} {\bibfnamefont {J.~I.}\ \bibnamefont {Cirac}},
  \bibinfo {author} {\bibfnamefont {M.}~\bibnamefont {Lewenstein}}, \ and\
  \bibinfo {author} {\bibfnamefont {P.}~\bibnamefont {Zoller}},\ }\href
  {\doibase 10.1103/PhysRevLett.77.5320} {\bibfield  {journal} {\bibinfo
  {journal} {Phys. Rev. Lett.}\ }\textbf {\bibinfo {volume} {77}},\ \bibinfo
  {pages} {5320} (\bibinfo {year} {1996})}\BibitemShut {NoStop}%
\bibitem [{\citenamefont {Shi}\ and\ \citenamefont
  {Zheng}(1997)}]{GaussianAnsatzAttractiveShi}%
  \BibitemOpen
  \bibfield  {author} {\bibinfo {author} {\bibfnamefont {H.}~\bibnamefont
  {Shi}}\ and\ \bibinfo {author} {\bibfnamefont {W.-M.}\ \bibnamefont
  {Zheng}},\ }\href {\doibase 10.1103/PhysRevA.55.2930} {\bibfield  {journal}
  {\bibinfo  {journal} {Phys. Rev. A}\ }\textbf {\bibinfo {volume} {55}},\
  \bibinfo {pages} {2930} (\bibinfo {year} {1997})}\BibitemShut {NoStop}%
\bibitem [{\citenamefont {Stoof}(1997)}]{GaussianAnsatzAttractiveStoof}%
  \BibitemOpen
  \bibfield  {author} {\bibinfo {author} {\bibfnamefont {H.~T.~C.}\
  \bibnamefont {Stoof}},\ }\href {\doibase 10.1007/BF02181289} {\bibfield
  {journal} {\bibinfo  {journal} {Journal of Statistical Physics}\ }\textbf
  {\bibinfo {volume} {87}},\ \bibinfo {pages} {1353} (\bibinfo {year}
  {1997})}\BibitemShut {NoStop}%
\bibitem [{\citenamefont {Parola}, \citenamefont {Salasnich},\ and\
  \citenamefont {Reatto}(1998)}]{GaussianAnsatzParola}%
  \BibitemOpen
  \bibfield  {author} {\bibinfo {author} {\bibfnamefont {A.}~\bibnamefont
  {Parola}}, \bibinfo {author} {\bibfnamefont {L.}~\bibnamefont {Salasnich}}, \
  and\ \bibinfo {author} {\bibfnamefont {L.}~\bibnamefont {Reatto}},\ }\href
  {\doibase 10.1103/PhysRevA.57.R3180} {\bibfield  {journal} {\bibinfo
  {journal} {Phys. Rev. A}\ }\textbf {\bibinfo {volume} {57}},\ \bibinfo
  {pages} {R3180} (\bibinfo {year} {1998})}\BibitemShut {NoStop}%
\bibitem [{\citenamefont {Fetter}(1997)}]{GaussianAnsatzFetter}%
  \BibitemOpen
  \bibfield  {author} {\bibinfo {author} {\bibfnamefont {A.~L.}\ \bibnamefont
  {Fetter}},\ }\href {\doibase 10.1007/BF02395929} {\bibfield  {journal}
  {\bibinfo  {journal} {Journal of Low Temperature Physics}\ }\textbf {\bibinfo
  {volume} {106}},\ \bibinfo {pages} {643} (\bibinfo {year}
  {1997})}\BibitemShut {NoStop}%
\bibitem [{\citenamefont {Dalfovo}\ \emph {et~al.}(1999)\citenamefont
  {Dalfovo}, \citenamefont {Giorgini}, \citenamefont {Pitaevskii},\ and\
  \citenamefont {Stringari}}]{BECReview}%
  \BibitemOpen
  \bibfield  {author} {\bibinfo {author} {\bibfnamefont {F.}~\bibnamefont
  {Dalfovo}}, \bibinfo {author} {\bibfnamefont {S.}~\bibnamefont {Giorgini}},
  \bibinfo {author} {\bibfnamefont {L.~P.}\ \bibnamefont {Pitaevskii}}, \ and\
  \bibinfo {author} {\bibfnamefont {S.}~\bibnamefont {Stringari}},\ }\href
  {\doibase 10.1103/RevModPhys.71.463} {\bibfield  {journal} {\bibinfo
  {journal} {Rev. Mod. Phys.}\ }\textbf {\bibinfo {volume} {71}},\ \bibinfo
  {pages} {463} (\bibinfo {year} {1999})}\BibitemShut {NoStop}%
\bibitem [{\citenamefont {Thomas}(1927)}]{TFThomas}%
  \BibitemOpen
  \bibfield  {author} {\bibinfo {author} {\bibfnamefont {L.~H.}\ \bibnamefont
  {Thomas}},\ }\href {\doibase 10.1017/S0305004100011683} {\bibfield  {journal}
  {\bibinfo  {journal} {Mathematical Proceedings of the Cambridge Philosophical
  Society}\ }\textbf {\bibinfo {volume} {23}},\ \bibinfo {pages} {542}
  (\bibinfo {year} {1927})}\BibitemShut {NoStop}%
\bibitem [{\citenamefont {Fermi}(1927)}]{TFFermi}%
  \BibitemOpen
  \bibfield  {author} {\bibinfo {author} {\bibfnamefont {E.}~\bibnamefont
  {Fermi}},\ }\href {\doibase 10.1017/S0305004100011683} {\bibfield  {journal}
  {\bibinfo  {journal} {Rend. Accad. Naz. Lincei}\ }\textbf {\bibinfo {volume}
  {6}},\ \bibinfo {pages} {602} (\bibinfo {year} {1927})}\BibitemShut {NoStop}%
\bibitem [{\citenamefont {Ruprecht}\ \emph {et~al.}(1995)\citenamefont
  {Ruprecht}, \citenamefont {Holland}, \citenamefont {Burnett},\ and\
  \citenamefont {Edwards}}]{AttractiveSphericalCollapse}%
  \BibitemOpen
  \bibfield  {author} {\bibinfo {author} {\bibfnamefont {P.~A.}\ \bibnamefont
  {Ruprecht}}, \bibinfo {author} {\bibfnamefont {M.~J.}\ \bibnamefont
  {Holland}}, \bibinfo {author} {\bibfnamefont {K.}~\bibnamefont {Burnett}}, \
  and\ \bibinfo {author} {\bibfnamefont {M.}~\bibnamefont {Edwards}},\ }\href
  {\doibase 10.1103/PhysRevA.51.4704} {\bibfield  {journal} {\bibinfo
  {journal} {Phys. Rev. A}\ }\textbf {\bibinfo {volume} {51}},\ \bibinfo
  {pages} {4704} (\bibinfo {year} {1995})}\BibitemShut {NoStop}%
\bibitem [{\citenamefont {Sakellari}, \citenamefont {Proukakis},\ and\
  \citenamefont {Adams}(2004)}]{CriticNumberDoubleWell}%
  \BibitemOpen
  \bibfield  {author} {\bibinfo {author} {\bibfnamefont {E.}~\bibnamefont
  {Sakellari}}, \bibinfo {author} {\bibfnamefont {N.~P.}\ \bibnamefont
  {Proukakis}}, \ and\ \bibinfo {author} {\bibfnamefont {C.~S.}\ \bibnamefont
  {Adams}},\ }\href {http://stacks.iop.org/0953-4075/37/i=18/a=009} {\bibfield
  {journal} {\bibinfo  {journal} {Journal of Physics B: Atomic, Molecular and
  Optical Physics}\ }\textbf {\bibinfo {volume} {37}},\ \bibinfo {pages} {3681}
  (\bibinfo {year} {2004})}\BibitemShut {NoStop}%
\bibitem [{\citenamefont {Wadati}\ and\ \citenamefont
  {Tsurumi}(1998)}]{CritNumberAnalytic}%
  \BibitemOpen
  \bibfield  {author} {\bibinfo {author} {\bibfnamefont {M.}~\bibnamefont
  {Wadati}}\ and\ \bibinfo {author} {\bibfnamefont {T.}~\bibnamefont
  {Tsurumi}},\ }\href {\doibase https://doi.org/10.1016/S0375-9601(98)00583-0}
  {\bibfield  {journal} {\bibinfo  {journal} {Physics Letters A}\ }\textbf
  {\bibinfo {volume} {247}},\ \bibinfo {pages} {287 } (\bibinfo {year}
  {1998})}\BibitemShut {NoStop}%
\bibitem [{\citenamefont {{Griffin}}, \citenamefont {{Nikuni}},\ and\
  \citenamefont {{Zaremba}}(2009)}]{YellowBook}%
  \BibitemOpen
  \bibfield  {author} {\bibinfo {author} {\bibfnamefont {A.}~\bibnamefont
  {{Griffin}}}, \bibinfo {author} {\bibfnamefont {T.}~\bibnamefont {{Nikuni}}},
  \ and\ \bibinfo {author} {\bibfnamefont {E.}~\bibnamefont {{Zaremba}}},\
  }\href@noop {} {\emph {\bibinfo {title} {Bose-Condensed Gases at Finite
  Temperatures, by Allan Griffin , Tetsuro Nikuni , Eugene Zaremba, Cambridge,
  UK: Cambridge University Press, 2009}}}\ (\bibinfo  {publisher} {Cambridge
  University Press},\ \bibinfo {year} {2009})\BibitemShut {NoStop}%
\bibitem [{\citenamefont {Pitaevskii}\ and\ \citenamefont
  {Stringari}(2016)}]{PitaevskiiBook}%
  \BibitemOpen
  \bibfield  {author} {\bibinfo {author} {\bibfnamefont {L.}~\bibnamefont
  {Pitaevskii}}\ and\ \bibinfo {author} {\bibfnamefont {S.}~\bibnamefont
  {Stringari}},\ }\href@noop {} {\emph {\bibinfo {title} {Bose-Einstein
  Condensation and Superfluidity}}}\ (\bibinfo  {publisher} {Oxford University
  Press},\ \bibinfo {year} {2016})\BibitemShut {NoStop}%
\bibitem [{\citenamefont {Fried}\ \emph {et~al.}(1998)\citenamefont {Fried},
  \citenamefont {Killian}, \citenamefont {Willmann}, \citenamefont {Landhuis},
  \citenamefont {Moss}, \citenamefont {Kleppner},\ and\ \citenamefont
  {Greytak}}]{HydrogenBEC}%
  \BibitemOpen
  \bibfield  {author} {\bibinfo {author} {\bibfnamefont {D.~G.}\ \bibnamefont
  {Fried}}, \bibinfo {author} {\bibfnamefont {T.~C.}\ \bibnamefont {Killian}},
  \bibinfo {author} {\bibfnamefont {L.}~\bibnamefont {Willmann}}, \bibinfo
  {author} {\bibfnamefont {D.}~\bibnamefont {Landhuis}}, \bibinfo {author}
  {\bibfnamefont {S.~C.}\ \bibnamefont {Moss}}, \bibinfo {author}
  {\bibfnamefont {D.}~\bibnamefont {Kleppner}}, \ and\ \bibinfo {author}
  {\bibfnamefont {T.~J.}\ \bibnamefont {Greytak}},\ }\href {\doibase
  10.1103/PhysRevLett.81.3811} {\bibfield  {journal} {\bibinfo  {journal}
  {Phys. Rev. Lett.}\ }\textbf {\bibinfo {volume} {81}},\ \bibinfo {pages}
  {3811} (\bibinfo {year} {1998})}\BibitemShut {NoStop}%
\bibitem [{\citenamefont {van~der Stam}\ \emph {et~al.}(2007)\citenamefont
  {van~der Stam}, \citenamefont {van Ooijen}, \citenamefont {Meppelink},
  \citenamefont {Vogels},\ and\ \citenamefont {van~der Straten}}]{LargeNaBEC}%
  \BibitemOpen
  \bibfield  {author} {\bibinfo {author} {\bibfnamefont {K.~M.~R.}\
  \bibnamefont {van~der Stam}}, \bibinfo {author} {\bibfnamefont {E.~D.}\
  \bibnamefont {van Ooijen}}, \bibinfo {author} {\bibfnamefont
  {R.}~\bibnamefont {Meppelink}}, \bibinfo {author} {\bibfnamefont {J.~M.}\
  \bibnamefont {Vogels}}, \ and\ \bibinfo {author} {\bibfnamefont
  {P.}~\bibnamefont {van~der Straten}},\ }\href {\doibase 10.1063/1.2424439}
  {\bibfield  {journal} {\bibinfo  {journal} {Review of Scientific
  Instruments}\ }\textbf {\bibinfo {volume} {78}},\ \bibinfo {pages} {013102}
  (\bibinfo {year} {2007})}\BibitemShut {NoStop}%
\bibitem [{\citenamefont {Naik}\ and\ \citenamefont {Raman}(2005)}]{Naik2005}%
  \BibitemOpen
  \bibfield  {author} {\bibinfo {author} {\bibfnamefont {D.~S.}\ \bibnamefont
  {Naik}}\ and\ \bibinfo {author} {\bibfnamefont {C.}~\bibnamefont {Raman}},\
  }\href {\doibase 10.1103/PhysRevA.71.033617} {\bibfield  {journal} {\bibinfo
  {journal} {Phys. Rev. A}\ }\textbf {\bibinfo {volume} {71}},\ \bibinfo
  {pages} {033617} (\bibinfo {year} {2005})}\BibitemShut {NoStop}%
\bibitem [{\citenamefont {Ruostekoski}\ \emph {et~al.}(1998)\citenamefont
  {Ruostekoski}, \citenamefont {Collett}, \citenamefont {Graham},\ and\
  \citenamefont {Walls}}]{MacroscopicSuperpositionsRuostekoski}%
  \BibitemOpen
  \bibfield  {author} {\bibinfo {author} {\bibfnamefont {J.}~\bibnamefont
  {Ruostekoski}}, \bibinfo {author} {\bibfnamefont {M.~J.}\ \bibnamefont
  {Collett}}, \bibinfo {author} {\bibfnamefont {R.}~\bibnamefont {Graham}}, \
  and\ \bibinfo {author} {\bibfnamefont {D.~F.}\ \bibnamefont {Walls}},\ }\href
  {\doibase 10.1103/PhysRevA.57.511} {\bibfield  {journal} {\bibinfo  {journal}
  {Phys. Rev. A}\ }\textbf {\bibinfo {volume} {57}},\ \bibinfo {pages} {511}
  (\bibinfo {year} {1998})}\BibitemShut {NoStop}%
\bibitem [{\citenamefont {Gordon}\ and\ \citenamefont
  {Savage}(1999)}]{MacroscopicStatesBEC1999}%
  \BibitemOpen
  \bibfield  {author} {\bibinfo {author} {\bibfnamefont {D.}~\bibnamefont
  {Gordon}}\ and\ \bibinfo {author} {\bibfnamefont {C.~M.}\ \bibnamefont
  {Savage}},\ }\href {\doibase 10.1103/PhysRevA.59.4623} {\bibfield  {journal}
  {\bibinfo  {journal} {Phys. Rev. A}\ }\textbf {\bibinfo {volume} {59}},\
  \bibinfo {pages} {4623} (\bibinfo {year} {1999})}\BibitemShut {NoStop}%
\bibitem [{\citenamefont {Louis}, \citenamefont {Brydon},\ and\ \citenamefont
  {Savage}(2001)}]{MacroscopicStatesBECs2001}%
  \BibitemOpen
  \bibfield  {author} {\bibinfo {author} {\bibfnamefont {P.~J.~Y.}\
  \bibnamefont {Louis}}, \bibinfo {author} {\bibfnamefont {P.~M.~R.}\
  \bibnamefont {Brydon}}, \ and\ \bibinfo {author} {\bibfnamefont {C.~M.}\
  \bibnamefont {Savage}},\ }\href {\doibase 10.1103/PhysRevA.64.053613}
  {\bibfield  {journal} {\bibinfo  {journal} {Phys. Rev. A}\ }\textbf {\bibinfo
  {volume} {64}},\ \bibinfo {pages} {053613} (\bibinfo {year}
  {2001})}\BibitemShut {NoStop}%
\bibitem [{\citenamefont {Ruostekoski}(2001)}]{CatStatesRuostekoski}%
  \BibitemOpen
  \bibfield  {author} {\bibinfo {author} {\bibfnamefont {J.}~\bibnamefont
  {Ruostekoski}},\ }in\ \href@noop {} {\emph {\bibinfo {booktitle} {Directions
  in Quantum Optics}}},\ \bibinfo {editor} {edited by\ \bibinfo {editor}
  {\bibfnamefont {H.~J.}\ \bibnamefont {Carmichael}}, \bibinfo {editor}
  {\bibfnamefont {R.~J.}\ \bibnamefont {Glauber}}, \ and\ \bibinfo {editor}
  {\bibfnamefont {M.~O.}\ \bibnamefont {Scully}}}\ (\bibinfo  {publisher}
  {Springer Berlin Heidelberg},\ \bibinfo {address} {Berlin, Heidelberg},\
  \bibinfo {year} {2001})\ pp.\ \bibinfo {pages} {77--87}\BibitemShut {NoStop}%
\bibitem [{\citenamefont {Mahmud}, \citenamefont {Perry},\ and\ \citenamefont
  {Reinhardt}(2003)}]{CatsBECsPhaseImprint}%
  \BibitemOpen
  \bibfield  {author} {\bibinfo {author} {\bibfnamefont {K.~W.}\ \bibnamefont
  {Mahmud}}, \bibinfo {author} {\bibfnamefont {H.}~\bibnamefont {Perry}}, \
  and\ \bibinfo {author} {\bibfnamefont {W.~P.}\ \bibnamefont {Reinhardt}},\
  }\href {http://stacks.iop.org/0953-4075/36/i=17/a=102} {\bibfield  {journal}
  {\bibinfo  {journal} {Journal of Physics B: Atomic, Molecular and Optical
  Physics}\ }\textbf {\bibinfo {volume} {36}},\ \bibinfo {pages} {L265}
  (\bibinfo {year} {2003})}\BibitemShut {NoStop}%
\bibitem [{\citenamefont {Micheli}\ \emph {et~al.}(2003)\citenamefont
  {Micheli}, \citenamefont {Jaksch}, \citenamefont {Cirac},\ and\ \citenamefont
  {Zoller}}]{CatStatesPhase2003}%
  \BibitemOpen
  \bibfield  {author} {\bibinfo {author} {\bibfnamefont {A.}~\bibnamefont
  {Micheli}}, \bibinfo {author} {\bibfnamefont {D.}~\bibnamefont {Jaksch}},
  \bibinfo {author} {\bibfnamefont {J.~I.}\ \bibnamefont {Cirac}}, \ and\
  \bibinfo {author} {\bibfnamefont {P.}~\bibnamefont {Zoller}},\ }\href
  {\doibase 10.1103/PhysRevA.67.013607} {\bibfield  {journal} {\bibinfo
  {journal} {Phys. Rev. A}\ }\textbf {\bibinfo {volume} {67}},\ \bibinfo
  {pages} {013607} (\bibinfo {year} {2003})}\BibitemShut {NoStop}%
\bibitem [{\citenamefont {Dunningham}\ \emph {et~al.}(2006)\citenamefont
  {Dunningham}, \citenamefont {Burnett}, \citenamefont {Roth},\ and\
  \citenamefont {Phillips}}]{MacroscopicSuperpositionsDunningham}%
  \BibitemOpen
  \bibfield  {author} {\bibinfo {author} {\bibfnamefont {J.~A.}\ \bibnamefont
  {Dunningham}}, \bibinfo {author} {\bibfnamefont {K.}~\bibnamefont {Burnett}},
  \bibinfo {author} {\bibfnamefont {R.}~\bibnamefont {Roth}}, \ and\ \bibinfo
  {author} {\bibfnamefont {W.~D.}\ \bibnamefont {Phillips}},\ }\href
  {http://stacks.iop.org/1367-2630/8/i=9/a=182} {\bibfield  {journal} {\bibinfo
   {journal} {New Journal of Physics}\ }\textbf {\bibinfo {volume} {8}},\
  \bibinfo {pages} {182} (\bibinfo {year} {2006})}\BibitemShut {NoStop}%
\bibitem [{\citenamefont {Smerzi}\ and\ \citenamefont
  {Trombettoni}(2003)}]{NonLinearTBApprox}%
  \BibitemOpen
  \bibfield  {author} {\bibinfo {author} {\bibfnamefont {A.}~\bibnamefont
  {Smerzi}}\ and\ \bibinfo {author} {\bibfnamefont {A.}~\bibnamefont
  {Trombettoni}},\ }\href {\doibase 10.1103/PhysRevA.68.023613} {\bibfield
  {journal} {\bibinfo  {journal} {Phys. Rev. A}\ }\textbf {\bibinfo {volume}
  {68}},\ \bibinfo {pages} {023613} (\bibinfo {year} {2003})}\BibitemShut
  {NoStop}%
\bibitem [{\citenamefont {Dalton}(2011)}]{DALTON2011668}%
  \BibitemOpen
  \bibfield  {author} {\bibinfo {author} {\bibfnamefont {B.}~\bibnamefont
  {Dalton}},\ }\href {\doibase https://doi.org/10.1016/j.aop.2010.10.006}
  {\bibfield  {journal} {\bibinfo  {journal} {Annals of Physics}\ }\textbf
  {\bibinfo {volume} {326}},\ \bibinfo {pages} {668 } (\bibinfo {year}
  {2011})}\BibitemShut {NoStop}%
\bibitem [{\citenamefont {Fuentes-Schuller}\ and\ \citenamefont
  {Barberis-Blostein}(2007)}]{AnalyticalBECs1}%
  \BibitemOpen
  \bibfield  {author} {\bibinfo {author} {\bibfnamefont {I.}~\bibnamefont
  {Fuentes-Schuller}}\ and\ \bibinfo {author} {\bibfnamefont {P.}~\bibnamefont
  {Barberis-Blostein}},\ }\href {http://stacks.iop.org/1751-8121/40/i=27/a=F04}
  {\bibfield  {journal} {\bibinfo  {journal} {Journal of Physics A:
  Mathematical and Theoretical}\ }\textbf {\bibinfo {volume} {40}},\ \bibinfo
  {pages} {F601} (\bibinfo {year} {2007})}\BibitemShut {NoStop}%
\bibitem [{\citenamefont {Gersch}\ and\ \citenamefont
  {Knollman}(1963)}]{BoseHubbard}%
  \BibitemOpen
  \bibfield  {author} {\bibinfo {author} {\bibfnamefont {H.~A.}\ \bibnamefont
  {Gersch}}\ and\ \bibinfo {author} {\bibfnamefont {G.~C.}\ \bibnamefont
  {Knollman}},\ }\href {\doibase 10.1103/PhysRev.129.959} {\bibfield  {journal}
  {\bibinfo  {journal} {Phys. Rev.}\ }\textbf {\bibinfo {volume} {129}},\
  \bibinfo {pages} {959} (\bibinfo {year} {1963})}\BibitemShut {NoStop}%
\bibitem [{\citenamefont {Fisher}\ \emph {et~al.}(1989)\citenamefont {Fisher},
  \citenamefont {Weichman}, \citenamefont {Grinstein},\ and\ \citenamefont
  {Fisher}}]{BoseHubbardLattice}%
  \BibitemOpen
  \bibfield  {author} {\bibinfo {author} {\bibfnamefont {M.~P.~A.}\
  \bibnamefont {Fisher}}, \bibinfo {author} {\bibfnamefont {P.~B.}\
  \bibnamefont {Weichman}}, \bibinfo {author} {\bibfnamefont {G.}~\bibnamefont
  {Grinstein}}, \ and\ \bibinfo {author} {\bibfnamefont {D.~S.}\ \bibnamefont
  {Fisher}},\ }\href {\doibase 10.1103/PhysRevB.40.546} {\bibfield  {journal}
  {\bibinfo  {journal} {Phys. Rev. B}\ }\textbf {\bibinfo {volume} {40}},\
  \bibinfo {pages} {546} (\bibinfo {year} {1989})}\BibitemShut {NoStop}%
\bibitem [{\citenamefont {Jaksch}\ \emph
  {et~al.}(1998{\natexlab{a}})\citenamefont {Jaksch}, \citenamefont {Bruder},
  \citenamefont {Cirac}, \citenamefont {Gardiner},\ and\ \citenamefont
  {Zoller}}]{BoseHubbardBEC1}%
  \BibitemOpen
  \bibfield  {author} {\bibinfo {author} {\bibfnamefont {D.}~\bibnamefont
  {Jaksch}}, \bibinfo {author} {\bibfnamefont {C.}~\bibnamefont {Bruder}},
  \bibinfo {author} {\bibfnamefont {J.~I.}\ \bibnamefont {Cirac}}, \bibinfo
  {author} {\bibfnamefont {C.~W.}\ \bibnamefont {Gardiner}}, \ and\ \bibinfo
  {author} {\bibfnamefont {P.}~\bibnamefont {Zoller}},\ }\href {\doibase
  10.1103/PhysRevLett.81.3108} {\bibfield  {journal} {\bibinfo  {journal}
  {Phys. Rev. Lett.}\ }\textbf {\bibinfo {volume} {81}},\ \bibinfo {pages}
  {3108} (\bibinfo {year} {1998}{\natexlab{a}})}\BibitemShut {NoStop}%
\bibitem [{\citenamefont {Haigh}, \citenamefont {Ferris},\ and\ \citenamefont
  {Olsen}(2010)}]{DemonstratingNOON}%
  \BibitemOpen
  \bibfield  {author} {\bibinfo {author} {\bibfnamefont {T.}~\bibnamefont
  {Haigh}}, \bibinfo {author} {\bibfnamefont {A.}~\bibnamefont {Ferris}}, \
  and\ \bibinfo {author} {\bibfnamefont {M.}~\bibnamefont {Olsen}},\ }\href
  {\doibase https://doi.org/10.1016/j.optcom.2010.04.067} {\bibfield  {journal}
  {\bibinfo  {journal} {Optics Communications}\ }\textbf {\bibinfo {volume}
  {283}},\ \bibinfo {pages} {3540 } (\bibinfo {year} {2010})}\BibitemShut
  {NoStop}%
\bibitem [{\citenamefont {Opanchuk}\ \emph {et~al.}(2016)\citenamefont
  {Opanchuk}, \citenamefont {Rosales-Z\'arate}, \citenamefont {Teh},\ and\
  \citenamefont {Reid}}]{QuantifyingNOONCoherence}%
  \BibitemOpen
  \bibfield  {author} {\bibinfo {author} {\bibfnamefont {B.}~\bibnamefont
  {Opanchuk}}, \bibinfo {author} {\bibfnamefont {L.}~\bibnamefont
  {Rosales-Z\'arate}}, \bibinfo {author} {\bibfnamefont {R.~Y.}\ \bibnamefont
  {Teh}}, \ and\ \bibinfo {author} {\bibfnamefont {M.~D.}\ \bibnamefont
  {Reid}},\ }\href {\doibase 10.1103/PhysRevA.94.062125} {\bibfield  {journal}
  {\bibinfo  {journal} {Phys. Rev. A}\ }\textbf {\bibinfo {volume} {94}},\
  \bibinfo {pages} {062125} (\bibinfo {year} {2016})}\BibitemShut {NoStop}%
\bibitem [{\citenamefont {Jack}(2002)}]{3BodyDecoherence}%
  \BibitemOpen
  \bibfield  {author} {\bibinfo {author} {\bibfnamefont {M.~W.}\ \bibnamefont
  {Jack}},\ }\href {\doibase 10.1103/PhysRevLett.89.140402} {\bibfield
  {journal} {\bibinfo  {journal} {Phys. Rev. Lett.}\ }\textbf {\bibinfo
  {volume} {89}},\ \bibinfo {pages} {140402} (\bibinfo {year}
  {2002})}\BibitemShut {NoStop}%
\bibitem [{\citenamefont {Bilardello}, \citenamefont {Trombettoni},\ and\
  \citenamefont {Bassi}(2017)}]{CSLBECs}%
  \BibitemOpen
  \bibfield  {author} {\bibinfo {author} {\bibfnamefont {M.}~\bibnamefont
  {Bilardello}}, \bibinfo {author} {\bibfnamefont {A.}~\bibnamefont
  {Trombettoni}}, \ and\ \bibinfo {author} {\bibfnamefont {A.}~\bibnamefont
  {Bassi}},\ }\href {\doibase 10.1103/PhysRevA.95.032134} {\bibfield  {journal}
  {\bibinfo  {journal} {Phys. Rev. A}\ }\textbf {\bibinfo {volume} {95}},\
  \bibinfo {pages} {032134} (\bibinfo {year} {2017})}\BibitemShut {NoStop}%
\bibitem [{\citenamefont {Fedichev}, \citenamefont {Reynolds},\ and\
  \citenamefont {Shlyapnikov}(1996)}]{ThreeBodyRate}%
  \BibitemOpen
  \bibfield  {author} {\bibinfo {author} {\bibfnamefont {P.~O.}\ \bibnamefont
  {Fedichev}}, \bibinfo {author} {\bibfnamefont {M.~W.}\ \bibnamefont
  {Reynolds}}, \ and\ \bibinfo {author} {\bibfnamefont {G.~V.}\ \bibnamefont
  {Shlyapnikov}},\ }\href {\doibase 10.1103/PhysRevLett.77.2921} {\bibfield
  {journal} {\bibinfo  {journal} {Phys. Rev. Lett.}\ }\textbf {\bibinfo
  {volume} {77}},\ \bibinfo {pages} {2921} (\bibinfo {year}
  {1996})}\BibitemShut {NoStop}%
\bibitem [{\citenamefont {Search}, \citenamefont {Zhang},\ and\ \citenamefont
  {Meystre}(2004)}]{InhibitingThreeBody}%
  \BibitemOpen
  \bibfield  {author} {\bibinfo {author} {\bibfnamefont {C.~P.}\ \bibnamefont
  {Search}}, \bibinfo {author} {\bibfnamefont {W.}~\bibnamefont {Zhang}}, \
  and\ \bibinfo {author} {\bibfnamefont {P.}~\bibnamefont {Meystre}},\ }\href
  {\doibase 10.1103/PhysRevLett.92.140401} {\bibfield  {journal} {\bibinfo
  {journal} {Phys. Rev. Lett.}\ }\textbf {\bibinfo {volume} {92}},\ \bibinfo
  {pages} {140401} (\bibinfo {year} {2004})}\BibitemShut {NoStop}%
\bibitem [{\citenamefont {Sch\"utzhold}\ and\ \citenamefont
  {Gnanapragasam}(2010)}]{InhibitingThreeBodyZeno}%
  \BibitemOpen
  \bibfield  {author} {\bibinfo {author} {\bibfnamefont {R.}~\bibnamefont
  {Sch\"utzhold}}\ and\ \bibinfo {author} {\bibfnamefont {G.}~\bibnamefont
  {Gnanapragasam}},\ }\href {\doibase 10.1103/PhysRevA.82.022120} {\bibfield
  {journal} {\bibinfo  {journal} {Phys. Rev. A}\ }\textbf {\bibinfo {volume}
  {82}},\ \bibinfo {pages} {022120} (\bibinfo {year} {2010})}\BibitemShut
  {NoStop}%
\bibitem [{\citenamefont {Mehta}, \citenamefont {Esry},\ and\ \citenamefont
  {Greene}(2007)}]{ThreeBodyOneD}%
  \BibitemOpen
  \bibfield  {author} {\bibinfo {author} {\bibfnamefont {N.~P.}\ \bibnamefont
  {Mehta}}, \bibinfo {author} {\bibfnamefont {B.~D.}\ \bibnamefont {Esry}}, \
  and\ \bibinfo {author} {\bibfnamefont {C.~H.}\ \bibnamefont {Greene}},\
  }\href {\doibase 10.1103/PhysRevA.76.022711} {\bibfield  {journal} {\bibinfo
  {journal} {Phys. Rev. A}\ }\textbf {\bibinfo {volume} {76}},\ \bibinfo
  {pages} {022711} (\bibinfo {year} {2007})}\BibitemShut {NoStop}%
\bibitem [{\citenamefont {Helfrich}\ and\ \citenamefont
  {Hammer}(2011)}]{ThreeBodyTwoD}%
  \BibitemOpen
  \bibfield  {author} {\bibinfo {author} {\bibfnamefont {K.}~\bibnamefont
  {Helfrich}}\ and\ \bibinfo {author} {\bibfnamefont {H.-W.}\ \bibnamefont
  {Hammer}},\ }\href {\doibase 10.1103/PhysRevA.83.052703} {\bibfield
  {journal} {\bibinfo  {journal} {Phys. Rev. A}\ }\textbf {\bibinfo {volume}
  {83}},\ \bibinfo {pages} {052703} (\bibinfo {year} {2011})}\BibitemShut
  {NoStop}%
\bibitem [{\citenamefont {D'Incao}, \citenamefont {Anis},\ and\ \citenamefont
  {Esry}(2015)}]{ThreeBodyTwoD2015}%
  \BibitemOpen
  \bibfield  {author} {\bibinfo {author} {\bibfnamefont {J.~P.}\ \bibnamefont
  {D'Incao}}, \bibinfo {author} {\bibfnamefont {F.}~\bibnamefont {Anis}}, \
  and\ \bibinfo {author} {\bibfnamefont {B.~D.}\ \bibnamefont {Esry}},\ }\href
  {\doibase 10.1103/PhysRevA.91.062710} {\bibfield  {journal} {\bibinfo
  {journal} {Phys. Rev. A}\ }\textbf {\bibinfo {volume} {91}},\ \bibinfo
  {pages} {062710} (\bibinfo {year} {2015})}\BibitemShut {NoStop}%
\bibitem [{\citenamefont {Dalvit}, \citenamefont {Dziarmaga},\ and\
  \citenamefont {Zurek}(2000)}]{CatsThermal}%
  \BibitemOpen
  \bibfield  {author} {\bibinfo {author} {\bibfnamefont {D.~A.~R.}\
  \bibnamefont {Dalvit}}, \bibinfo {author} {\bibfnamefont {J.}~\bibnamefont
  {Dziarmaga}}, \ and\ \bibinfo {author} {\bibfnamefont {W.~H.}\ \bibnamefont
  {Zurek}},\ }\href {\doibase 10.1103/PhysRevA.62.013607} {\bibfield  {journal}
  {\bibinfo  {journal} {Phys. Rev. A}\ }\textbf {\bibinfo {volume} {62}},\
  \bibinfo {pages} {013607} (\bibinfo {year} {2000})}\BibitemShut {NoStop}%
\bibitem [{\citenamefont {Schelle}(2009)}]{ThermalInteractionsThesis}%
  \BibitemOpen
  \bibfield  {author} {\bibinfo {author} {\bibfnamefont {A.}~\bibnamefont
  {Schelle}},\ }\emph {\bibinfo {title} {{Environment-induced dynamics in a
  dilute Bose-Einstein condensate}}},\ \href
  {https://tel.archives-ouvertes.fr/tel-00438496} {Ph.D. thesis},\ \bibinfo
  {school} {{Universit{\'e} Pierre et Marie Curie - Paris VI}} (\bibinfo {year}
  {2009})\BibitemShut {NoStop}%
\bibitem [{\citenamefont {Gardiner}\ and\ \citenamefont
  {Zoller}(1998)}]{QuantumKineticTheoryIII}%
  \BibitemOpen
  \bibfield  {author} {\bibinfo {author} {\bibfnamefont {C.~W.}\ \bibnamefont
  {Gardiner}}\ and\ \bibinfo {author} {\bibfnamefont {P.}~\bibnamefont
  {Zoller}},\ }\href {\doibase 10.1103/PhysRevA.58.536} {\bibfield  {journal}
  {\bibinfo  {journal} {Phys. Rev. A}\ }\textbf {\bibinfo {volume} {58}},\
  \bibinfo {pages} {536} (\bibinfo {year} {1998})}\BibitemShut {NoStop}%
\bibitem [{\citenamefont {Jaksch}\ \emph
  {et~al.}(1998{\natexlab{b}})\citenamefont {Jaksch}, \citenamefont {Gardiner},
  \citenamefont {Gheri},\ and\ \citenamefont
  {Zoller}}]{QuantumKineticTheoryIV}%
  \BibitemOpen
  \bibfield  {author} {\bibinfo {author} {\bibfnamefont {D.}~\bibnamefont
  {Jaksch}}, \bibinfo {author} {\bibfnamefont {C.~W.}\ \bibnamefont
  {Gardiner}}, \bibinfo {author} {\bibfnamefont {K.~M.}\ \bibnamefont {Gheri}},
  \ and\ \bibinfo {author} {\bibfnamefont {P.}~\bibnamefont {Zoller}},\ }\href
  {\doibase 10.1103/PhysRevA.58.1450} {\bibfield  {journal} {\bibinfo
  {journal} {Phys. Rev. A}\ }\textbf {\bibinfo {volume} {58}},\ \bibinfo
  {pages} {1450} (\bibinfo {year} {1998}{\natexlab{b}})}\BibitemShut {NoStop}%
\bibitem [{\citenamefont {Leanhardt}\ \emph {et~al.}(2003)\citenamefont
  {Leanhardt}, \citenamefont {Pasquini}, \citenamefont {Saba}, \citenamefont
  {Schirotzek}, \citenamefont {Shin}, \citenamefont {Kielpinski}, \citenamefont
  {Pritchard},\ and\ \citenamefont {Ketterle}}]{05nKBEC}%
  \BibitemOpen
  \bibfield  {author} {\bibinfo {author} {\bibfnamefont {A.~E.}\ \bibnamefont
  {Leanhardt}}, \bibinfo {author} {\bibfnamefont {T.~A.}\ \bibnamefont
  {Pasquini}}, \bibinfo {author} {\bibfnamefont {M.}~\bibnamefont {Saba}},
  \bibinfo {author} {\bibfnamefont {A.}~\bibnamefont {Schirotzek}}, \bibinfo
  {author} {\bibfnamefont {Y.}~\bibnamefont {Shin}}, \bibinfo {author}
  {\bibfnamefont {D.}~\bibnamefont {Kielpinski}}, \bibinfo {author}
  {\bibfnamefont {D.~E.}\ \bibnamefont {Pritchard}}, \ and\ \bibinfo {author}
  {\bibfnamefont {W.}~\bibnamefont {Ketterle}},\ }\href {\doibase
  10.1126/science.1088827} {\bibfield  {journal} {\bibinfo  {journal}
  {Science}\ }\textbf {\bibinfo {volume} {301}},\ \bibinfo {pages} {1513}
  (\bibinfo {year} {2003})}\BibitemShut {NoStop}%
\bibitem [{\citenamefont {Bali}\ \emph {et~al.}(1999)\citenamefont {Bali},
  \citenamefont {O'Hara}, \citenamefont {Gehm}, \citenamefont {Granade},\ and\
  \citenamefont {Thomas}}]{VacuumRateAndHeating}%
  \BibitemOpen
  \bibfield  {author} {\bibinfo {author} {\bibfnamefont {S.}~\bibnamefont
  {Bali}}, \bibinfo {author} {\bibfnamefont {K.~M.}\ \bibnamefont {O'Hara}},
  \bibinfo {author} {\bibfnamefont {M.~E.}\ \bibnamefont {Gehm}}, \bibinfo
  {author} {\bibfnamefont {S.~R.}\ \bibnamefont {Granade}}, \ and\ \bibinfo
  {author} {\bibfnamefont {J.~E.}\ \bibnamefont {Thomas}},\ }\href {\doibase
  10.1103/PhysRevA.60.R29} {\bibfield  {journal} {\bibinfo  {journal} {Phys.
  Rev. A}\ }\textbf {\bibinfo {volume} {60}},\ \bibinfo {pages} {R29} (\bibinfo
  {year} {1999})}\BibitemShut {NoStop}%
\bibitem [{\citenamefont {{Paw{\l}owski}}\ and\ \citenamefont {{Rz{\c
  a}{\.z}ewski}}(2010)}]{VacuumMaterEq}%
  \BibitemOpen
  \bibfield  {author} {\bibinfo {author} {\bibfnamefont {K.}~\bibnamefont
  {{Paw{\l}owski}}}\ and\ \bibinfo {author} {\bibfnamefont {K.}~\bibnamefont
  {{Rz{\c a}{\.z}ewski}}},\ }\href {\doibase 10.1103/PhysRevA.81.013620}
  {\bibfield  {journal} {\bibinfo  {journal} {Phys. Rev. A}\ }\textbf {\bibinfo
  {volume} {81}},\ \bibinfo {pages} {013620} (\bibinfo {year}
  {2010})}\BibitemShut {NoStop}%
\bibitem [{\citenamefont {Pichler}, \citenamefont {Daley},\ and\ \citenamefont
  {Zoller}(2010)}]{DecoherencePhotonEmission}%
  \BibitemOpen
  \bibfield  {author} {\bibinfo {author} {\bibfnamefont {H.}~\bibnamefont
  {Pichler}}, \bibinfo {author} {\bibfnamefont {A.~J.}\ \bibnamefont {Daley}},
  \ and\ \bibinfo {author} {\bibfnamefont {P.}~\bibnamefont {Zoller}},\ }\href
  {\doibase 10.1103/PhysRevA.82.063605} {\bibfield  {journal} {\bibinfo
  {journal} {Phys. Rev. A}\ }\textbf {\bibinfo {volume} {82}},\ \bibinfo
  {pages} {063605} (\bibinfo {year} {2010})}\BibitemShut {NoStop}%
\bibitem [{\citenamefont {Ferrini}\ \emph {et~al.}(2011)\citenamefont
  {Ferrini}, \citenamefont {Spehner}, \citenamefont {Minguzzi},\ and\
  \citenamefont {Hekking}}]{PhaseLaserNoiseBEC}%
  \BibitemOpen
  \bibfield  {author} {\bibinfo {author} {\bibfnamefont {G.}~\bibnamefont
  {Ferrini}}, \bibinfo {author} {\bibfnamefont {D.}~\bibnamefont {Spehner}},
  \bibinfo {author} {\bibfnamefont {A.}~\bibnamefont {Minguzzi}}, \ and\
  \bibinfo {author} {\bibfnamefont {F.~W.~J.}\ \bibnamefont {Hekking}},\ }\href
  {\doibase 10.1103/PhysRevA.84.043628} {\bibfield  {journal} {\bibinfo
  {journal} {Phys. Rev. A}\ }\textbf {\bibinfo {volume} {84}},\ \bibinfo
  {pages} {043628} (\bibinfo {year} {2011})}\BibitemShut {NoStop}%
\bibitem [{\citenamefont {Tiecke}\ \emph {et~al.}(2003)\citenamefont {Tiecke},
  \citenamefont {Kemmann}, \citenamefont {Buggle}, \citenamefont {Shvarchuck},
  \citenamefont {von Klitzing},\ and\ \citenamefont
  {Walraven}}]{BFieldDoubleWellTrap}%
  \BibitemOpen
  \bibfield  {author} {\bibinfo {author} {\bibfnamefont {T.~G.}\ \bibnamefont
  {Tiecke}}, \bibinfo {author} {\bibfnamefont {M.}~\bibnamefont {Kemmann}},
  \bibinfo {author} {\bibfnamefont {C.}~\bibnamefont {Buggle}}, \bibinfo
  {author} {\bibfnamefont {I.}~\bibnamefont {Shvarchuck}}, \bibinfo {author}
  {\bibfnamefont {W.}~\bibnamefont {von Klitzing}}, \ and\ \bibinfo {author}
  {\bibfnamefont {J.~T.~M.}\ \bibnamefont {Walraven}},\ }\href
  {http://stacks.iop.org/1464-4266/5/i=2/a=368} {\bibfield  {journal} {\bibinfo
   {journal} {Journal of Optics B: Quantum and Semiclassical Optics}\ }\textbf
  {\bibinfo {volume} {5}},\ \bibinfo {pages} {S119} (\bibinfo {year}
  {2003})}\BibitemShut {NoStop}%
\bibitem [{\citenamefont {Schumm}(2006)}]{SchummThesis}%
  \BibitemOpen
  \bibfield  {author} {\bibinfo {author} {\bibfnamefont {T.}~\bibnamefont
  {Schumm}},\ }\emph {\bibinfo {title} {{Bose-Einstein condensates in magnetic
  double well potentials}}},\ \href
  {https://pastel.archives-ouvertes.fr/tel-00129501} {Ph.D. thesis},\ \bibinfo
  {school} {{Universit{\'e} Paris Sud - Paris XI}} (\bibinfo {year}
  {2006})\BibitemShut {NoStop}%
\bibitem [{\citenamefont {Ferrini}\ \emph {et~al.}(2010)\citenamefont
  {Ferrini}, \citenamefont {Spehner}, \citenamefont {Minguzzi},\ and\
  \citenamefont {Hekking}}]{MagNoiseInDoubleWell}%
  \BibitemOpen
  \bibfield  {author} {\bibinfo {author} {\bibfnamefont {G.}~\bibnamefont
  {Ferrini}}, \bibinfo {author} {\bibfnamefont {D.}~\bibnamefont {Spehner}},
  \bibinfo {author} {\bibfnamefont {A.}~\bibnamefont {Minguzzi}}, \ and\
  \bibinfo {author} {\bibfnamefont {F.~W.~J.}\ \bibnamefont {Hekking}},\ }\href
  {\doibase 10.1103/PhysRevA.82.033621} {\bibfield  {journal} {\bibinfo
  {journal} {Phys. Rev. A}\ }\textbf {\bibinfo {volume} {82}},\ \bibinfo
  {pages} {033621} (\bibinfo {year} {2010})}\BibitemShut {NoStop}%
\bibitem [{\citenamefont {Ghirardi}, \citenamefont {Pearle},\ and\
  \citenamefont {Rimini}(1990)}]{CSL}%
  \BibitemOpen
  \bibfield  {author} {\bibinfo {author} {\bibfnamefont {G.~C.}\ \bibnamefont
  {Ghirardi}}, \bibinfo {author} {\bibfnamefont {P.}~\bibnamefont {Pearle}}, \
  and\ \bibinfo {author} {\bibfnamefont {A.}~\bibnamefont {Rimini}},\ }\href
  {\doibase 10.1103/PhysRevA.42.78} {\bibfield  {journal} {\bibinfo  {journal}
  {Phys. Rev. A}\ }\textbf {\bibinfo {volume} {42}},\ \bibinfo {pages} {78}
  (\bibinfo {year} {1990})}\BibitemShut {NoStop}%
\bibitem [{\citenamefont {Ng}\ and\ \citenamefont
  {Geller}(1969)}]{ErrorFunctionIdentities}%
  \BibitemOpen
  \bibfield  {author} {\bibinfo {author} {\bibfnamefont {E.~W.}\ \bibnamefont
  {Ng}}\ and\ \bibinfo {author} {\bibfnamefont {M.}~\bibnamefont {Geller}},\
  }\href@noop {} {\bibfield  {journal} {\bibinfo  {journal} {Journal of
  Research of the National Bureau of Standards B}\ }\textbf {\bibinfo {volume}
  {73}},\ \bibinfo {pages} {1} (\bibinfo {year} {1969})}\BibitemShut {NoStop}%
\bibitem [{\citenamefont {Wang}(1988)}]{SpheroidGravPotExt}%
  \BibitemOpen
  \bibfield  {author} {\bibinfo {author} {\bibfnamefont {W.~X.}\ \bibnamefont
  {Wang}},\ }\href {http://stacks.iop.org/0305-4470/21/i=22/a=026} {\bibfield
  {journal} {\bibinfo  {journal} {Journal of Physics A: Mathematical and
  General}\ }\textbf {\bibinfo {volume} {21}},\ \bibinfo {pages} {4245}
  (\bibinfo {year} {1988})}\BibitemShut {NoStop}%
\bibitem [{\citenamefont {Wang}(1989)}]{SpheroidGravPotInside}%
  \BibitemOpen
  \bibfield  {author} {\bibinfo {author} {\bibfnamefont {W.~X.}\ \bibnamefont
  {Wang}},\ }\href {http://stacks.iop.org/0305-4470/22/i=9/a=031} {\bibfield
  {journal} {\bibinfo  {journal} {Journal of Physics A: Mathematical and
  General}\ }\textbf {\bibinfo {volume} {22}},\ \bibinfo {pages} {1459}
  (\bibinfo {year} {1989})}\BibitemShut {NoStop}%
\end{thebibliography}%

\end{document}